\documentclass[iop,twocolumn,numberedappendix]{emulateapj}
\pdfoutput=1

\usepackage{amsmath,amsfonts,amssymb,mathrsfs,graphicx,color}
\usepackage{appendix}
\usepackage{ulem}
\usepackage{natbib}

\graphicspath{{./figures/}}

\newcommand{\ie}{{\it i.e.}}

\newcommand{\eg}{{\it e.g.}}

\newcommand{\eq}{Eq.}
\newcommand{\eqs}{Eqs.}

\newcommand{\fig}{Fig.}
\newcommand{\Fig}{Fig.}

\newcommand{\Sec}{Section}

\newcommand{\App}{Appendix}

\newcommand{\Tab}{Table}

\newcommand{\equ}[1]{\eq~(\ref{equ:#1})}
\newcommand{\figu}[1]{\fig~\ref{fig:#1}}

\newcommand{\bi}{\begin{itemize}}
\newcommand{\ei}{\end{itemize}}

\def\apj{Astrophys.\ J.}
\def\apjl{Astrophys.\ J.}
\def\apjs{Astrophys.\ J.\ Supp.}
\def\aap{Astron.\ Astrophys.}

\shorttitle{Multi-messenger light curves from gamma-ray bursts in the internal shock model}
\shortauthors{Bustamante, Heinze, Murase \& Winter}

\begin{document}

\title{Multi-messenger light curves from gamma-ray bursts in the internal shock model}

\author{Mauricio Bustamante\altaffilmark{1,2}, Jonas Heinze\altaffilmark{3}, Kohta Murase\altaffilmark{4,5} \& Walter Winter\altaffilmark{3}}
\affil{
$^{1}$ Center for Cosmology and AstroParticle Physics (CCAPP), The Ohio State University, Columbus, OH 43210, USA\\
$^{2}$ Department of Physics, The Ohio State University, Columbus, OH 43210, USA\\
$^{3}$ Deutsches Elektronen-Synchrotron (DESY), Platanenallee 6, 15738 Zeuthen, Germany\\
$^{4}$ Center for Particle and Gravitational Astrophysics, The Pennsylvania State University, University Park, PA16802, USA\\
$^{5}$ Department of Astronomy \& Astrophysics, The Pennsylvania State University, University Park, PA16802, USA\\
bustamanteramirez.1@osu.edu, jonas.heinze@desy.de, walter.winter@desy.de, murase@psu.edu
}

% \date{\today}
% \date{January 27, 2017}

\begin{abstract}

Gamma-ray bursts (GRBs) are promising as sources of neutrinos and cosmic rays. In the internal shock scenario, blobs of plasma emitted from a central engine collide within a relativistic jet and form shocks, leading to particle acceleration and emission. Motivated by present experimental constraints and sensitivities, we improve the predictions of particle emission by investigating time-dependent effects from multiple shocks.
We produce synthetic light curves with different variability timescales that stem from properties of the central engine. For individual GRBs, qualitative conclusions about model parameters, neutrino production efficiency, and delays in high-energy gamma rays can be deduced from inspection of the gamma-ray light curves. GRBs with fast time variability without additional prominent pulse structure tend to be efficient neutrino emitters, whereas GRBs with fast variability modulated by a broad pulse structure can be inefficient neutrino emitters and produce delayed high-energy gamma-ray signals. Our results can be applied to quantitative tests of the GRB origin of ultra-high-energy cosmic rays, and have the potential to impact current and future multi-messenger searches. 

\end{abstract}

\keywords{Gamma-ray burst: general --- Neutrinos --- Astroparticle physics --- Methods: numerical}

\maketitle

%%%%%%%%%%%%%%%%%%%%%%%%%%%%%%%%%%%%%%%%%%%%%%%%%%%%%%%%%%%%%%%%%%%%%%%%%%%%%%%%%%%
%%%%%%%%%%%%%%%%%%%%%%%%%%%%%%%%%%%%%%%%%%%%%%%%%%%%%%%%%%%%%%%%%%%%%%%%%%%%%%%%%%%

\section{Introduction}

The most energetic particles discovered --- ultra-high-energy cosmic rays (UHECRs), above $10^9$ GeV, and high-energy astrophysical neutrinos, above 100 TeV --- are detected with regularity, but their sources remain unknown. The extreme physical conditions required to reach these energies restrict the potential source classes. Gamma-ray bursts (GRBs) are one such class. They are the most luminous transient electromagnetic astrophysical phenomena: they emit gamma rays up to $\sim 100$ GeV, concentrated in only a few tens of seconds. Their observed high luminosities and inferred high particle densities make them prime candidate sources of UHECRs and neutrinos. 

One of the leading explanations of GRB emission is the fireball model~\citep{Rees:1992ek}. In it, a central engine --- likely, a black hole --- injects plasma blobs with different relativistic speeds into the jet flow.  When they collide with one another, they create mildly relativistic shocks in which their kinetic energy is transformed into internal energy that is radiated away as high-energy particles~\citep{Rees:1994nw,Paczynski:1994uv}. If the jets contain enough baryons, then protons~\citep{Milgrom:1995um,Waxman:1995vg,Vietri:1995hs} and nuclei~\citep{Murase:2008mr,Wang:2007xj} can be accelerated by the shocks into UHECRs which, upon interacting with source photons, create pions and other mesons. Their decays produce PeV gamma rays and neutrinos~\citep{Waxman:1997ti}. 

In recent years, IceCube has started testing the hypothesis of GRBs as sources of high-energy neutrinos~\citep{Abbasi:2012zw,Aartsen:2014aqy,Aartsen:2016qcr}. The simplest versions of the fireball model are now in tension with the data. In these, the neutrino flux prediction for a burst is based on the emission from a single representative plasma blob collision in the jet --- this is known as the one-zone approach.

While high-luminosity GRBs cannot account for the diffuse high-energy astrophysical neutrino signal detected by IceCube~\citep{Tamborra:2015qza,Aartsen:2016qcr}, they persist as interesting objects because of the following two reasons. First, they can still be the sources of UHECRs. This can happen if UHECR emission and neutrino emission are not tightly correlated~\citep{Baerwald:2013pu}, a possibility that we consider in the present work. Second, they are attractive targets for neutrino telescopes, as timing and directional gamma-ray information can be used to reduce neutrino backgrounds. They are ideal sites to look for joint gamma-ray and neutrino signals.

In this paper, we focus on the connection between gamma rays, UHECRs, and neutrinos in individual GRBs. Unlike most of the existing literature, we consider the multi-messenger emission from a multitude of different plasma collisions along the jet, each one occurring under different physical conditions --- this is known as the multi-zone approach. 

We construct synthetic GRB light curves --- curves of photon rate versus time --- that successfully reproduce generic features of real ones. We identify the properties of the central engine that lead to such light curves, and study the consequences for different messengers. For example, light curves dominated by broad pulses overlaid with fast variability imply a ``disciplined'' engine that ejects shells with little spread in speed at any given time. In such cases, the neutrino production efficiency can be low, and high-energy gamma-ray signals can reach Earth delayed with respect to low-energy signals. Conversely, light curves with no broad pulse structure hint at likely efficient neutrino emitters. We also test the robustness of the assumptions going into the minimal neutrino flux estimate found in~\citet{Bustamante:2014oka}.

This paper is organized as follows. In \Sec~\ref{sec:GRBEmissionModels}, we comment on GRB emission models. In \Sec~\ref{sec:SimulatingCollisions}, we introduce our multi-zone simulation and a sample of GRBs computed with it. In \Sec~\ref{sec:SyntheticLightCurves}, we compute the gamma-ray and neutrino light curves for each of them. In \Sec~\ref{sec:MultiMessengerEmission} we calculate their associated quasi-diffuse neutrino fluxes and show that different types of particles are emitted from different regions in the jet. In \Sec~\ref{sec:delay}, we show that, for some GRBs, delays between the light curves in different energy bands are possible, and why. We summarize and conclude in \Sec~\ref{sec:Conclusions}. Appendix~\ref{sec:model} contains a detailed presentation of the multi-zone collision model used in our simulations. Appendix~\ref{sec:alt} contains a discussion of the impact of alternative assumptions for the collision dynamics.

%%%%%%%%%%%%%%%%%%%%%%%%%%%%%%%%%%%%%%%%%%%%%%%%%%%%%%%%%%%%%%%%%%%%%%%%%%%%%%%%%%%
%%%%%%%%%%%%%%%%%%%%%%%%%%%%%%%%%%%%%%%%%%%%%%%%%%%%%%%%%%%%%%%%%%%%%%%%%%%%%%%%%%%

\section{One-zone vs.\ multi-zone emission}\label{sec:GRBEmissionModels}

GRBs are conceivably fueled by matter accretion, either in the collapse of a massive star --- long-duration bursts, lasting $> 2$ s --- or in the merging of two neutron stars or a neutron star and a black hole --- short-duration bursts, lasting $< 2$ s. The central engine, likely a newly formed black hole, emits two relativistic matter jets in opposite directions. When one of them points towards Earth, we might detect it as a GRB. We focus on long-duration bursts because they have higher gamma-ray luminosity ($\sim 10^{52}$ erg s$^{-1}$), represent $\sim 75\%$ of observed bursts, and their emission mechanism has been more deeply studied. 

In the internal shock scenario of the fireball model, internal plasma collisions within the jet account for particle emission during the initial, or prompt, phase of the burst, which typically lasts 10--100 s. Ultimately, the jet reaches the circumburst medium, triggering a late emission phase --- the afterglow. We will focus exclusively on the prompt phase, where PeV neutrino emission is well-motivated. This is the energy range where existing water-Cherenkov neutrino telescopes, like IceCube and ANTARES, are most sensitive. However, the internal shock scenario, while successful, is not without issues, which we introduce below.

The GRB prompt emission mechanism has been under debate for years (see reviews by ~\citet{Piran:1999kx,Meszaros:2006rc,Kumar:2014upa,Meszaros:2015zka,Pe'er:2015rfa}). 
In the classical model, prompt gamma-ray emission in the MeV range is attributed to optically-thin synchrotron emission from non-thermal electrons accelerated at internal shocks~\citep{Rees:1994nw}, possibly supplemented by a thermal component from the fireball. Indeed, there are observational indications of such a thermal component
(see, \eg, \citet{Pe'er:2012ja,2011ApJ...727L..33G}).
However, this classical internal shock model has difficulties in explaining the data, including the low-energy spectral index~\citep{Preece:1998jy}, radiation efficiency~\citep{Kumar:1999cv,Zhang:2006uj}, and spectral energy relations~\citep{Amati:2002ny,Yonetoku:2003gi}. 

Among alternative theories, photospheric emission models have become popular after the \textit{Fermi} satellite was launched~\citep{Thompson:1994zh,Rees:2004gt,Pe'er:2005kz,Ioka:2007qk,Giannios:2007yj,Beloborodov:2009be,Lazzati:2013ym}. In them, the bulk of the prompt emission is attributed to quasi-thermal emission from below the photosphere, \ie, from the region where gamma rays are unable to escape due to high optical depth to electron-photon scattering.
Amid different versions of photospheric emission models, dissipative photosphere models have been discussed both theoretically and observationally, due to their appealing implications~\citep{Beloborodov:2012ys,Hascoet:2013zz,2013ApJ...764..143V}. Some authors consider the superposition of pure thermal emission to explain the prompt emission spectrum~\citep{Lundman:2012ic}.  

Magnetic reconnection models have also been largely of interest~\citep{Meszaros:1996ww,Lyutikov:2003ih,Bosnjak:2011pt,McKinney:2010bn,Zhang:2010jt}, due to certain advantages~\citep{Kumar:2014upa}, though detailed microphysics is much more uncertain. For example, they can explain the high polarization observed in some GRBs~\citep{Yonetoku:2012wz,Gotz:2009ak} and the absence of bright thermal emission~\citep{2009ApJ...700L..65Z,Gao:2014bfa}.

Even though the classical internal shock model has several theoretical issues, internal shocks may still play an important role in the dissipation of the outflowing kinetic energy. For instance, some photospheric emission models require sub-photospheric dissipation that is often attributed to internal shocks~\citep{Rees:2004gt}. Magnetic reconnection may also be driven by shocks beyond the photosphere~\citep{Zhang:2010jt}. Even in the optically-thin internal shock model, the observed spectra can be explained by synchrotron emission with some modifications such as stochastic acceleration of electrons in the shock downstream~\citep{Bykov:1996vm,Murase:2011cx,Asano:2015oia}.

In the internal shock model, shell collisions inside the jet are responsible for the shape of the GRB light curves. They typically have a fast time variability $t_\text{v}$, of $\sim$ 10--100 ms ~\citep{Bhat:2013fsa,Golkhou:2015lsa}, frequently superposed on top of slower pulse structure~\citep{Nakar:2001iz,RamirezRuiz:1999fr}. Since the light curve reflects the particularities of the central emitter and jet propagation~\citep{Nakar:2002gd,Zhang:2016mgs,Loopez-Camara:2016obp}, no two are equal, though they can be classified on the basis of their morphology. 

One-zone collision models assume average shell properties --- such as an average speed or Lorentz factor $\langle \Gamma \rangle$ --- derived from gamma-ray observations. Neutrino emission is computed for a single representative collision~\citep{Dermer:2003zv,Guetta:2003wi} and is scaled by the total number of collisions --- roughly, $T_{90} / t_\text{v} \sim 1000$, with $T_{90} \sim 10$ s the burst duration  --- to yield the flux for the whole burst, implying that all collisions are identical (see, \eg, \citet{Asano:2005wb}).

However, it is unrealistic that all collisions occur at the same radius. The main reason is that all shells would have to be emitted with precisely tuned speeds. A different reason comes from the fact that energy deposited in the afterglow cannot be too large; therefore, a sizable part of the energy must be dissipated during the preceding prompt phase. Because dissipation of kinetic energy into radiation scales with the difference of the Lorentz factors of the colliding shells, this implies that the differences must be large. In fact, there is considerable spread in the inferred values of bulk Lorentz factors of GRBs observed by the {\it Fermi} satellite~\citep{Ackermann:2012cma}. Assuming for the Lorentz factors a broad or bi-modal distribution leads to efficient dissipation of kinetic energy in the prompt phase~\citep{Kumar:1999cv,Zhang:2006uj}. As a consequence, collisions occur at many positions throughout the jet, from below the photosphere --- where gamma rays cannot escape --- to the circumburst medium, and that particle emission will be different for each collision.

The impact of a distribution\footnote{\citet{Guetta:2003wi,Becker:2005ej,Stamatikos:2006jy,Baerwald:2011ee} studied the complementary problem of calculating, in the one-zone approach, the expected neutrino flux from individual GRBs using their observed electromagnetic properties, and the impact of distributions of model parameter values on the quasi-diffuse flux.} of Lorentz factors or emission radii have been addressed qualitatively~\citep{Murase:2005hy} and quantitatively~\citep{Guetta:2000ye,Guetta:2001cd,Bustamante:2014oka,Globus:2014fka}. For example, \citet{Bustamante:2014oka}  predicted a minimal quasi-diffuse prompt neutrino flux from super-photospheric collisions, at the level of $\sim 10^{-11} \, \mathrm{GeV \, cm^{-2} \, s^{-1} \, sr^{-1}}$ per flavor, which is presently below the sensitivity of IceCube. In later sections, we discuss the relationship among different types of messengers in multiple-collision models, focusing on how the properties of the central engine impact the GRB light curves and neutrino emission efficiency.
 
IceCube recently discovered a diffuse astrophysical flux of neutrinos between about 30 TeV and 2 PeV~\citep{Aartsen:2013bka,Aartsen:2013jdh,Aartsen:2013eka,Aartsen:2014gkd,Aartsen:2015rwa,Aartsen:2015xup}. However, no individual neutrinos are associated to known GRBs~\citep{Abbasi:2012zw,Aartsen:2014aqy,Aartsen:2016qcr}.  
The resulting upper limit, extrapolated from a catalog of bursts, is about an order of magnitude below the diffuse flux. Thus, it is unlikely that classical high-luminosity GRBs are the main origin of the observed IceCube neutrinos~\citep{Murase:2013ffa,Laha:2013eev,Bustamante:2014oka}. (However, low-luminosity classes of GRBs can describe the diffuse neutrino flux without violating the stacking limits~\citep{Murase:2006mm,Gupta:2006jm}, subject to their assumed source density.)

Nevertheless, this does not preclude GRBs from being the dominant sources of UHECRs.
So far, results of these observational searches have been interpreted mostly in the context of one-zone models~\citep{Waxman:1997ti,Guetta:2003wi,Li:2011ah,Hummer:2011ms,He:2012tq}. While the plain one-zone ansatz with commonly assumed parameters favorable for the UHECR explanation is starting to be challenged by IceCube data~\citep{Hummer:2011ms,He:2012tq}, it is not fully ruled out yet.

Another important aspect is that the relationship between neutrino and cosmic-ray emission depends on the UHECR escape mechanism~\citep{Baerwald:2013pu}. It has been shown that, depending on the parameter values, a fraction of cosmic rays must directly escape from the sources without neutrino production~\citep{Baerwald:2013pu}. Earlier approaches focused on the emission of neutrons only~\citep{Ahlers:2011jj}, which leads to a stringent relation between the neutrino and cosmic-ray fluxes. Consequently, bounds on GRB neutrinos translate into bounds on the GRB parameter space; see~\citet{Baerwald:2014zga} for the bounds in the one-zone model. Further quantitative tests are necessary to test the possibility that UHECRs mainly come from GRBs, especially taking into account the effects from multiple collisions.

%%%%%%%%%%%%%%%%%%%%%%%%%%%%%%%%%%%%%%%%%%%%%%%%%%%%%%%%%%%%%%%%%%%%%%%%%%%%%%%%%%%
%%%%%%%%%%%%%%%%%%%%%%%%%%%%%%%%%%%%%%%%%%%%%%%%%%%%%%%%%%%%%%%%%%%%%%%%%%%%%%%%%%%

\section{Simulating collisions in a GRB jet}\label{sec:SimulatingCollisions}

\begin{figure}[t]
 \centering
 \includegraphics[width=\columnwidth]{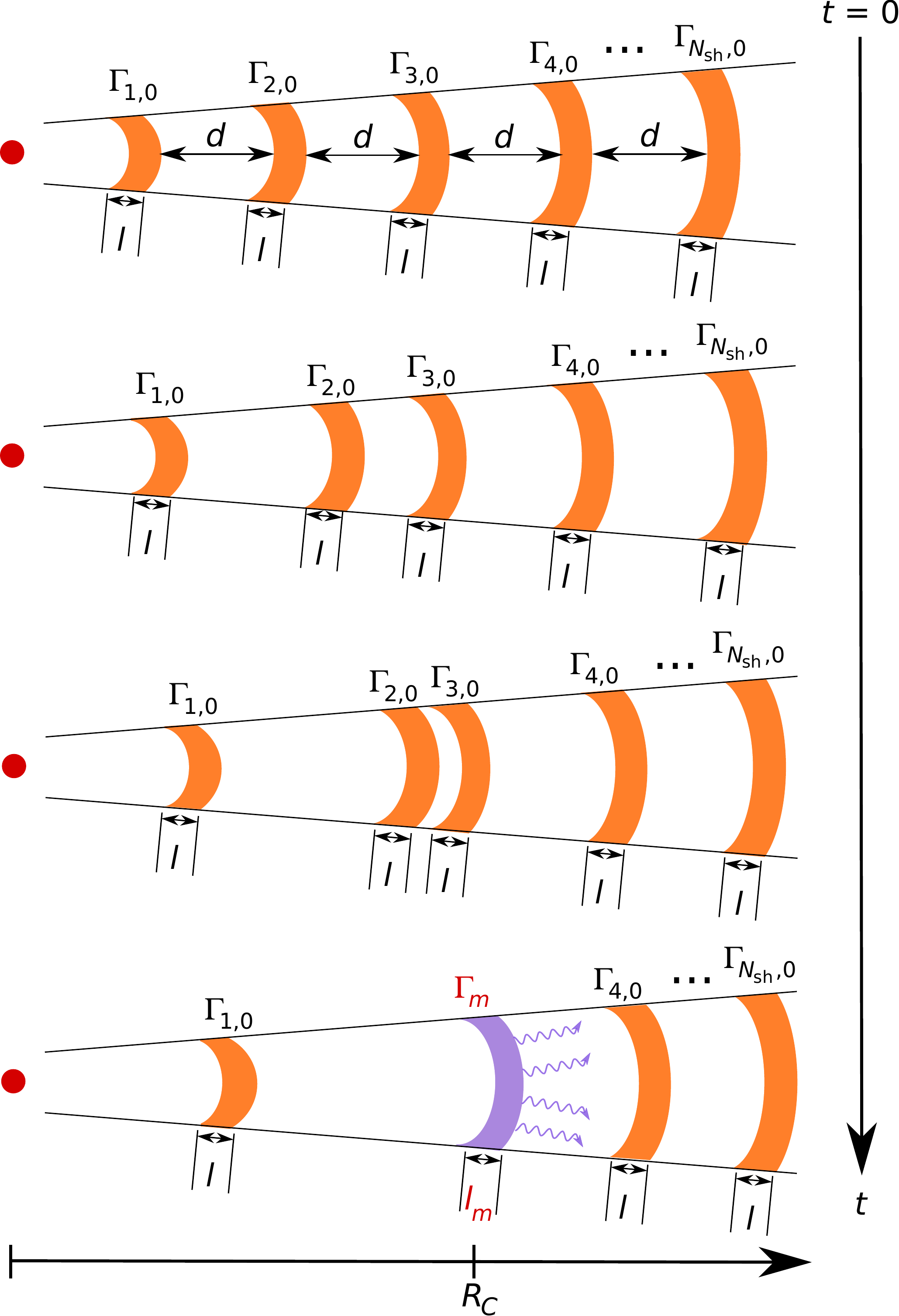}
 \caption{\label{fig:schematic_evol} Illustration of the initialization, evolution, and collision of plasma shells in our simulations. See main text and Appendix~\ref{sec:model} for details.}
\end{figure}

We study the prompt emission phase of the GRB in the internal shock scenario by simulating collisions of many propagating plasma blobs. Our simulations are based on~\citet{Kobayashi:1997jk,Daigne:1998xc}.  See Appendix~\ref{sec:model} for details.

Figure~\ref{fig:schematic_evol} illustrates the evolution of the blobs along the jet. To simplify the computations, we take them to be spherical shells. The simulation starts with the central engine emitting $N_{\mathrm{sh}} \sim 1000$ shells. Each one propagates at a different relativistic speed; typically, they have Lorentz factors $\Gamma \sim 100$. After some time, they catch up to one another, collide inelastically, and merge into new shells. At collisionless shocks generated during the collisions, a fraction of the kinetic energy of the colliding shells is used for particle acceleration and creation, and becomes the internal energy of the newly created shells. The new shells cool instantly by emitting high-energy particles, continue propagating in the jet, and may collide again. Each simulation in this work comprises about 1000 collisions, as almost all initial shells collide.

Non-thermal electrons receive a fraction $\epsilon_e$ of the internal energy of the new shell, protons receive a fraction $\epsilon_p$, and magnetic fields receive a fraction $\epsilon_B$.  Because electrons cool fast via synchrotron emission, $\epsilon_e$ can also be regarded as the energy fraction of radiated gamma rays.  We assume that there is energy equipartition between electrons and magnetic fields, and ten times more energy in protons\ \citep{Abbasi:2009ig, Hummer:2011ms}, \ie, $\epsilon_e = \epsilon_B = 1/12$ and $\epsilon_p / \epsilon_e = 10$; see \Sec\ \ref{section:LightCurves}.  Thanks to the fast cooling, $\epsilon_p / \epsilon_e$ can be regarded as the non-thermal baryon loading factor, defined as the ratio of cosmic-ray energy to radiation energy\ \citep{Murase:2005hy}.

The volume of the shell (see \Sec~\ref{section:FireballEvolution}) grows with the distance $r$ to the central emitter $\propto r^2$ until the radial expansion of a shell becomes important at the shell spreading radius. Except in collisions, the number of particles in a shell is conserved. So, as a shell propagates and expands the density of particles in it falls $\propto r^{-2}$.  This affects where in the jet different types of particles are produced and emitted: neutrinos come predominantly from close to the photosphere, where densities are high; ultra-high-energy cosmic rays come from intermediate collision radii; and high-energy gamma rays escape abundantly from large collision radii~\citep{Bustamante:2014oka}.

In a shell collision, if both protons and electrons are accelerated up to sufficiently high energies, $p\gamma$ interactions create pions via the $\Delta^+(1232)$ resonance and other processes. High-energy gamma rays are produced in $\pi^0 \to \gamma \gamma$ (and by synchrotron and inverse-Compton radiation by electrons), while high-energy neutrinos are produced in $\pi^+ \to \mu^+ \nu_\mu \to \bar{\nu}_\mu e^+ \nu_e \nu_\mu$ (and, via additional production channels, by its charge-conjugated version). The pion production efficiency is, on average, $20\%$; this is the fraction of proton energy received by pions. It is distributed roughly evenly among the pion decay products, so each neutrino carries $\sim 5\%$ of the proton energy. Therefore, production of PeV neutrinos requires 20 PeV protons.

At the source, for the protons we assume a power-law spectrum $\propto E_p^{\prime -2} \exp\left( -E_p^\prime / E^\prime_{p,\max} \right)$, with $E_p^\prime$ the proton energy, as expected from Fermi acceleration (primed quantities are in the shock rest frame). The maximum proton energy $E^\prime_{p,\max}$ is computed by balancing the acceleration rate, and the synchrotron, adiabatic, and photohadronic energy-loss rates~\citep{Baerwald:2013pu}. 

For the target photons, we assume a broken power-law spectrum, \equ{PhotonSpectrum}, which resembles the gamma-ray spectrum observed at Earth. This assumption is certainly justified beyond the photosphere, from where the photons can escape. Note that we neither generate the photon spectrum from first principles, nor explain its origin, which typically includes radiation processes such as inverse Compton scattering, synchrotron emission, and bremsstrahlung. The Band function can be reproduced by invoking various effects, such as slow heating of electrons~\citep{Bykov:1996vm,Murase:2011cx,Asano:2015oia}. In this sense, our approach is model-independent; this comes at the expense of insight on the radiation processes at work and their detailed individual implications on secondary production.
To calculate the secondary particle production, we use the state-of-the-art NeuCosmA~\citep{Hummer:2011ms} software (see \Sec~\ref{sec:neutrinos}).

UHECRs are emitted via two mechanisms: either $p\gamma$ interactions transform protons into neutrons which escape the merged shell, and later beta-decay back into protons~\citep{Mannheim:1998wp}, or protons directly leak out of the shell without interacting. The latter situation occurs if the proton Larmor radius --- determined by the physical conditions in the shell --- exceeds the shell width at the highest energies; see~\citet{Baerwald:2013pu} for a detailed discussion. This direct escape produces a hard proton spectrum (it is dubbed a ``high-pass filter'' in \citet{Globus:2014fka}). Direct proton escape dominates over neutron escape only when the photon densities are low. Therefore, neutrino production associated to neutron escape is higher than that associated to direct proton escape~\citep{Mannheim:1998wp}. 

Gamma rays are produced throughout the jet. However, they only escape if they are produced above the ``photosphere'', \ie, the radius below which Thomson scattering occurs frequently enough to effectively trap them (see \Sec~\ref{sec:neutrinos}). We adopt the pragmatic point of view that solid predictions about GRB dynamics and secondary production must be grounded in gamma-ray observations, which come exclusively from above the photosphere. The drawback of this assumption is that we cannot accurately calculate secondary production below the photosphere, since the sub-photospheric photons are subject to significant thermalization and the observed spectrum becomes quasi-thermal~\citep{Beloborodov:2012ys,Hascoet:2013zz,2013ApJ...764..143V}. For our calculations of the gamma-ray flux, we will use only super-photospheric collisions. 
For neutrinos, we will additionally estimate possible sub-photospheric contributions. Photomeson production is more efficient below the photosphere, so that neutrino emission around the photosphere can be dominant, especially in photospheric emission models~\citep{Murase:2008sp,Wang:2008zm,Murase:2013hh,Bartos:2013hf,Bustamante:2014oka}.

\begin{figure*}[t]
 \centering
 \includegraphics[width=0.23\textwidth]{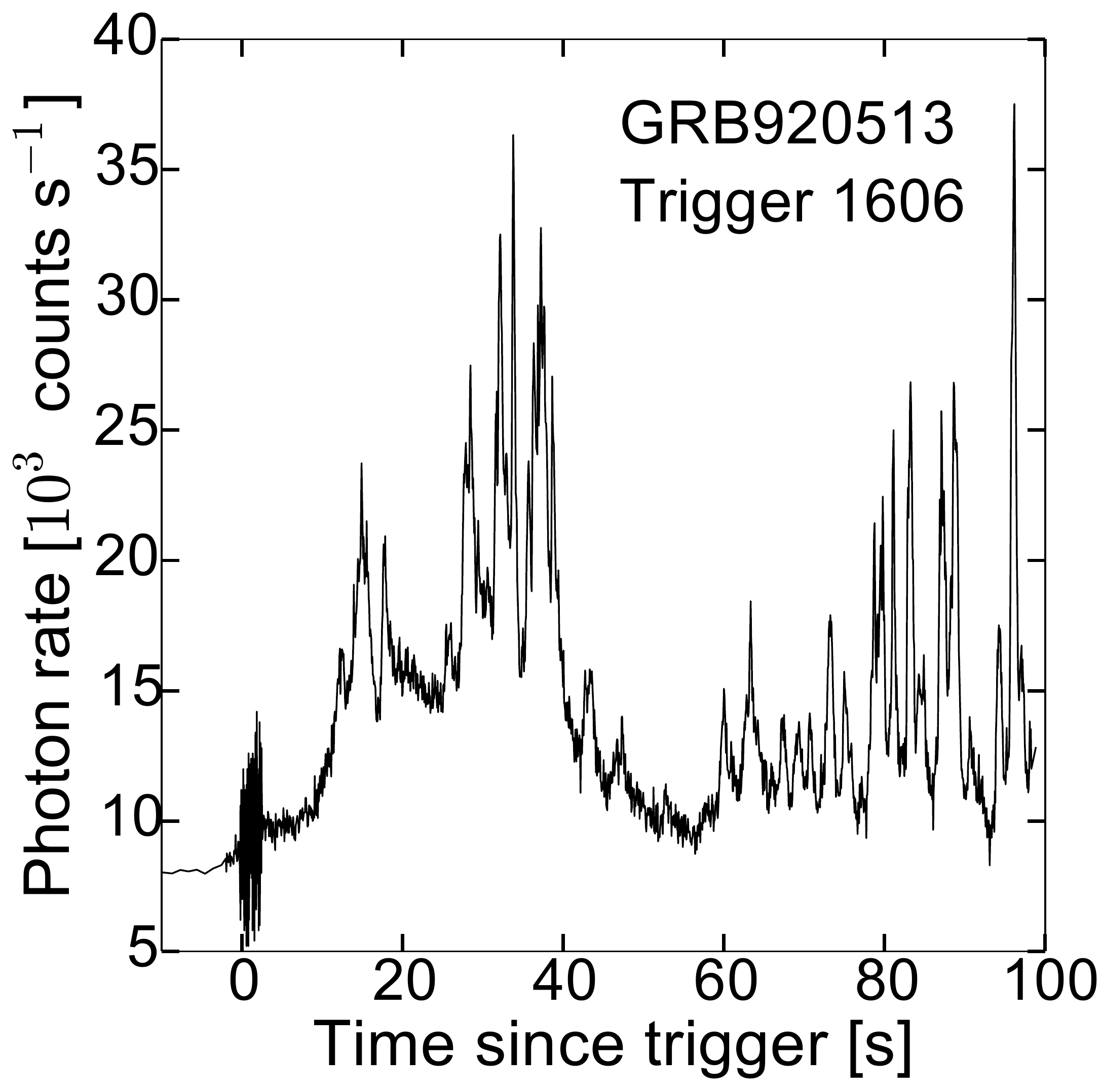}
 \includegraphics[width=0.23\textwidth]{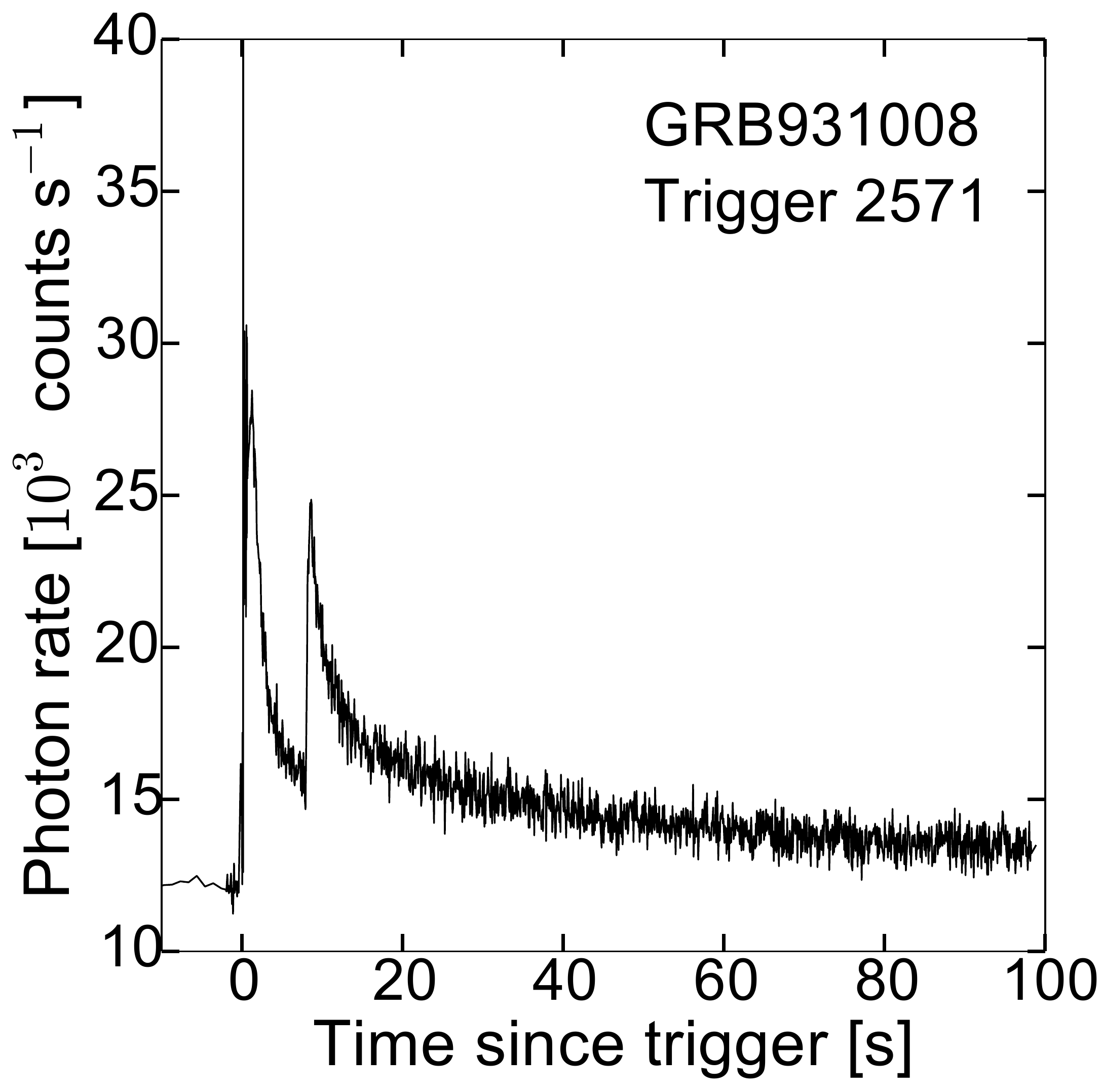}
 \includegraphics[width=0.23\textwidth]{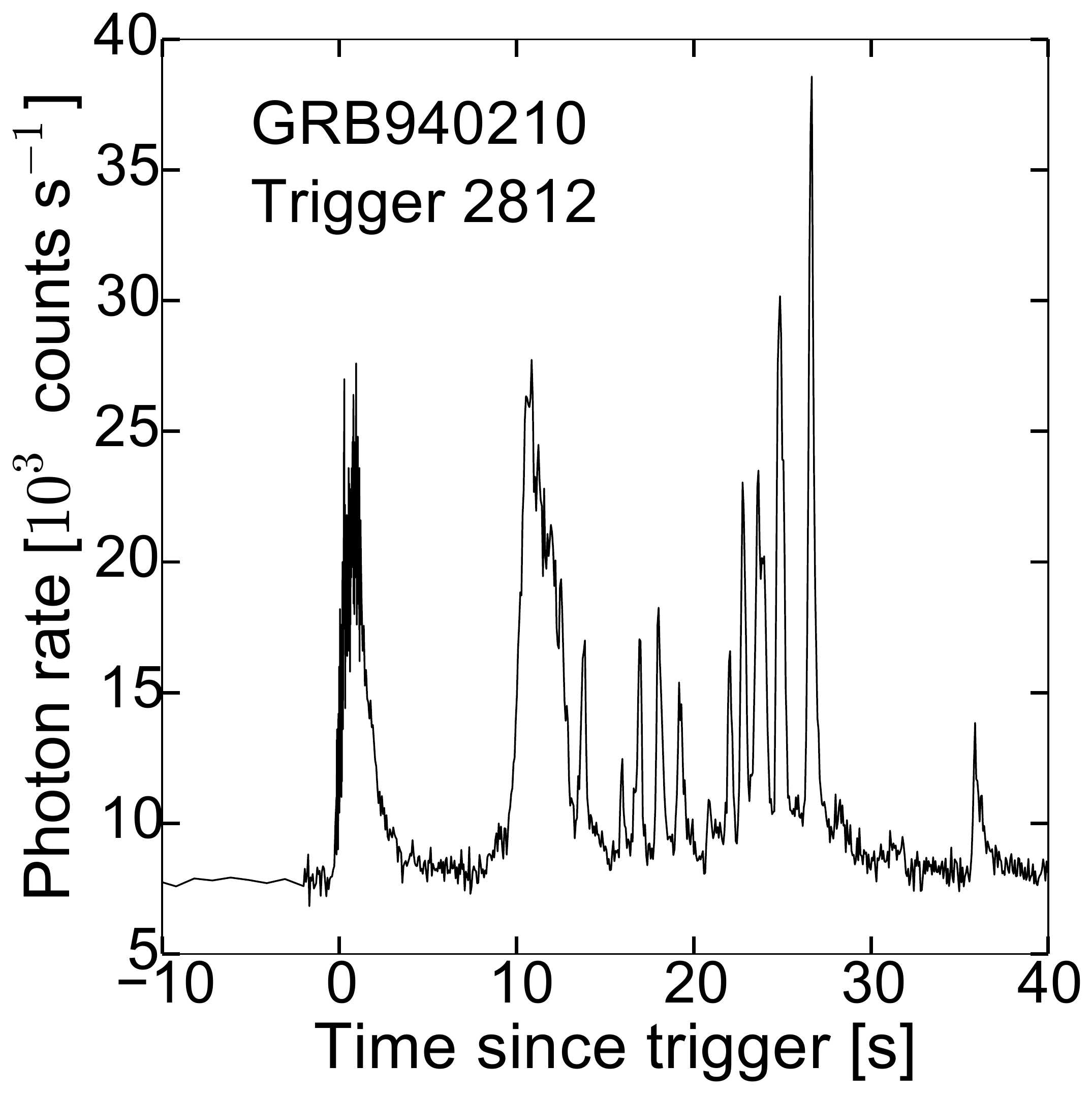}
 \includegraphics[width=0.23\textwidth]{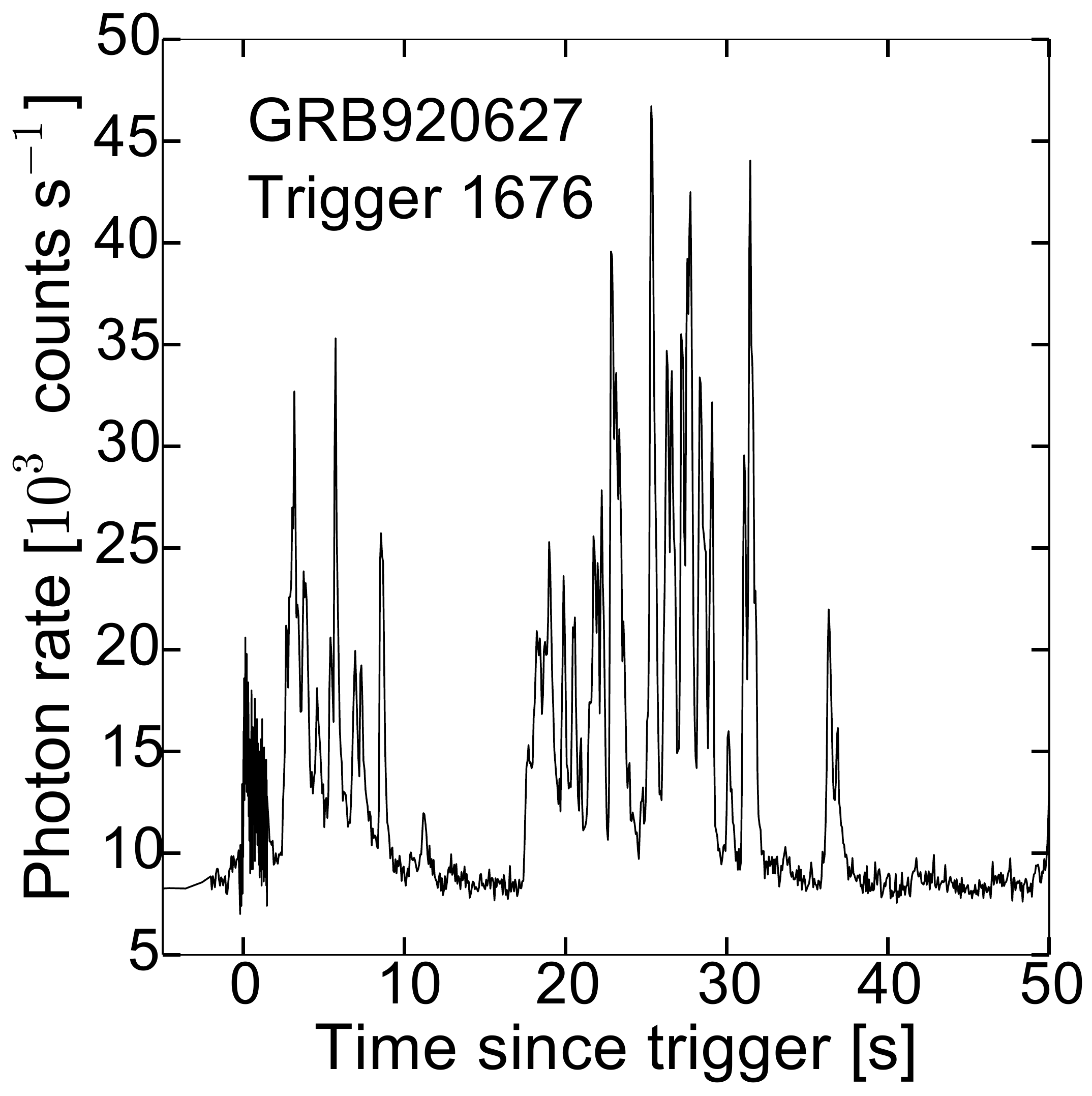} 
 \caption{\label{fig:lcurveexamples} Selected GRB light curves, detected by BATSE~\citep{Paciesas:1999tp} in the $> 20$ keV range (channels 1--4); from left to right: GRB920513, GRB931008, GRB940210, and GRB920627. The implied time resolution is 64 ms.}
\end{figure*}

For super-photospheric collisions, the gamma rays are still attenuated in energy and produce particle cascades, via electron-positron pair-production processes inside shells (see below).
After gamma rays leave the source, they scatter off cosmological photon fields --- microwave, optical, infrared --- and their energies are degraded down to a few hundred GeV or lower. UHECRs also lose energy through photohadronic interactions and inelastic scattering off cosmological photons. Neutrinos are affected only by the adiabatic cosmological expansion, which changes their energies in, at most, a factor of a few. 

We normalize the gamma-ray emission of our simulated bursts to typical time-integrated GRB gamma-ray energies, $E_{\gamma,\text{norm}}^\text{iso} = 10^{53}$ erg (in the source reference frame). This is equated to the sum energy emitted by all collisions, sub- and super-photospheric (\equ{fgamma}). In our simulated bursts, about $50 \%$ of the energy is liberated in super-photospheric collisions.

%%%%%%%%%%%%%%%%%%%%%%%%%%%%%%%%%%%%%%%%%%%%%%%%%%%%%%%%%%%%%%%%%%%%%%%%%%%%%%%%%%%
%%%%%%%%%%%%%%%%%%%%%%%%%%%%%%%%%%%%%%%%%%%%%%%%%%%%%%%%%%%%%%%%%%%%%%%%%%%%%%%%%%%

\section{Synthetic light curves}\label{sec:SyntheticLightCurves}

GRB light curves are highly irregular. They are extremely variable, and a fraction of GRBs have minimum variability timescales of ms~\citep{Golkhou:2015lsa}. The time resolution of the detector imposes a lower bound on the variability timescale that can be inferred. Pulses typically have a duration of $\sim1$~s with an asymmetric structure~\citep{Nakar:2001iz,RamirezRuiz:1999fr}, and some GRBs show distinct quiescent time between the pulses.

In the internal shock scenario, the shape of the GRB light curve is the result of shell collisions~\citep{Kobayashi:1997jk,Daigne:1998xc,Spada:1999fd,Kobayashi:2001iq,Beloborodov:2000nn,Daigne:2003tp,Aoi:2009ty}. When shells collide, a gamma-ray pulse and a neutrino pulse are emitted. The superposition of all pulses emitted during the evolution of the burst makes up the light curve. Different features in the light curves are due to differences in the distribution of collision radii, as determined by the initial speeds of the shells, imprinted on them by the central emitter. See \App~\ref{section:LightCurves} for an explanation of how we construct the light curves.

The simulation parameters from our benchmark model from~\citet{Bustamante:2014oka} are re-branded here as ``GRB 1''. Note that we slightly adjusted the collision model\footnote{The main difference compared to the earlier computation in\ \citet{Bustamante:2014oka} is that, before, we determined which two shells should collide next by computing the absolute value of the times needed for all contiguous shells to collide, and selecting the two shells with the minimum value, whereas now we include causality (so that slower shells cannot catch up with faster ones).  We also fixed a problem with the maintenance of the collision data lists (the shell speed was not always updated correctly).  As explained in the main text, these modifications do not change qualitatively the results for GRB 1 nor our conclusions.}; therefore, the results are not exactly the same as in~\citet{Bustamante:2014oka}. Most notably, collisions occur on average at slightly lower radii now. The initial values of shell Lorentz factors, $\Gamma_{k,0}$, are randomly sampled from a log-normal distribution defined by the characteristic value $\Gamma_0$ and the amplitude of fluctuations $A_\Gamma$:
\begin{equation}\label{equ:LogNormalDistribution}
\ln \left( \frac{\Gamma_{k,0}-1}{\Gamma_0-1} \right) = A_\Gamma \cdot x \; ,
\end{equation}
where the random variable $x$ follows a Gaussian distribution, $P(x) dx = \left( 2 \pi \right)^{-1} e^{-x^2/2} dx$.
When $A_\Gamma < 1$, the mean value $\langle\Gamma\rangle \approx \Gamma_0$ and the variance $\Delta \Gamma \approx A_\Gamma \Gamma_0$.  When $A_\Gamma>1$, the mean value and variance are significantly affected by fluctuations. Large $A_\Gamma \gtrsim 1$ are typically required for efficient conversion of kinetic energy to radiated energy~\citep{Kobayashi:2001iq}, since the latter is proportional to the difference in speeds between two colliding shells. In addition, $A_\Gamma \gtrsim 0.1$ is necessary for the number of collisions to be large, \ie, $N_{\text{coll}} \sim N_{\text{sh}}$. The light curve for GRB 1 is in \figu{lcurves}. It is dominated by fast variability, which is determined by the ratio of $T_{90}$ to $N_{\text{coll}}$, on the order of tens of ms, which is typical for GRBs.

\begin{table*}[t]
 \centering
 \caption{\label{tab:input}Description of simulated GRBs 1--6}
 \begin{tabular}{cccccccccp{7.7cm}}
  \hline
  \hline
  Model & $\Gamma_{0,1}$ & $A_{\Gamma,1}$ & $\Gamma_{0,2}$ & $A_{\Gamma,2}$ & $T_p$ &  $N_\text{up}$ & $N_\text{down}$ & $E_{\gamma,\text{norm}}^\text{iso}$ [erg] & Description \\
  \hline
  1 & 500 & 1.0 & \--- & \--- & \--- & \--- & \--- & $10^{53}$ & Fixed $\Gamma$ and $A_\Gamma$; benchmark from~\citet{Bustamante:2014oka} \\
  2 & 500 & 1.0 & 50 & 0.1 & \--- & \--- & \--- &  $10^{53}$ & Linear speedup of $\Gamma$ \\
  3 & 50 & 0.1 & 500 & 0.1 & 0.34 & \--- & \--- & $10^{53}$ & Sawtooth $\Gamma$ (linear slowdown three times) with narrow distribution \\
  4 & 50 & 0.1 & 500 & 1.0 & 0.2 & \--- & \--- & $10^{53}$ & Oscillating $\Gamma$ (five periods) with increasing distribution width \\
  5 & 50 & 0.1 & 500 & 0.1 & 0.2 & \--- & \--- & $10^{53}$ & Oscillating $\Gamma$ (four periods) with lower amplitude increasing and narrow distribution \\
  6 &  50 & 0.1 & 500 & 1.0 & $1/8$ &  250 & 250 & $10^{53}$ &  Oscillating $\Gamma$ (four periods) with distribution widening up; in addition, engine intermittent: $N_\text{up}$ pulses followed by $N_\text{down}$ pulses; corresponds to increasing Lorentz factor during uptime\\
  \hline
  \hline
 \end{tabular}
 \tablecomments{Common values for all models: $N_\text{sh} = 1000$, $\delta t_{\text{eng}} = 10^{-2}$ s, $d = l = c \cdot \delta t_{\text{eng}}$, $r_\text{min} = 10^3$ km, $r_\text{dec} = 5.5 \cdot 10^{11}$ km, $z = 2$,  $\epsilon_e = \epsilon_B = 1/12$, $\epsilon_p = 5/6$, $\eta = 1.0$ (acceleration efficiency~\protect{\citep{Baerwald:2013pu}}). See \Tab~\ref{tab:ParameterDescription} for an explanation of each parameter.  The period $T_p$ for the oscillating cases refers to $\Gamma$ changing between $\Gamma_{0,1}$ and $\Gamma_{0,2}$, and $A_\Gamma$ changing between $A_{\Gamma,1}$ and $A_{\Gamma,2}$; $T_p$ is a fraction of the total number of emitted shells. This means that $\Gamma$ and $A_{\Gamma}$ change between first and second value with a factor $\sin^2 \left( k/(N_{\text{sh}} \cdot T_p ) \cdot \pi \right) $, where $k$ is the index of shell ($1 \le k \le N_{\text{sh}}$).  For GRB 5, the lower amplitude increases from  $\Gamma_{0,1}$ to $\Gamma_{0,2}$ linearly  with $k$.}
\end{table*}

\begin{figure*}[tp]
 \begin{center}
   \includegraphics[width=0.33\textwidth]{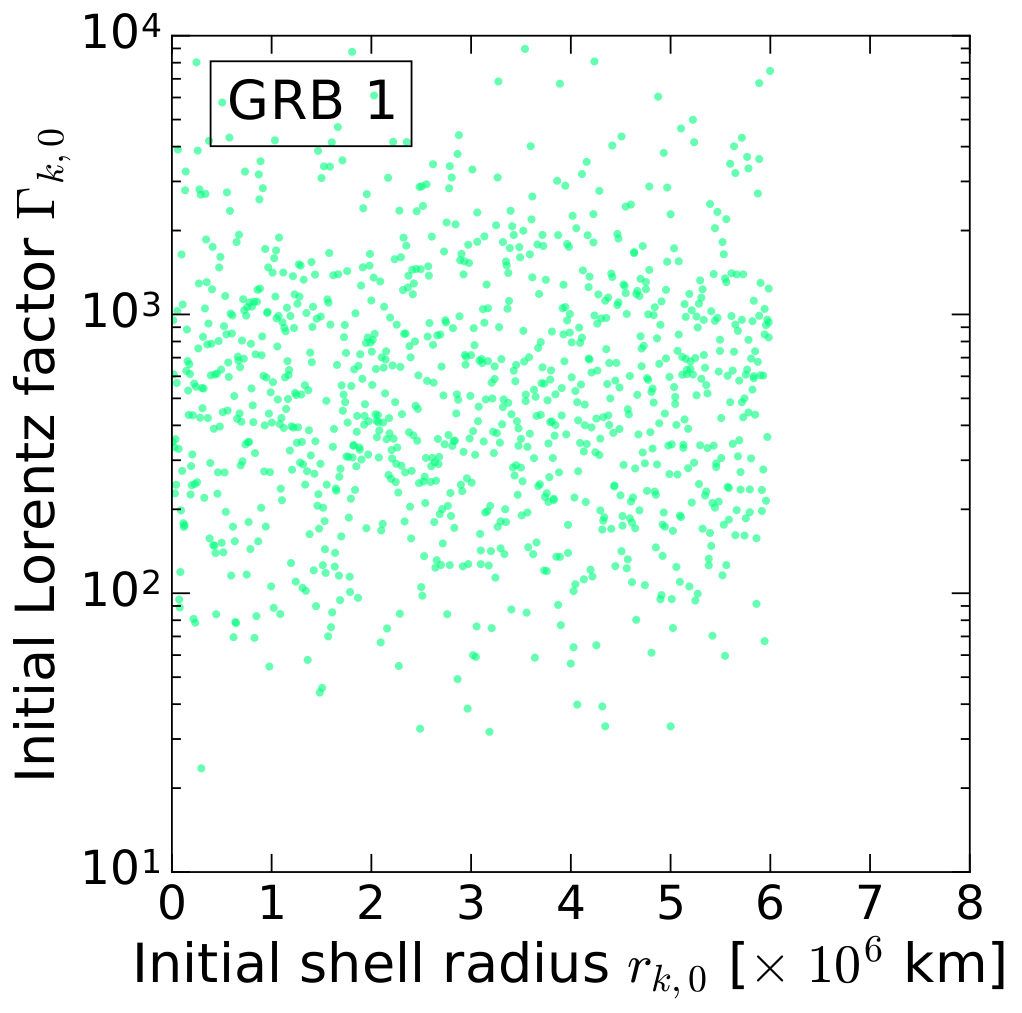}
   \includegraphics[width=0.33\textwidth]{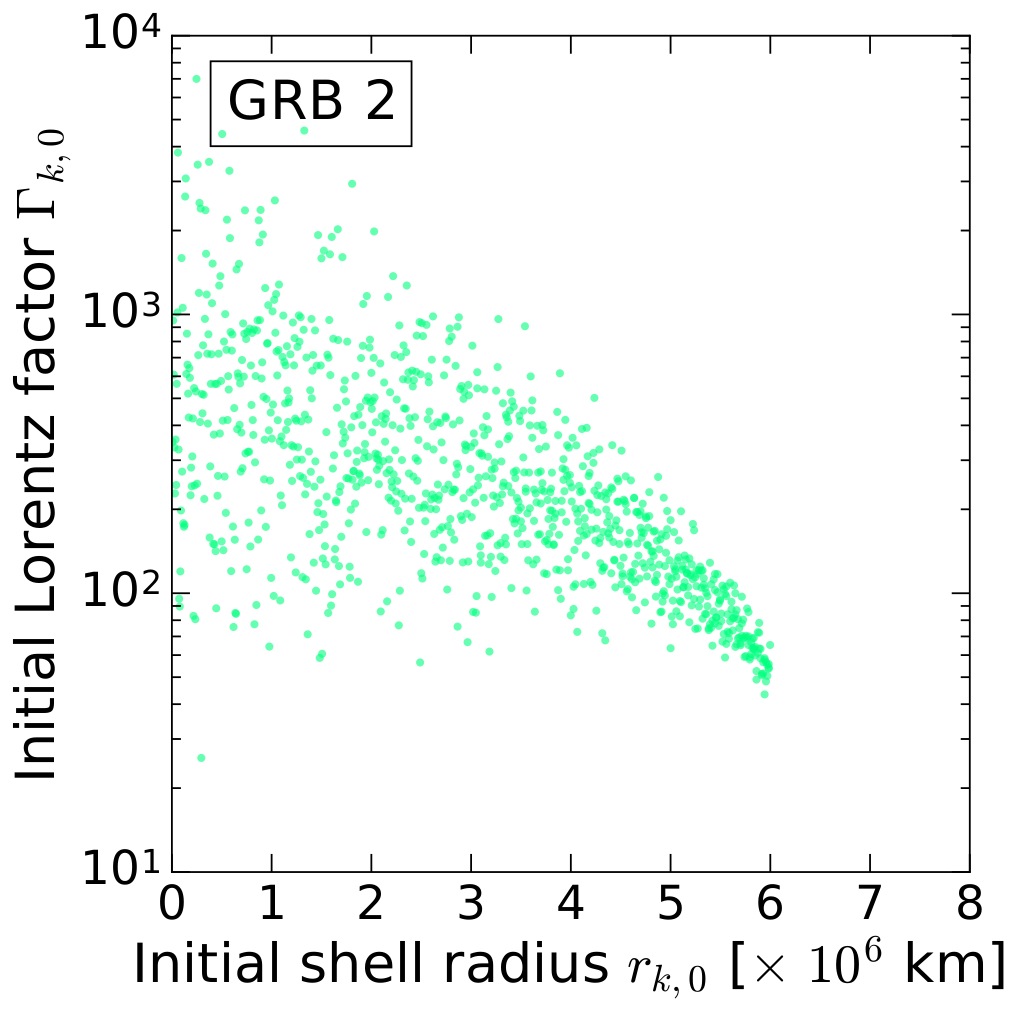}
   \includegraphics[width=0.33\textwidth]{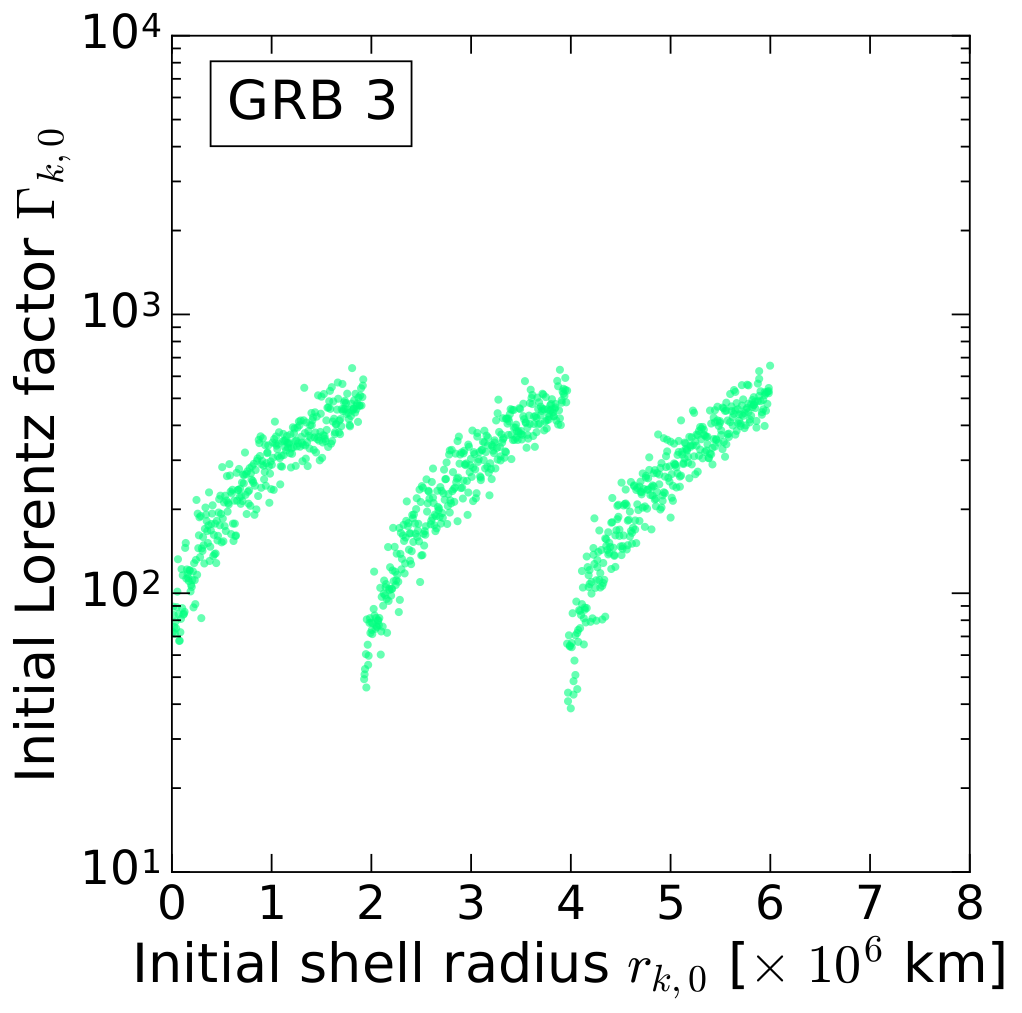}
   \includegraphics[width=0.33\textwidth]{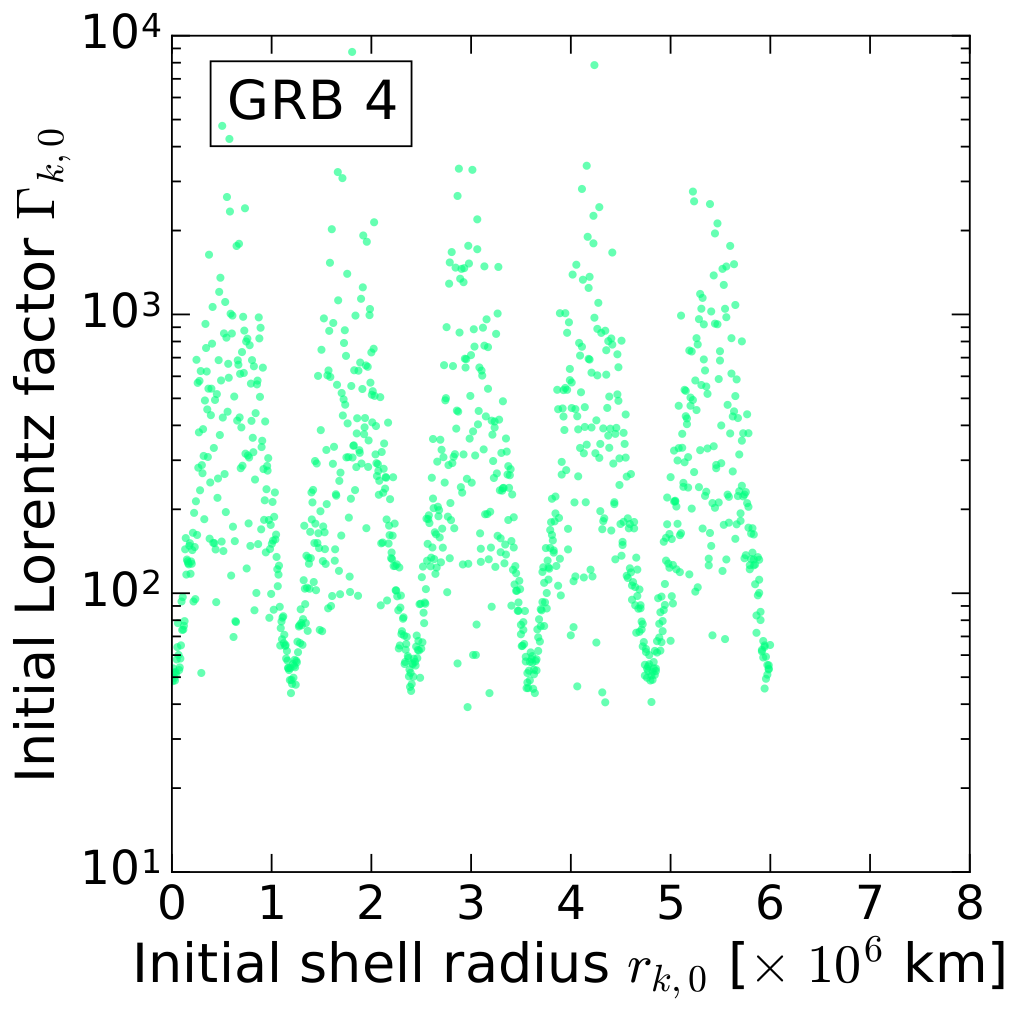}
   \includegraphics[width=0.33\textwidth]{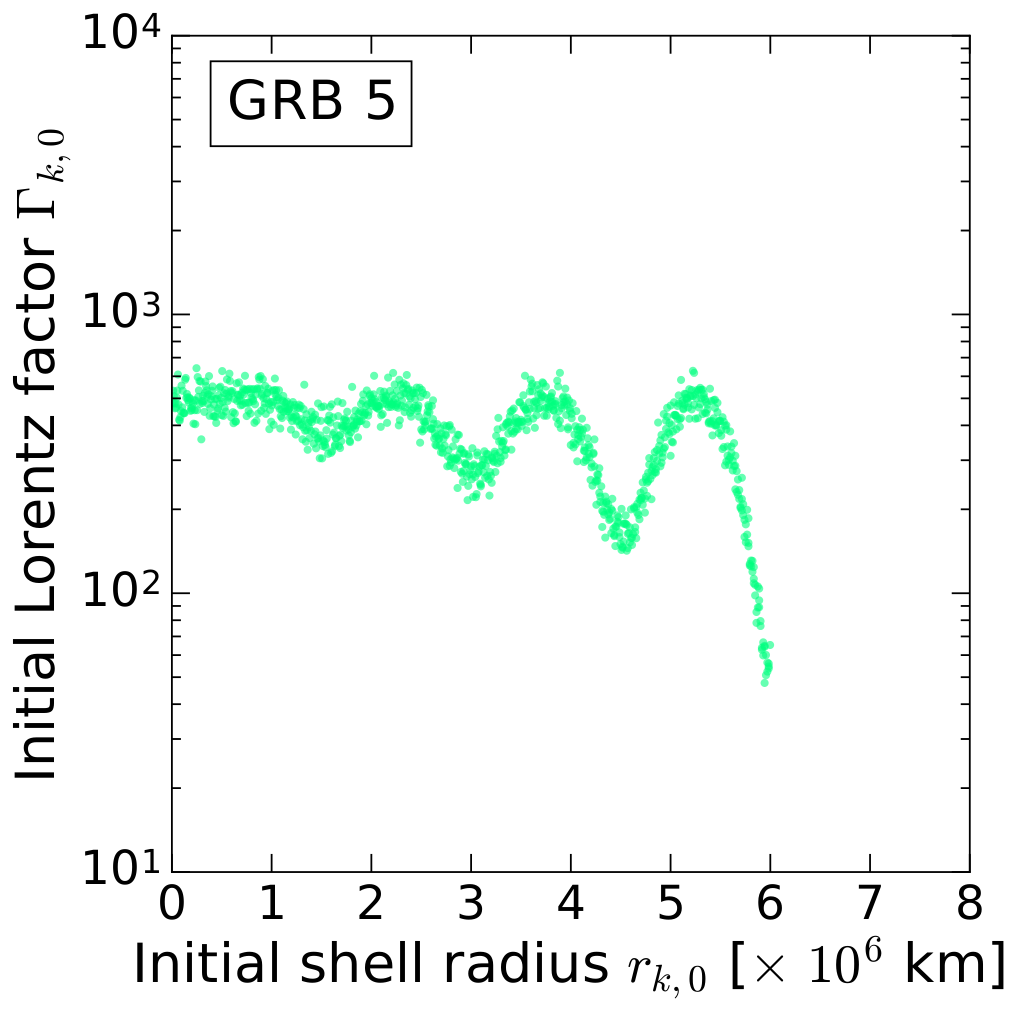}
   \includegraphics[width=0.33\textwidth]{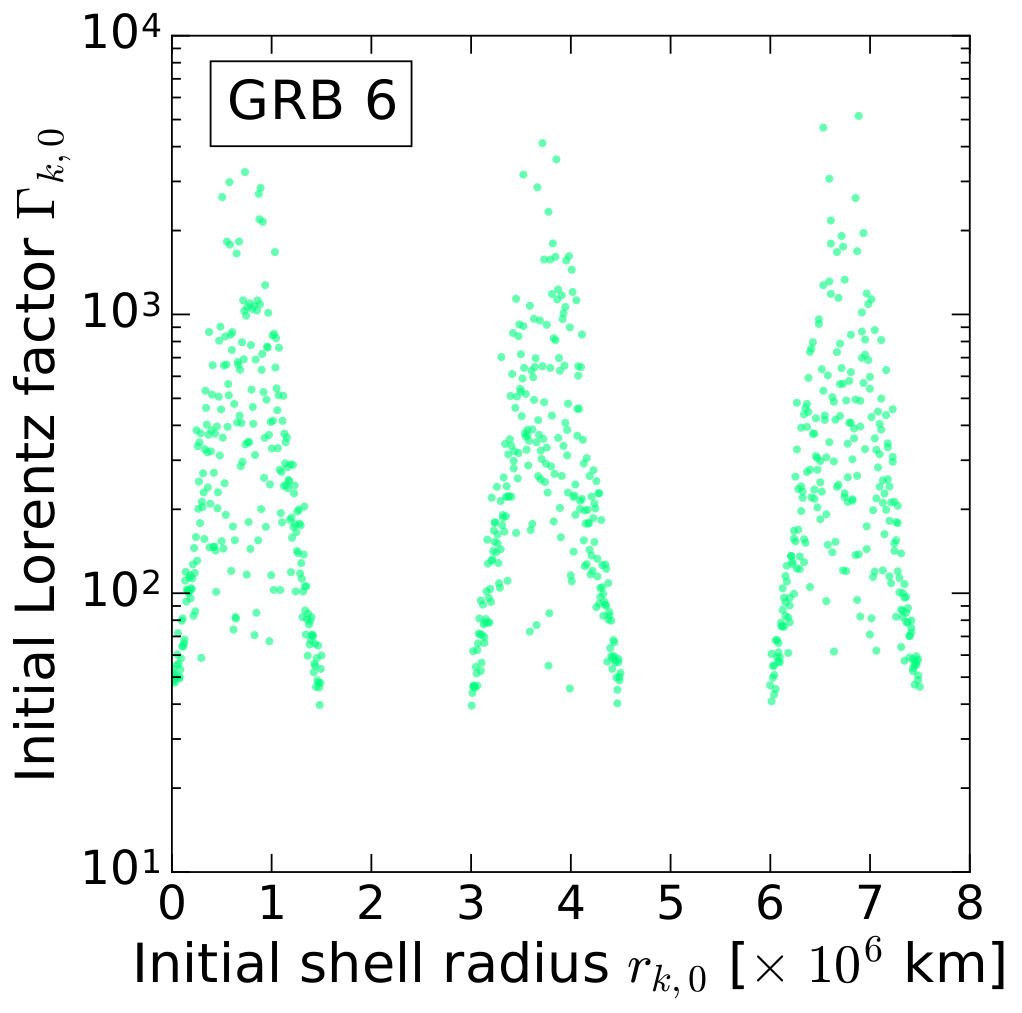}
 \end{center}
 \caption{\label{fig:init_grbs}Initial values of the Lorentz factors of the shells in GRBs 1--6, at the start of the simulations. See \Tab\ \ref{tab:input} for descriptions of the underlying distributions of initial Lorentz factors in each simulation.}
\end{figure*}

\begin{figure*}[t!]
 \begin{center}
  \includegraphics[width=\columnwidth]{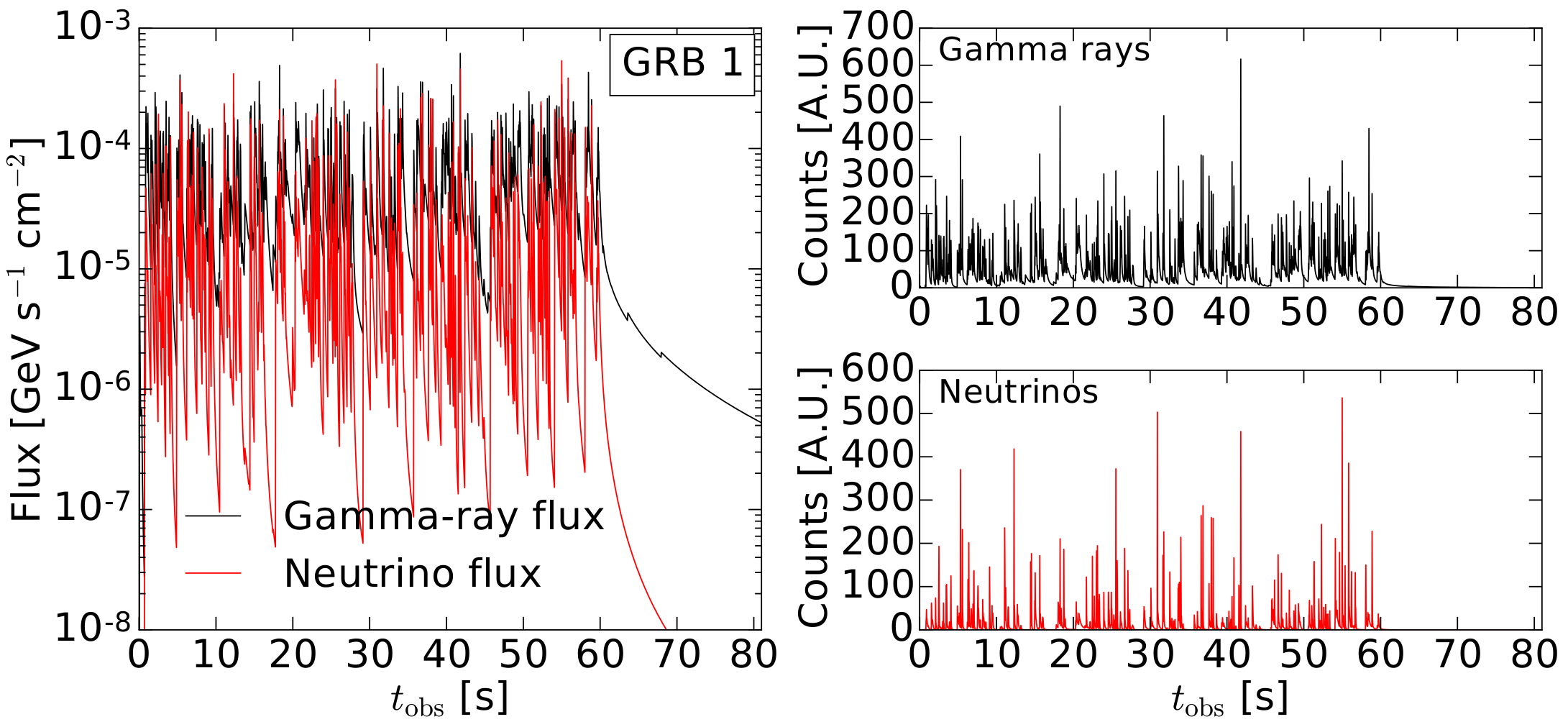}
  \includegraphics[width=\columnwidth]{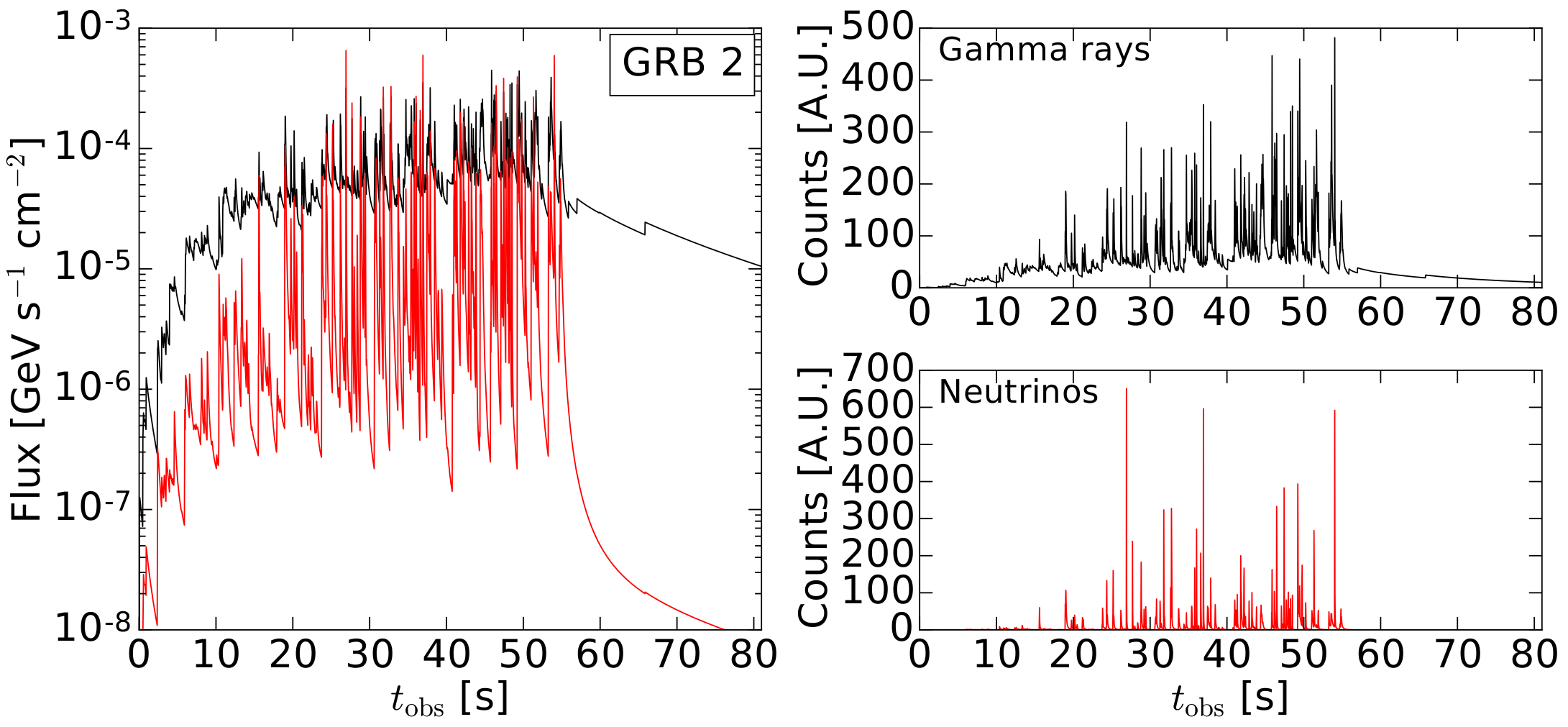}
  \includegraphics[width=\columnwidth]{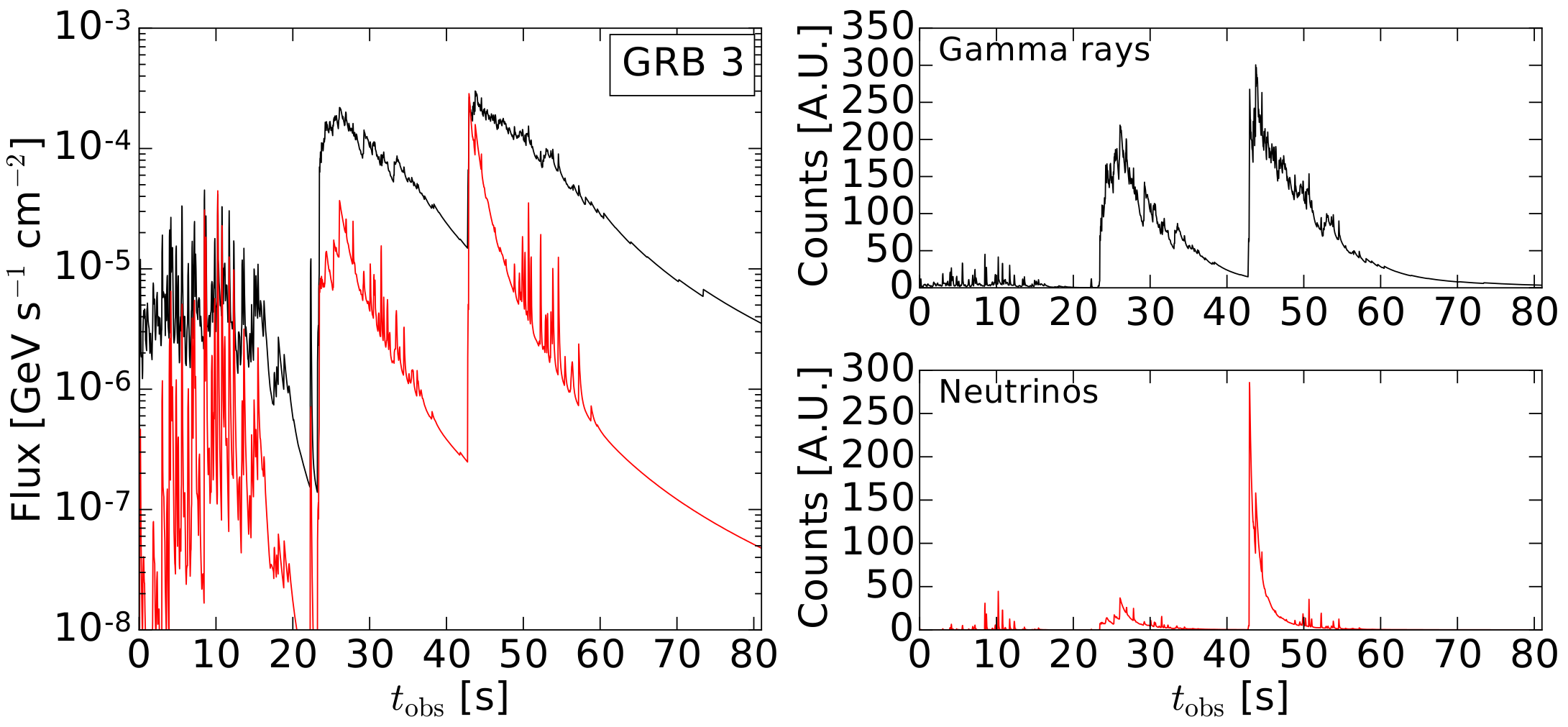}
  \includegraphics[width=\columnwidth]{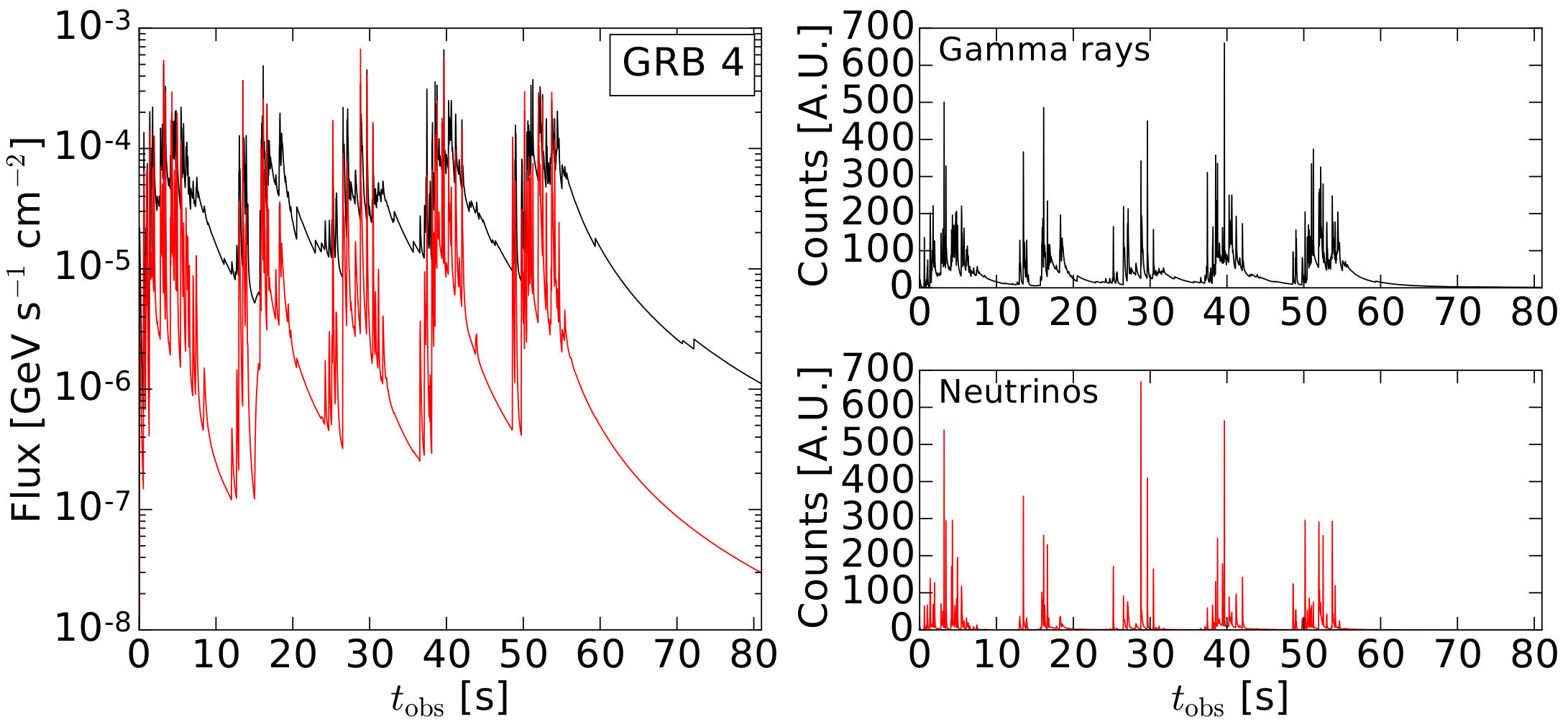}
  \includegraphics[width=\columnwidth]{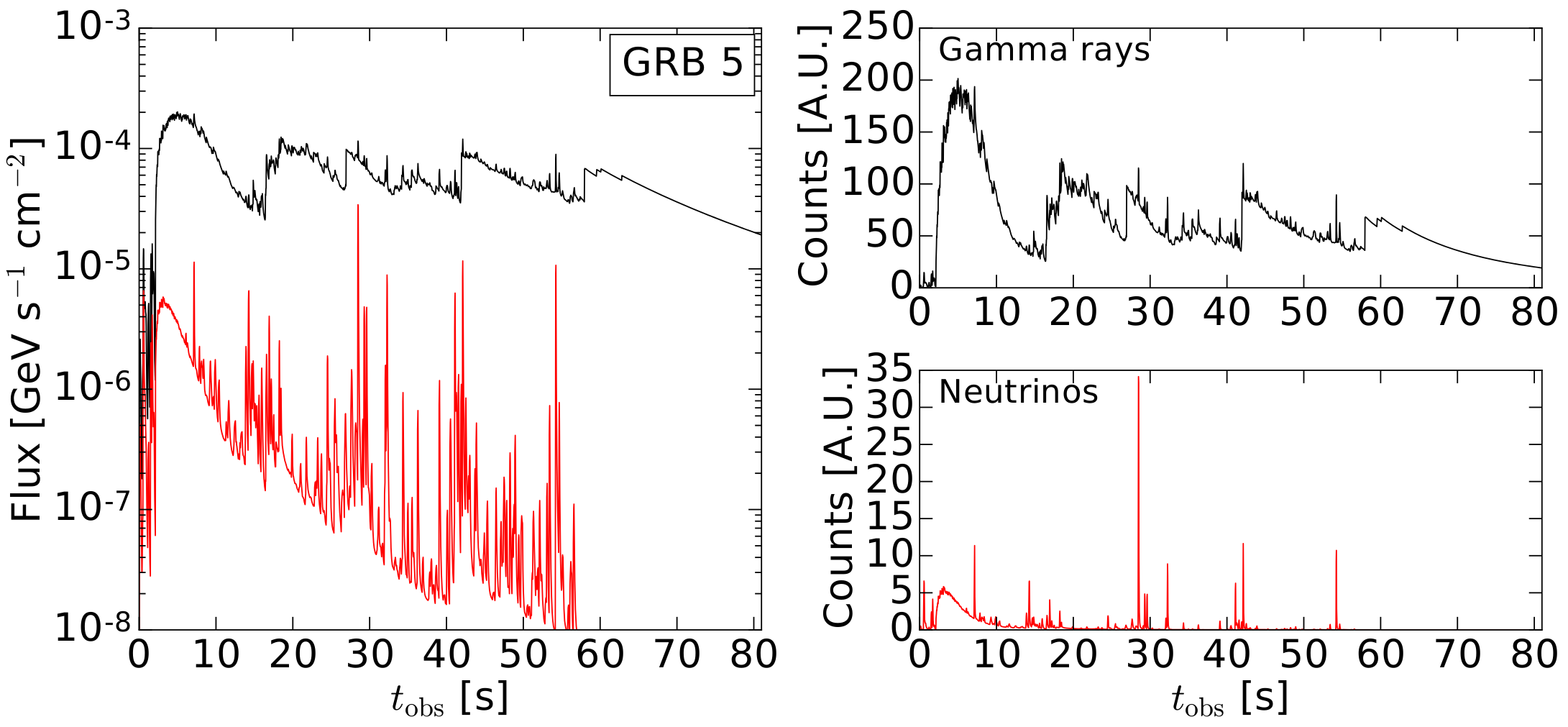}
  \includegraphics[width=\columnwidth]{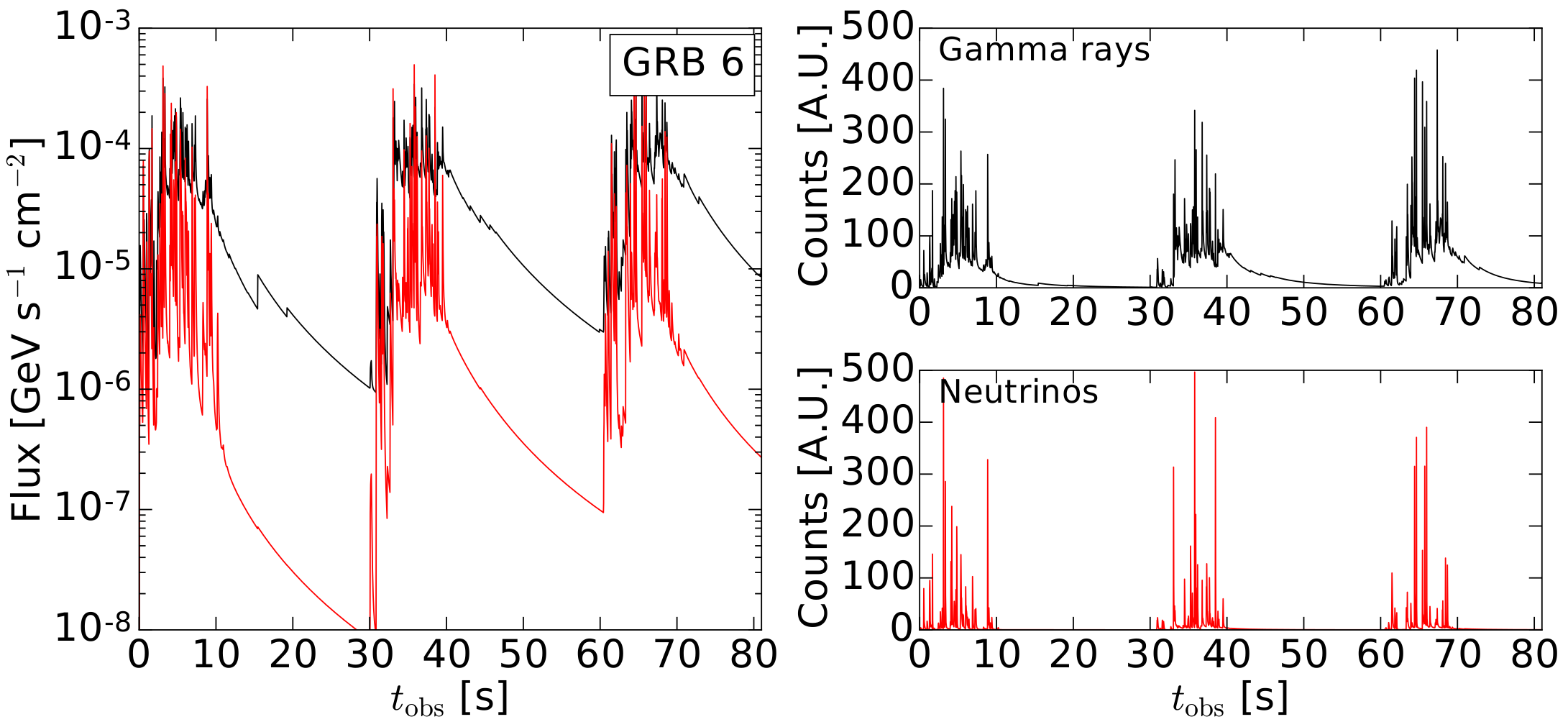}
  \end{center}
 \caption{\label{fig:lcurves} Synthetic gamma-ray and neutrino light curves for the simulated GRBs 1--6, from collisions beyond the photosphere. Photon and neutrino counts are in arbitrary units, obtained by multiplying the flux times a factor of $10^6$ GeV$^{-1}$ cm$^2$ s. }
\end{figure*}

Figure \ref{fig:lcurveexamples} shows selected real gamma-ray light curves that have slower structure overlaid with fast variability. In the internal shock scenario, the slower structure can be produced by modifying the behavior of the engine. We can change the widths of the initial shells, the separations between them or the spread $A_\Gamma$, emit shells intermittently, ramp up or down the Lorentz factors during shell emission, or make them oscillate (see also~\citet{Daigne:1998xc}). 

Table~\ref{tab:input} lists our sample simulations, GRBs 1--6, where we have implemented these options
We have chosen parameter values such that the associated synthetic light curves reproduce features similar to \figu{lcurveexamples}. 

Figure \ref{fig:init_grbs} shows the randomly sampled initial values of the Lorentz factors $\Gamma_{k,0}$ of the shells in GRBs 1--6, as a function of initial radius $r_{k,0}$.  See \Tab\ \ref{tab:input} for descriptions of the underlying distributions.

Figure~\ref{fig:lcurves} shows the synthetic gamma-ray and neutrino light curves for GRBs 1--6.
GRB 1 has fast time variability without prominent features. 
GRB 2 has a speedup in $\Gamma$ during shell emission; a single pulse is overlaid with fast variability. For a slowdown, the pulse occurs earlier and the light curve is time-inverted. However, the efficiency is much lower in this case, as the fast shells are emitted first.
GRB 3 has three pulses with linear slowdown. The second and third pulses collide with slow shells from the preceding pulse and therefore contribute more strongly to the light curve than the first pulse.  GRBs 4 and 6 have more oscillating periods.
GRB 5 is oscillating as well, but its lower amplitude increases linearly during emission.
Comparison with \figu{lcurveexamples} reveals that GRB 3 was inspired by GRB931008; GRBs 4 and 6, by case GRB920627; and GRB 5, by cases GRB920513 and GRB940210. 
Appendix\ \ref{sec:sim_add} contains four more simulation examples, with different engine assumptions but similar behavior to that of GRBs 1--6, showing that these are representative.

Most GRB light curves detected by {\it Fermi}~\citep{Ackermann:2013zfa,vonKienlin:2014nza} do not have prominent features like the ones in \figu{lcurveexamples}. The reason why the latter are featured in the literature more often than simpler single-pulse or fast-variability light curves is possibly a selection effect (\ie, they are more interesting to show and study). From that perspective, it is conceivable that GRBs 3--6 are not ``typical'' GRBs.

The light curves of GRBs 3 and 5 are qualitatively different from the others in one key aspect: they have a dominant, broad pulse structure overlaid with fast time variability, whereas in the other bursts the fast component is more relevant. This feature can be traced back to the input parameters in \Tab~\ref{tab:input}:
GRBs 3 and 5 have a ``disciplined'' central engine that emits shells within a narrow $\Gamma$ distribution ($A_\Gamma = 0.1$), while the average $\Gamma$ changes slowly. Therefore, most collisions occur at larger radii compared to the cases where the spread in $\Gamma$ is larger. We will see that this also affects the neutrino production efficiency in the case of GRB 5.

The internal shock model has been invoked as a successful model to explain irregular features of the GRB light curve~\citep{Kobayashi:1997jk,Daigne:1998xc,Spada:1999fd,Kobayashi:2001iq,Beloborodov:2000nn,Daigne:2003tp,Aoi:2009ty}. However, because the classical internal shock model is being challenged, light curve predictions of alternative models have also been extensively investigated. In particular, it has been shown that turbulence or magnetic reconnection models can better explain the observed structure of the GRB light curve~\citep{Narayan:2008xq,2009ApJ...695L..10L,Zhang:2013ycn,Beniamini:2015sua}. 
Some recent studies suggested that GRB light curves may consist of the superposition of slow and fast components, as inferred by a gradual depletion of the fast component at low energies~\citep{2006AA...447..499V}. Sub-structures of the observed GRB pulses can be easily accounted for by assuming relativistic motions in the bulk of a relativistic jet. Although their origins are unclear at present, such relativistic motions could be realized in the ICMART model~\citep{Zhang:2013ycn}. Compared to these explanations, our pulse structure comes from the properties of the engine rather than the jet.

Figure~\ref{fig:lcurves} shows there is no linear correlation between the heights of gamma-ray and neutrino pulses.
This is because the height of a neutrino pulse is, via the pion production efficiency, more sensitive to the collision radius than the height of a gamma-ray pulse, \ie, it depends on the proton and photon densities, which drop $\propto R_\text{C}^2$, where $R_\text{C}$ is the collision radius.

There are no long time delays between gamma-ray and neutrino pulses: they are within $T_{90}$ (see \equ{T90Def}) of each other. There may be, however, short delays.  For example, in GRB 3, the neutrino peak corresponding to the first large gamma-ray peak is suppressed. So the first gamma-ray detection will have occurred $\sim 20$~s, before the neutrino instrument triggers.  In GRB 4, the quiescent periods of gamma-ray and neutrino emission coincide, which may be exploited by neutrino telescopes to set further time window cuts.

In GRB 5, the fast time variability gives rise to neutrino spikes, whereas the longer pulses seen in gamma rays are hardly present in neutrinos. Indeed, the rise time of the spike of particle emission associated to one collision depends strongly on $\Gamma$ (see \equ{trise}) and, consequently, it is $\propto R_\text{C}$. Therefore, faster rises --- sharper structures --- are created by collisions at smaller radii, where the neutrino production efficiency is higher.

\begin{table*}[t]
 \begin{center}
  \caption{\label{tab:OutputParameters}Parameters output by simulated GRBs 1--6}
  \begin{tabular}{ccccccccc}
   \hline
   \hline
   Model & $N_\text{coll}$ & $t_\text{v}$ [ms] & $T_{90}$ [s] & $ E_{\gamma,\text{tot}}^\text{iso}$ [erg] & $E_{p,\text{tot}}^\text{iso}$ [erg] &  $E_{\nu,\text{tot}}^\text{iso}$ [erg] &
   $E_{\nu,\text{tot}}^\text{iso}/E_{\gamma,\text{tot}}^\text{iso}$ [$\%$] & $\varepsilon$ [$\%$]
   \\
   \tableline

   1 & 987 & 53  & 54.0   & $5.2 \cdot 10^{52}$ & $6.2 \cdot 10^{52}$ & $1.4 \cdot 10^{52}$ & 26.9 & 26.8 \\
   2 & 999 & 47  & 47.0   & $6.5 \cdot 10^{52}$ & $4.4 \cdot 10^{52}$ & $9.3 \cdot 10^{51}$ & 14.3 & 19.6 \\
   3 & 951 & 33  & 35.5   & $6.2 \cdot 10^{52}$ & $5.0 \cdot 10^{52}$ & $7.6 \cdot 10^{51}$ & 12.3 & 10.5 \\
   4 & 987 & 52  & 52.6   & $4.6 \cdot 10^{52}$ & $4.0 \cdot 10^{52}$ & $9.4 \cdot 10^{51}$ & 20.4 & 21.7 \\
   5 & 990 & 57  & 57.9   & $8.9 \cdot 10^{52}$ & $1.7 \cdot 10^{52}$ & $6.6 \cdot 10^{50}$ & 0.7 & 10.6 \\
   6 & 985 & 97  & 98.0   & $6.1 \cdot 10^{52}$ & $4.2 \cdot 10^{52}$ & $1.1 \cdot 10^{52}$ & 18.0 & 23.6 \\

   \hline
   \hline
  \end{tabular}
  \tablecomments{The parameters are: variability timescale ($t_\text{v}$), total energy emitted as gamma rays ($E_{\gamma,\text{tot}}^\text{iso}$), as protons ($E_{p,\text{tot}}^\text{iso}$), as neutrinos ($E_{\nu,\text{tot}}^\text{iso}$), ratio between neutrino and gamma-ray energies ($E_{\nu,\text{tot}}^\text{iso} / E_{\gamma,\text{tot}}^\text{iso}$), and overall emission efficiency $\varepsilon$. Energies are computed using super-photospheric collisions only. Only protons in the UHECR energy range, above $> 10^{10}$ GeV, are counted. The efficiency $\varepsilon$ is defined as the ratio of total energy dissipated by all types of particles (gamma rays, protons, neutrinos) to the total kinetic energy initially available.}
 \end{center}
\end{table*}

Table~\ref{tab:OutputParameters} lists the parameters output by GRBs 1--6.  GRBs 1, 2, 4, and 5 have time variabilities of $\sim 50 \, \text{ms}$ and durations of $\sim 50 \, \text{s}$. In GRB 3, the duration ($33 \, \text{s}$) and time variability ($36 \, \text{ms}$) are smaller, as mainly two of the three peaks contribute. In GRB 6, pulses are separated by a downtime and, therefore, duration and variability are larger ($98 \, \text{s}$ and $97 \, \text{ms}$, respectively).

The dissipation efficiency $\varepsilon$ of a burst in \Tab~\ref{tab:OutputParameters} is the ratio between total energy dissipated by all types of particles in super-photospheric collisions and total kinetic energy available at the start of the simulation. Most simulations\footnote{Bursts reach higher efficiencies ($\sim 40\%$) if they have a square-pulse (``box-like'') distribution of $\Gamma$. Since this distribution is unrealistic, we do not discuss it further.} have high $\varepsilon \approx 11$--$27\%$, which reasonably agrees with previous work~\citep{Beloborodov:2000nn}. The efficiency is lower in cases with narrow $\Gamma$ distribution, as expected.

The gamma-ray emission efficiency $\varepsilon_\gamma\approx\epsilon_e\varepsilon$ is about 10 times smaller since most of the dissipation energy is assumed to be carried by protons rather than electrons. Here, $\epsilon_e$ is the fraction of energy in electrons and photons; see \Sec\ \ref{section:LightCurves}. Such a value for the radiation efficiency may be too small compared to ones preferred by observations. However, if the internal energy is carried by thermal protons or confined cosmic rays, it is natural to expect the reconversion of the internal energy into the kinetic energy; see \App\ \ref{sec:alt}. It has been shown that this effect increases the gamma-ray emission efficiency, represented by the ratio of prompt gamma-ray energy to afterglow kinetic energy, calculated in an approach where shells reflects off each other after colliding, \ie, collisions are not perfectly inelastic~\citep{Kobayashi:2001iq}. Recent results on afterglow modeling also suggest a small value of the gamma-ray emission efficiency~\citep{Beniamini:2015eaa}.

%%%%%%%%%%%%%%%%%%%%%%%%%%%%%%%%%%%%%%%%%%%%%%%%%%%%%%%%%%%%%%%%%%%%%%%%%%%%%%%%%%%
%%%%%%%%%%%%%%%%%%%%%%%%%%%%%%%%%%%%%%%%%%%%%%%%%%%%%%%%%%%%%%%%%%%%%%%%%%%%%%%%%%%

\section{Multi-messenger emission}\label{sec:MultiMessengerEmission}

%%%%%%%%%%%%%%%%%%%%%%%%%%%%%%%%%%%%%%%%%%%%%%%%%%%%%%%%%%%%%%%%%%%%%%%%%%%%%%%%%%%

Here we discuss the emission of multiple messengers and their relation.

\subsection{Weak vs.\ strong neutrino emitters}\label{sec:WeakVsStrong}

\begin{figure}[t!]
 \begin{center}
  \includegraphics[width=\columnwidth]{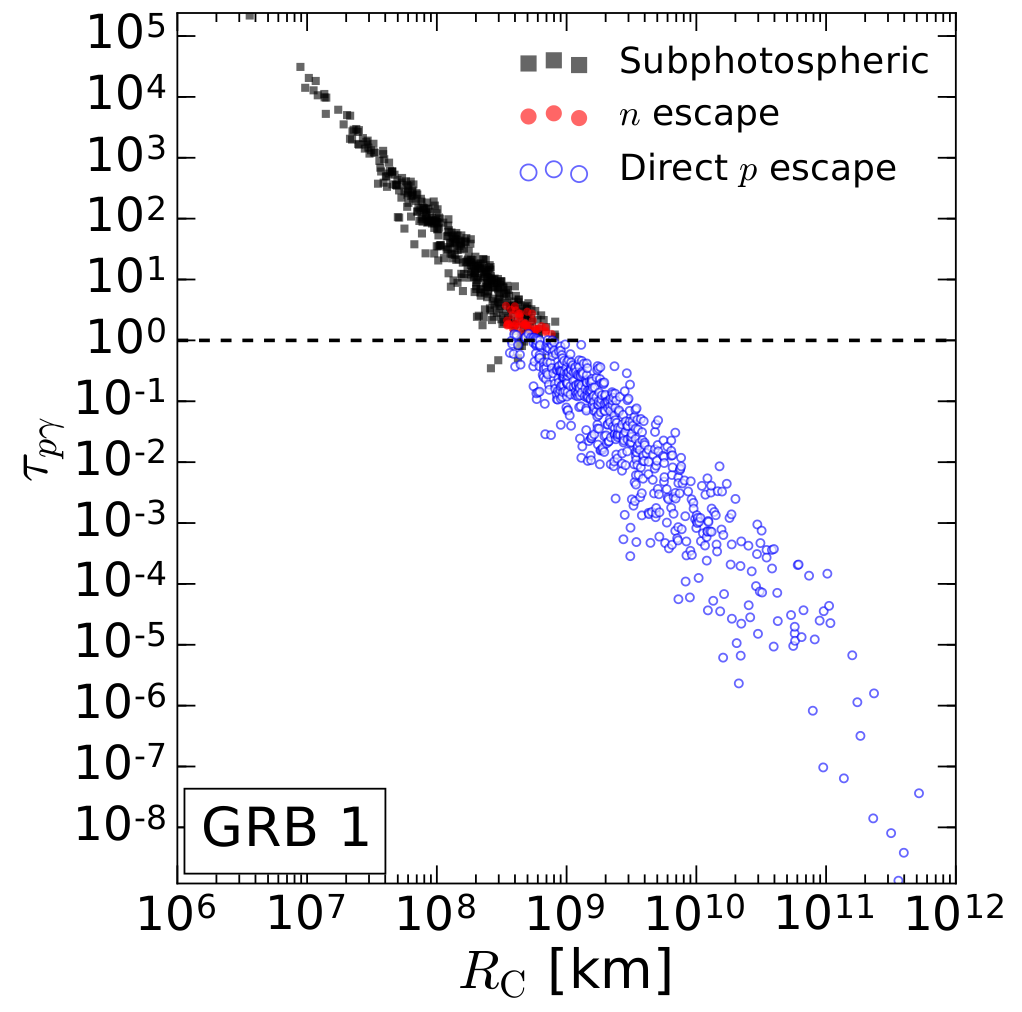} %GRB 1
  \includegraphics[width=\columnwidth]{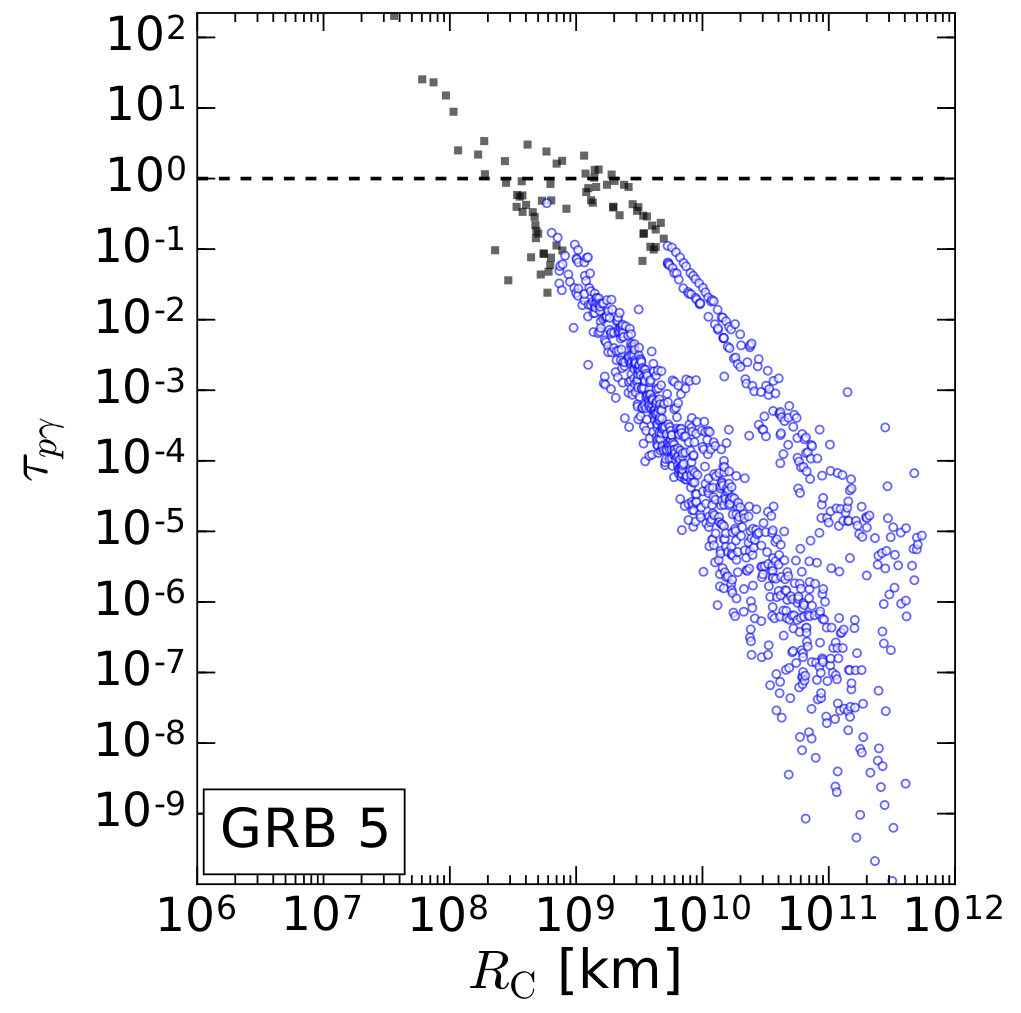} %GRB 5
 \end{center}
 \caption{\label{fig:optdepth} Optical depth $\tau_{p \gamma}$ for all collisions in GRBs 1 (top) and 5 (bottom) as a function of collision radius. The horizontal line corresponds to $\tau_{p \gamma}=1$. Black filled rectangles are sub-photospheric collisions, red filled dots are super-photospheric collisions where the dominant UHECR component is neutron escape, and blue unfilled dots are super-photospheric collisions where the dominant component is direct proton escape.}
\end{figure}

The time-integrated neutrino fluence of a simulated burst, for a baryon-rich jet, roughly scales as~\citep{Bustamante:2014oka}
\begin{equation}
\mathcal{F}_\nu \propto \frac{N_{\rm coll}(\tau_{p\gamma} \gtrsim 1)}{N_{\rm coll}} \cdot \varepsilon \cdot  E^{\mathrm{iso}}_{\gamma,\text{tot}} \; ,
\label{equ:fnu}
\end{equation}
assuming a fixed photon break energy (see \App~\ref{section:LightCurves}). The first factor gives the fraction of collisions with high optical depth $\tau_{p \gamma} \gtrsim 1$ to $p \gamma$ interactions, $\varepsilon$ is the energy dissipation efficiency, and $E^{\mathrm{iso}}_{\gamma,\text{tot}}$ is the total energy emitted as gamma rays in super-photospheric collisions. Unlike one-zone predictions~\citep{Waxman:1997ti,Guetta:2003wi,Li:2011ah,Hummer:2011ms,He:2012tq}, the fluence does not depend on the average Lorentz factor of the shells.

Figure \ref{fig:optdepth} shows $\tau_{p \gamma}$ as a function of $R_\text{C}$ for collisions in GRBs 1 and 5. In GRB 1, a strong neutrino emitter, about 60 in 1000 collision occurred above the photosphere --- so that they were optically thin to Thomson scattering --- {\it and} were still optically thick to photomeson production. The other simulations with broad $\Gamma$ distributions have similar results. Therefore, the first factor in \equ{fnu} is $\sim 0.05$ for strong neutrino emitters. The energy dissipation efficiency in \Tab~\ref{tab:OutputParameters} lies around $\varepsilon = 0.2$ for these  GRBs. 
As a result, for fixed $E^{\mathrm{iso}}_{\gamma,\text{tot}}$, the quasi-diffuse neutrino flux that we infer below~\citep{Bustamante:2014oka} is relatively robust.

The situation is different for GRBs 3 and 5. They have lower efficiencies $\varepsilon \approx 10\%$.  More importantly, they have no optically thick collisions close to the photosphere; see \figu{optdepth}, bottom, for GRB 5. The reason is that they emit shells with a variable but narrow Lorentz factor distribution (see \figu{init_grbs}) which tends to induce collisions at larger radii, where  photon densities are low and, therefore, so is neutrino production efficiency. In particular, this makes GRB 5 our weakest neutrino emitter, \ie, it has the lowest ratio of emitted neutrino energy to gamma-ray energy beyond the photosphere.

While the same effect should also make GRB 3 a weak neutrino emitter, its neutrino flux is still on a level with our other examples. The reason for this comes from its very specific initial shell setup. It consists of three narrow pulses, each with decreasing $\Gamma$. The collisions are therefore dominated by the first, fast shells of a pulse running into the preceding pulse --- these shells have largely different Lorentz factors; in particular, the differences are larger than in GRB 5.  Most of the neutrino emission comes from these first collisions, which happen below and slightly above the photosphere. 

We saw that GRBs with light curves dominated by fast variability are likely to be strong neutrino emitters. The reverse conclusion does however not hold. While both GRBs 3 and 5 have gamma-ray light curves with broad pulses overlaid with fast variability, only GRB 5 is a weak neutrino emitter. {\it Therefore, in the multi-zone internal shock model, we can tell, by inspection of the gamma-ray light curve alone, whether or not a GRB is likely to be a strong neutrino emitter. Conclusions about weak neutrino emitters require a closer inspection of the specific light curve morphology.}

\begin{figure}[t!]
 \begin{center}
  \includegraphics[width=\columnwidth]{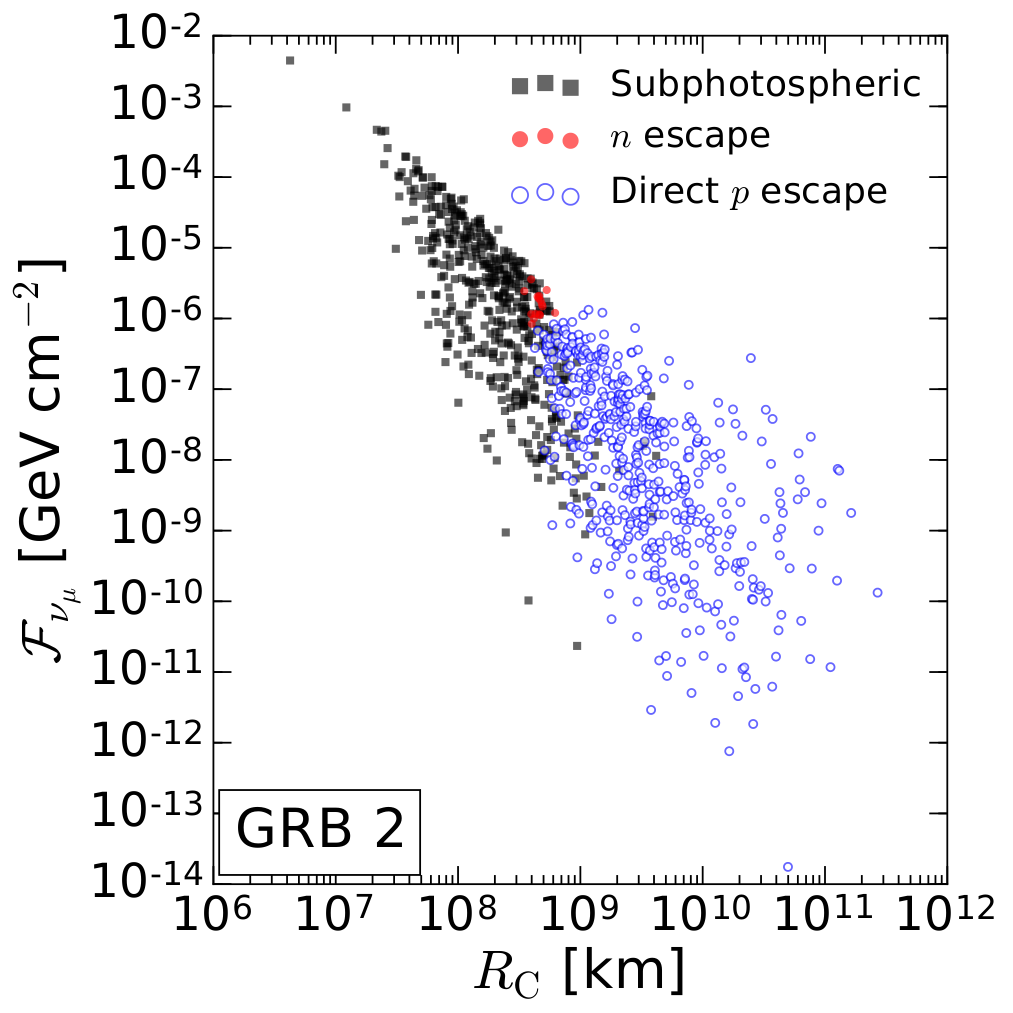}
  \includegraphics[width=\columnwidth]{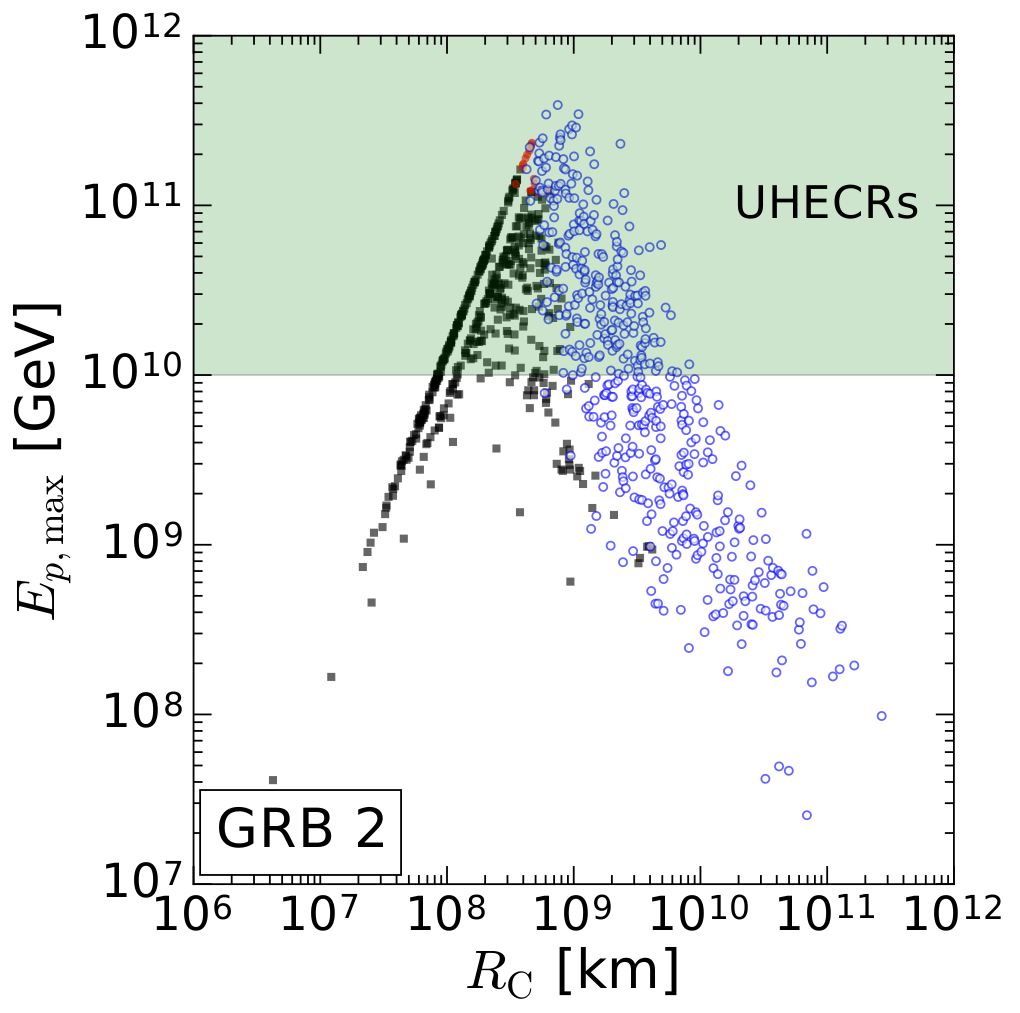}  
 \end{center}
 \caption{\label{fig:scatter} Muon-neutrino ($\nu_\mu + \bar{\nu}_\mu$) fluence (top) and maximum cosmic ray energy (in source frame, bottom) for collisions in GRB 2. The legend is the same as for \figu{optdepth}. In the bottom panel, the UHECR range $E_{p,\max} > 10^{10}$ GeV is shaded.}
\end{figure}

Figure \ref{fig:scatter}, top, shows the neutrino fluence for GRB 2. The neutrino fluence follows the behavior of $\tau_{p\gamma}$ from \figu{optdepth} (which is shown there for different examples). The average fluence per collision drops stronger than approximately $\propto R_\text{C}^{-2}$. In principle, the fluence from sub-photospheric collisions is high, due to the high extrapolated photon density; however, we do not use those collisions in our flux calculations.

%%%%%%%%%%%%%%%%%%%%%%%%%%%%%%%%%%%%%%%%%%%%%%%%%%%%%%%%%%%%%%%%%%%%%%%%%%%%%%%%%%%

\subsection{Quasi-diffuse neutrino flux}

\begin{figure*}[tp]
 \begin{center}
   \includegraphics[width=0.33\textwidth]{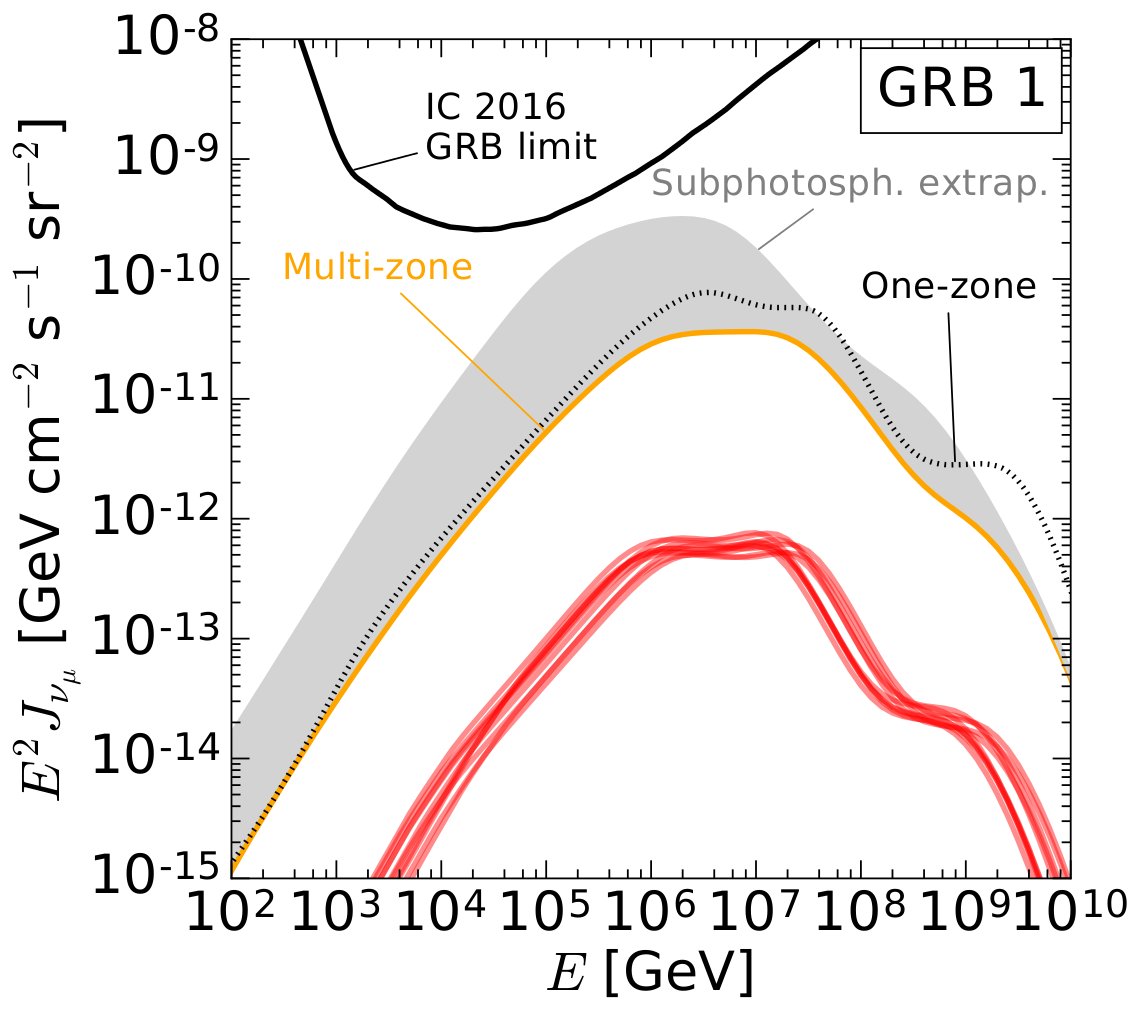}
   \includegraphics[width=0.33\textwidth]{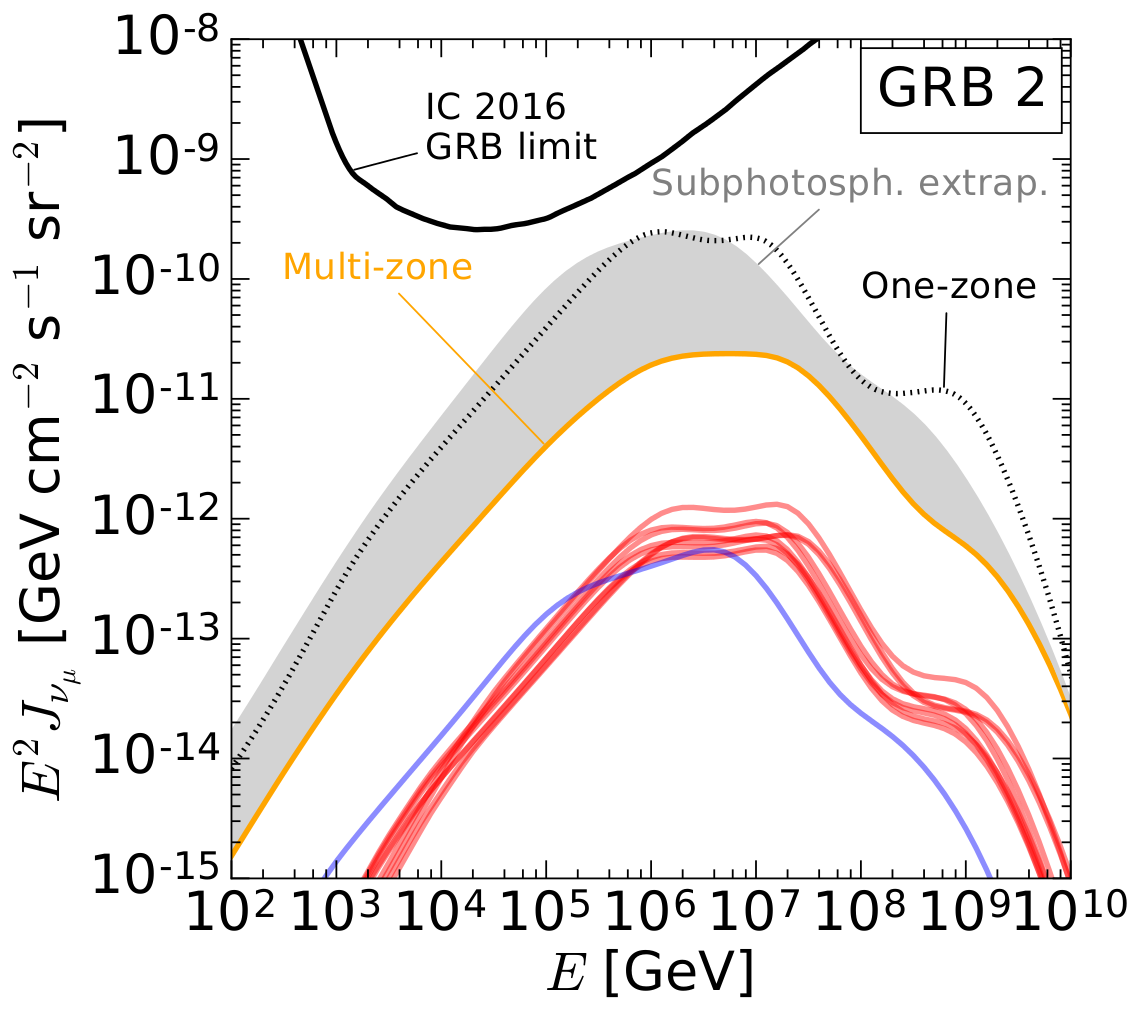}
   \includegraphics[width=0.33\textwidth]{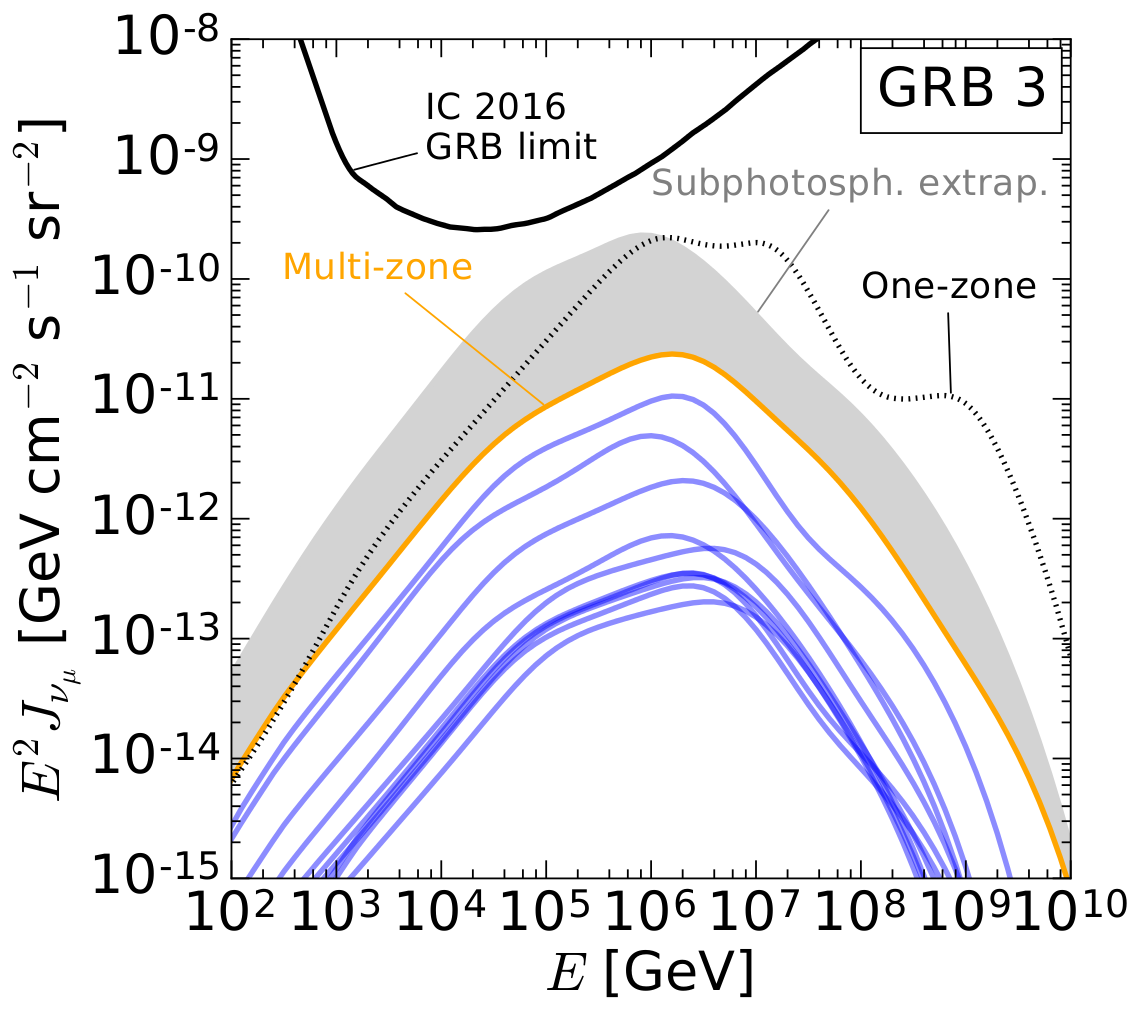}
   \includegraphics[width=0.33\textwidth]{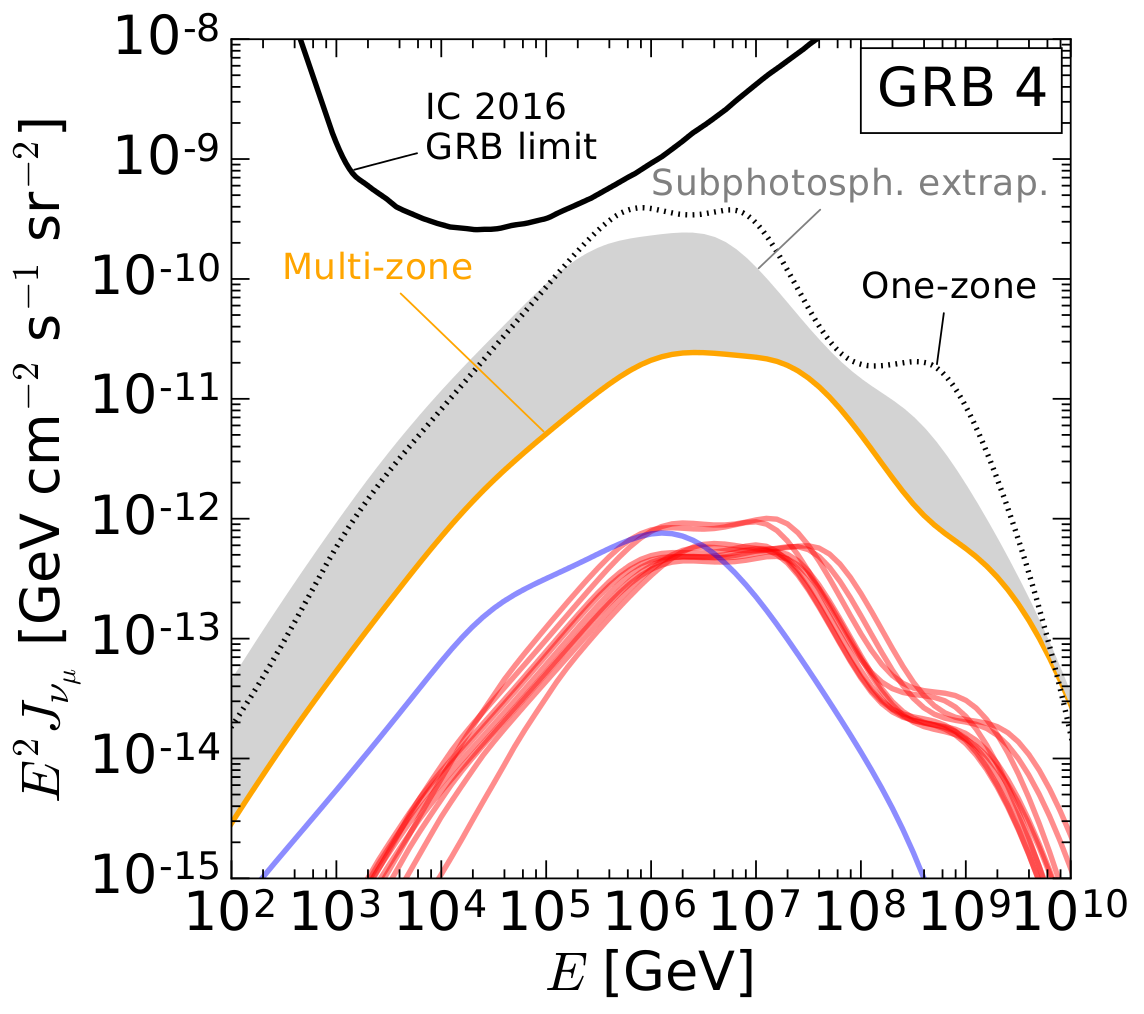}
   \includegraphics[width=0.33\textwidth]{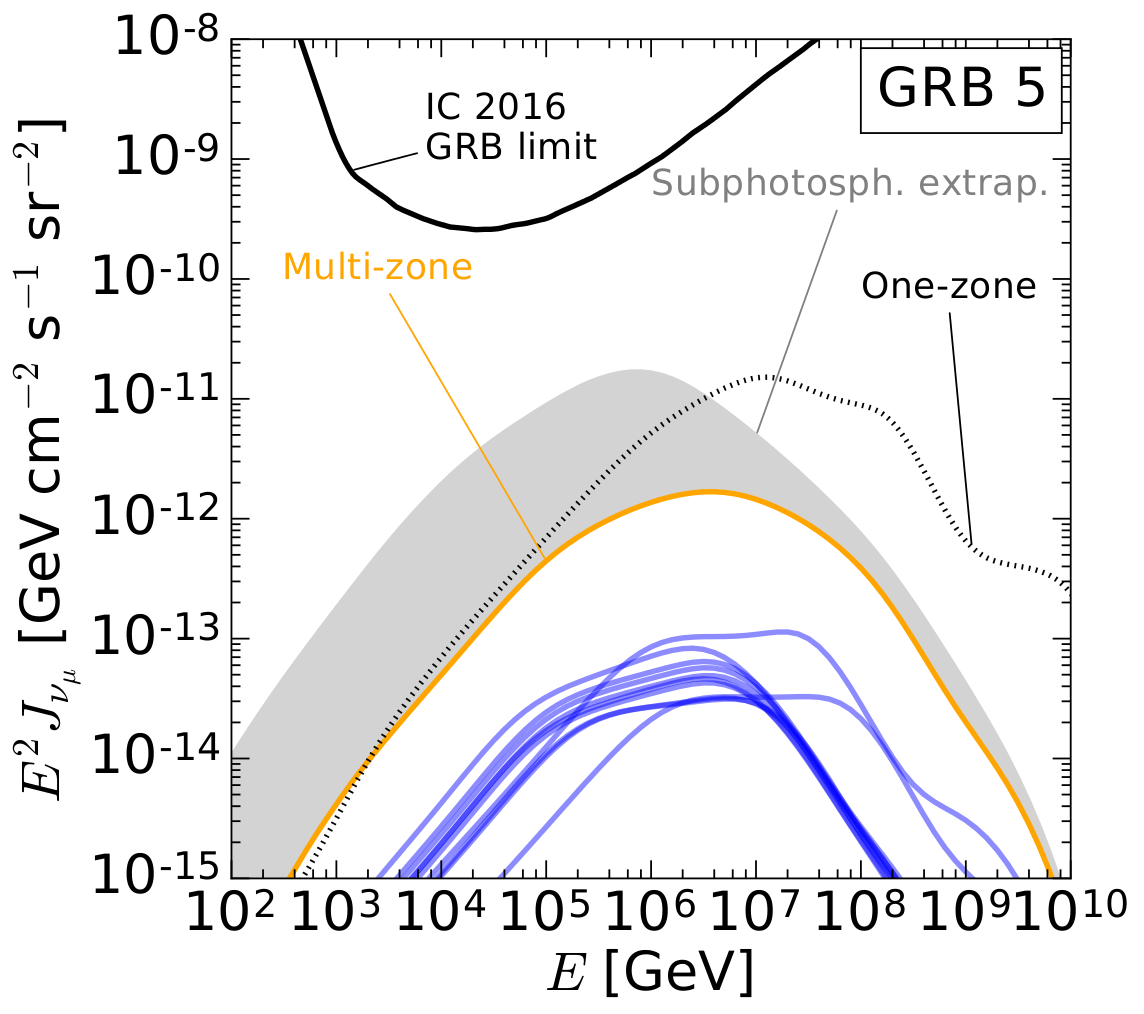}
   \includegraphics[width=0.33\textwidth]{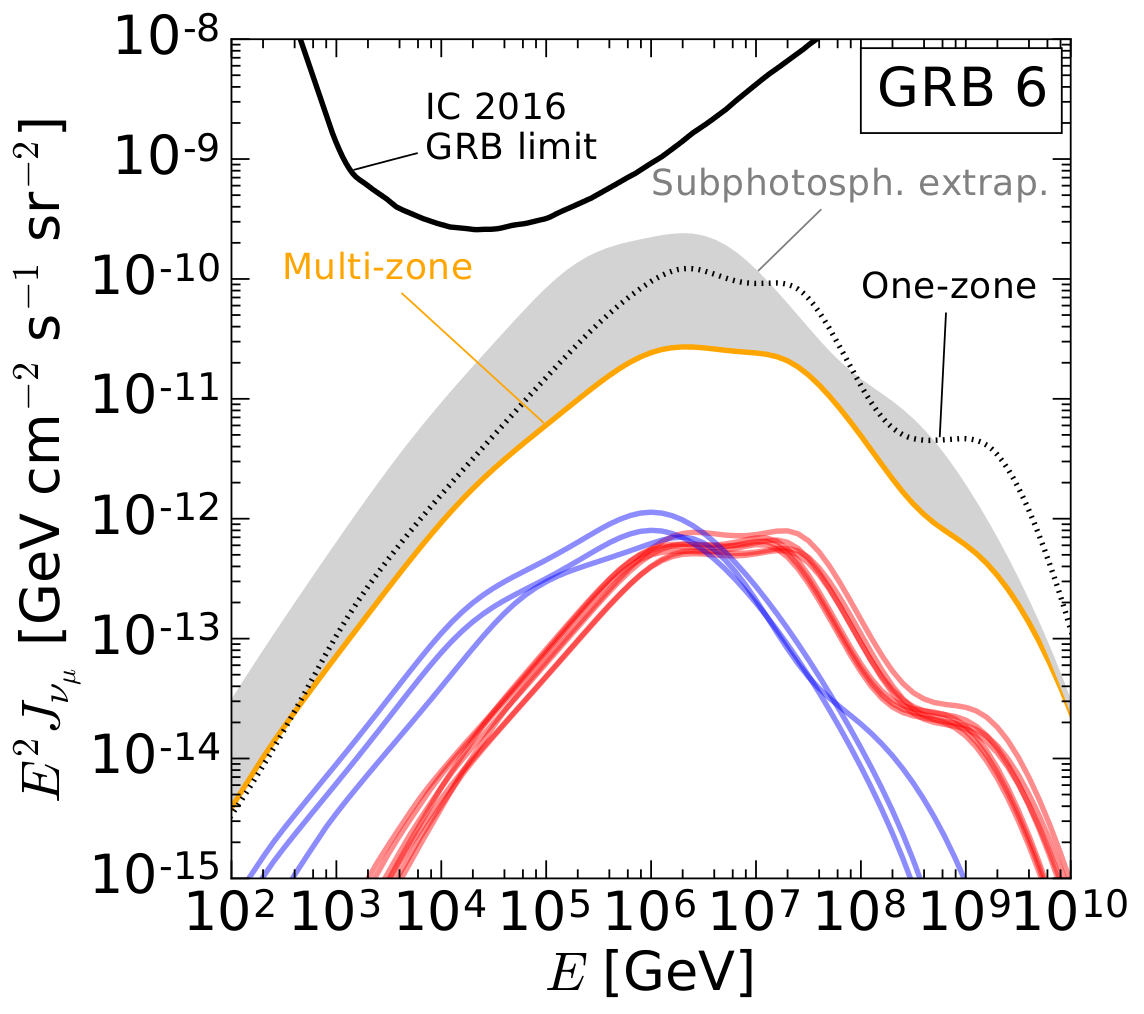}
 \end{center}
 \caption{\label{fig:neutrinos} All-sky quasi-diffuse $\nu_\mu + \bar{\nu}_\mu$  fluxes in our simulated multi-zone GRBs 1--6. Numerical one-zone predictions~\protect{\citep{Hummer:2011ms}} are included for comparison; they are calculated using the average burst parameters computed as in \Sec~\ref{section:LightCurves}. The shaded regions give the potential contribution from sub-photospheric collisions. The dominant contributions from individual collisions are shown as thin curves, corresponding to cases where the optical depth to photohadronic interactions $\tau_{p\gamma}$ is larger (red/light) or smaller (blue/dark) than unity.  The IceCube 2016 upper limit was calculated using their latest reported detector effective area and exposure in a stacked GRB search using tracks coming from the Northern Hemisphere\ \citep{Aartsen:2016qcr}.}
\end{figure*}

We derive the all-sky quasi-diffuse $\nu_\mu + \bar{\nu}_\mu$ flux $J_{\nu_\mu}$ associated to a particular simulated GRB by scaling its fluence by the rate of long-duration GRBs, $\dot{N} = 667$ per year, \ie, $J_{\nu_\mu} = \mathcal{F}_{\nu_\mu} \cdot \dot{N} \cdot \left( 4 \pi \right)^{-1}$. Since this flux does not contain contributions from sub-photospheric collisions, it is effectively a lower limit on the prompt GRB neutrino flux. For our original benchmark, GRB 1,~\citet{Bustamante:2014oka} found that the flux is robust against variations in burst parameters like $\delta t_\text{eng}$ (see \Tab\ \ref{tab:ParameterDescription}) and $N_\text{sh}$. We will discuss below how it depends on underlying assumptions, and what the corresponding fluxes are for GRBs 2--6.

Figure \ref{fig:neutrinos} shows the fluxes for GRBs 1--6.
For all but GRB 5, the flux is $E^2 \, J_{\nu_\mu} \approx 2 \cdot 10^{-11} \, \mathrm{GeV \, cm^{-2} \, s^{-1} \, sr^{-1}}$ around 1 PeV --- this is close to the value found for GRB 1 in~\citet{Bustamante:2014oka}. For GRB 5, the flux is somewhat lower, as it has fewer optically thick collisions, which is in agreement with \equ{fnu}. (For GRB 3, the same could be expected, but instead it has a higher neutrino flux, as explained in \Sec\ \ref{sec:WeakVsStrong}.) Most real light curves lack the non-trivial features seen in GRB 5 (and GRB 3), and are therefore likely strong neutrino emitters. So it is conceivable that most bursts are instead like GRBs 1, 2, 4, and 6, and that the quasi-diffuse flux lies indeed at the level predicted using those bursts.

IceCube has searched for correlations between neutrino arrival directions and positions of known GRBs~\citep{Abbasi:2012zw,Aartsen:2014aqy,Aartsen:2016qcr}. No significant signal from GRBs has been found, in consistency with our prediction.  \figu{neutrinos} includes the differential upper bound from~\citet{Abbasi:2012zw}.

One-zone and multi-zone models have similar (average) burst parameters and compute the flux of secondaries similarly. However, \figu{neutrinos} shows that fluxes calculated with the multi-zone model are typically lower than with the one-zone model~\citep{Hummer:2011ms} (see dashed curves in \figu{neutrinos}). The reason is that all shells are assumed to have the same collision radius in the one-zone model, which tends to be underestimated: since the neutrino production efficiency decreases non-linearly with the collision radius, the average value of the collision radius is, in general, not representative for the neutrino production. Nevertheless, we could define a single effective collision radius for neutrino production in the one-zone model that is different from radius for gamma rays; this would be done by folding in the production efficiency calculated with the fraction of collisions occurring at that radius. 
The average or representative jet parameters --- such as the typical collision radius --- are derived from gamma-ray observations (see \Sec~\ref{section:LightCurves}), which implies that parameters representative for gamma rays may not be representative for the other messengers.

%%%%%%%%%%%%%%%%%%%%%%%%%%%%%%%%%%%%%%%%%%%%%%%%%%%%%%%%%%%%%%%%%%%%%%%%%%%%%%%%%%%

\subsection{Cosmic rays}

From \Tab~\ref{tab:OutputParameters}, we can see that all of the GRBs 1--6 are relatively efficient cosmic-ray emitters, although the required energy output per GRB, of at least $10^{53} \, \mathrm{erg}$ in the discussed energy range, should likely be a factor of a few larger to explain UHECR observations.\footnote{For details, see \Sec~2 in~\citet{Baerwald:2014zga}, where also the dependence on the source evolution is discussed. Such an increase can be achieved either by a somewhat larger gamma-ray luminosity, or by a somewhat larger baryonic loading.} Within the presented model, it is conceivable that GRBs are the sources of UHECRs.

The connection between cosmic rays and neutrinos depends on how UHECRs escape the shells. Photohadronic interactions will transform protons into neutrons; neutrinos will also be produced. If {\it all} cosmic rays escape as neutrons (``neutron escape'') the connection is strong~\citep{Mannheim:1998wp,Ahlers:2011jj}: one neutrino of each flavor is expected per observed UHECR proton. This possibility can be clearly ruled out~\citep{Abbasi:2012zw,Baerwald:2014zga}. However, at the highest energies, when the proton Larmor radius exceeds the shell width, protons can directly escape the shells without interacting in it, which leads to a hard spectrum (``direct escape''). In addition, diffusion may lead to escape depending on properties of magnetic fields.

In each merged shell, one or another escape component dominates\footnote{This holds also for UHECR nuclei that have not been photodisintegrated~\citep{Globus:2014fka}. The survival of heavy nuclei is shown to be possible and their escaping flux may explain the observed UHECR flux~\citep{Murase:2008mr,Wang:2007xj}.}, depending on the properties of the shell~\citep{Baerwald:2013pu}. If we consider escape processes other than  neutron escape, which is implicitly assumed in most of the previous literature~\citep{Waxman:1997ti,Dermer:2003zv,Guetta:2003wi,Murase:2005hy}, the latest IceCube data cannot exclude GRBs as the sources of UHECRs even in a one-zone model, but constraints on the average shell parameters can be obtained from the efficient neutrino production~\citep{Baerwald:2014zga}. 

Figure \ref{fig:scatter}, bottom, shows the maximum proton energy $E_{p,\text{max}}$ to which protons are accelerated in collisions in GRB 2, as a function of collision radius. Below the photosphere (black boxes), proton synchrotron losses dominate and $E_{p,\text{max}}$ increases with $R_\text{C}$. Around $R_\text{C} \simeq 10^{8.5}$--$10^{10} \, \mathrm{km}$, protons reach $10^{10}$ GeV and higher. This is where UHECRs are emitted. At large $R_\text{C}$, falling magnetic fields yield lower acceleration rates and energies.
Neutron emission is correlated with efficient neutrino production, since neutrons and charged pions are produced together in $p\gamma$ interactions. However, this occurs only in a few collisions, in a narrow range of low collision radii, where proton and photon densities are high; see red dots. In effect, cosmic ray emission via neutron escape is limited by \equ{fnu}.
Most collisions occur at larger radii, so that the average collision radius for CR emission tends to be higher than for neutrinos (blue circles). There, particle densities are low enough for direct proton escape to dominate, without associated neutrino production. Given that only few collisions are neutron-dominated, the pure neutron escape assumptions for GRBs~\citep{Ahlers:2011jj} cannot be justified.  However, a quantitative statement requires further study beyond the scope of this work, as it depends on the relative contribution between neutron-dominated and direct escape-dominated shells.
For a discussion of UHECR nuclei, see~\citet{Bustamante:2014oka,Globus:2014fka}.

%%%%%%%%%%%%%%%%%%%%%%%%%%%%%%%%%%%%%%%%%%%%%%%%%%%%%%%%%%%%%%%%%%%%%%%%%%%%%%%%%%%

\subsection{Multi-messenger emission from different radii}

\begin{figure*}[tp]
 \begin{center}
   \includegraphics[width=0.33\textwidth]{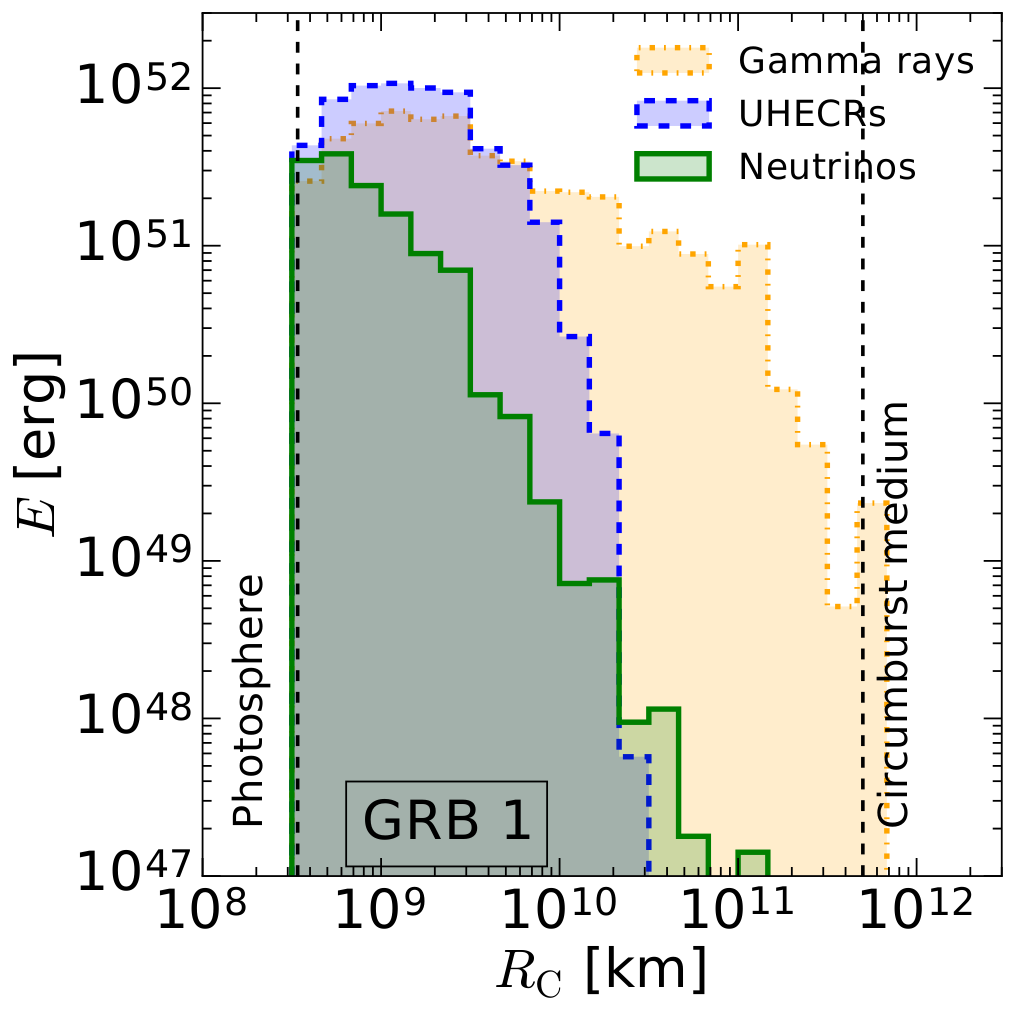}
   \includegraphics[width=0.33\textwidth]{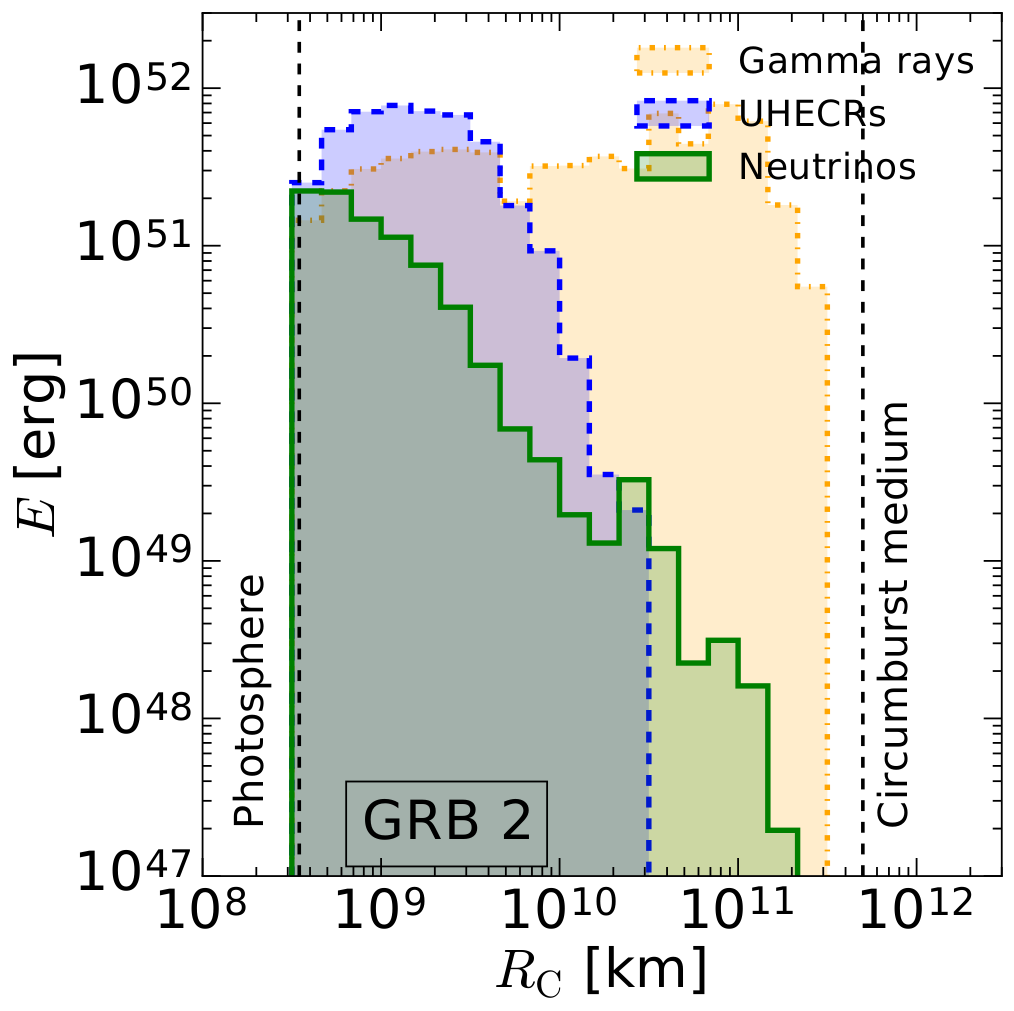}
   \includegraphics[width=0.33\textwidth]{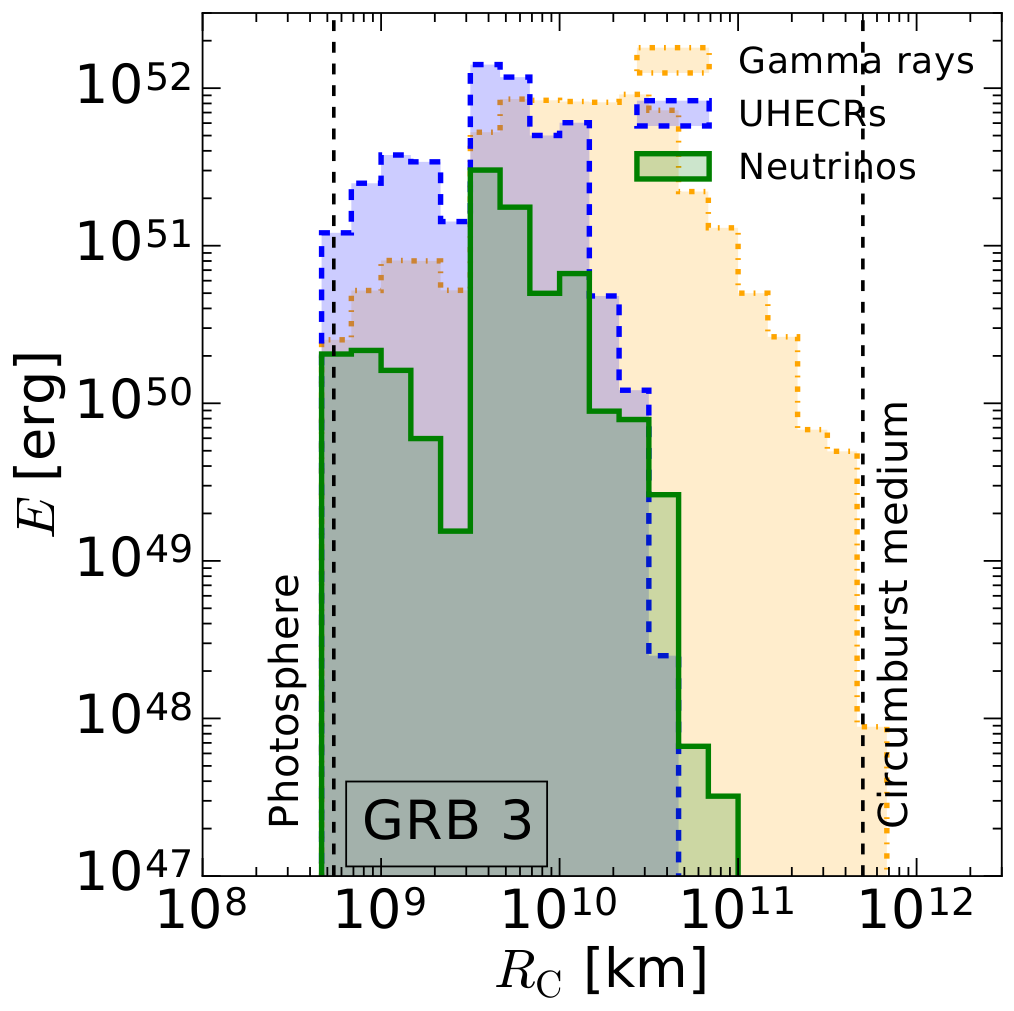}
   \includegraphics[width=0.33\textwidth]{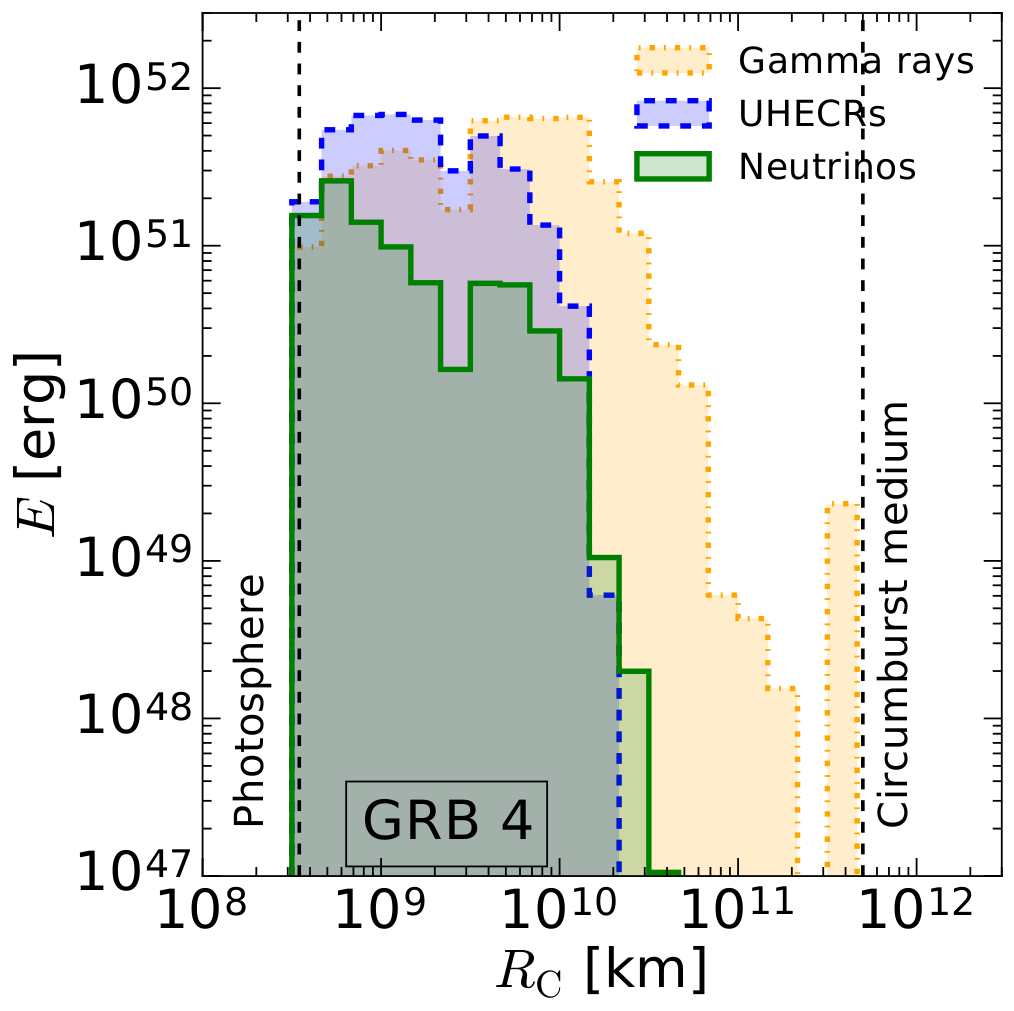}
   \includegraphics[width=0.33\textwidth]{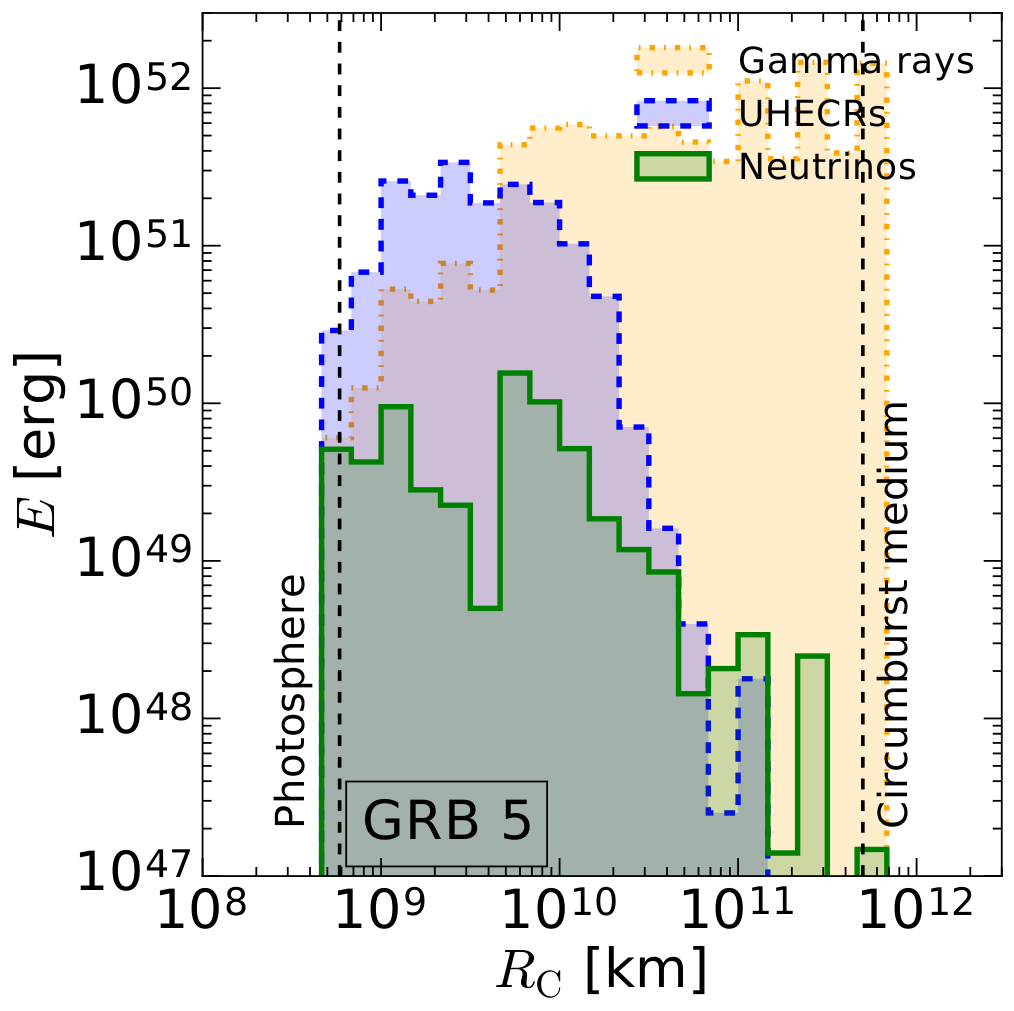}
   \includegraphics[width=0.33\textwidth]{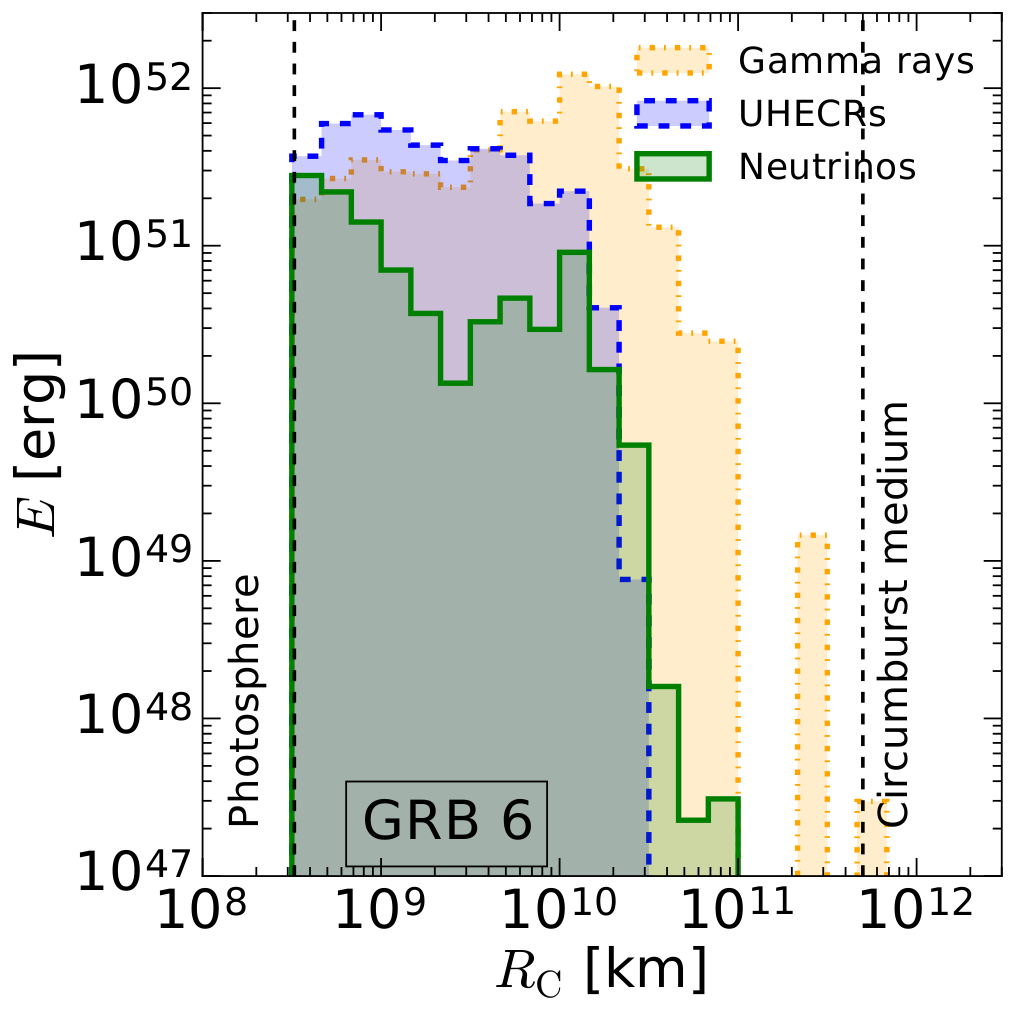}
 \end{center}
 \caption{\label{fig:energy} Energy output as a function of collision radius in neutrinos, UHECR protons ($E_p>10^{10} \, \text{GeV}$), and gamma rays. The approximate photospheric and (assumed) circumburst radius are marked, as well as the UHECR escape regions where either neutron escape or direct escape dominates.}
\end{figure*}

Figure~\ref{fig:energy} shows a key feature of the multi-zone GRB model that is not captured by the one-zone model: that neutrinos, gamma rays, and UHECR protons are emitted from different regions of the jet~\citep{Bustamante:2014oka}. This holds regardless of the difference in burst parameters among GRBs 1--6. Neutrinos are produced close to the photosphere, as discussed above. UHECRs tend to be produced at somewhat larger radii.  At low radii, UHECRs escape as neutrons; at larger radii, most UHECRs escape directly as protons.  Gamma rays tend to come from even larger radii.  While their production is more evenly distributed in collision radius, at low radii, pair-production ($\gamma \gamma \to e^+ e^-$) drives their energy down, so high-energy gamma rays mainly come from large radii.

%%%%%%%%%%%%%%%%%%%%%%%%%%%%%%%%%%%%%%%%%%%%%%%%%%%%%%%%%%%%%%%%%%%%%%%%%%%%%%%%%%%
%%%%%%%%%%%%%%%%%%%%%%%%%%%%%%%%%%%%%%%%%%%%%%%%%%%%%%%%%%%%%%%%%%%%%%%%%%%%%%%%%%%

\section{Delayed high-energy gamma rays}
\label{sec:delay}

\begin{figure*}[t]
 \begin{center}
   \includegraphics[width=\textwidth]{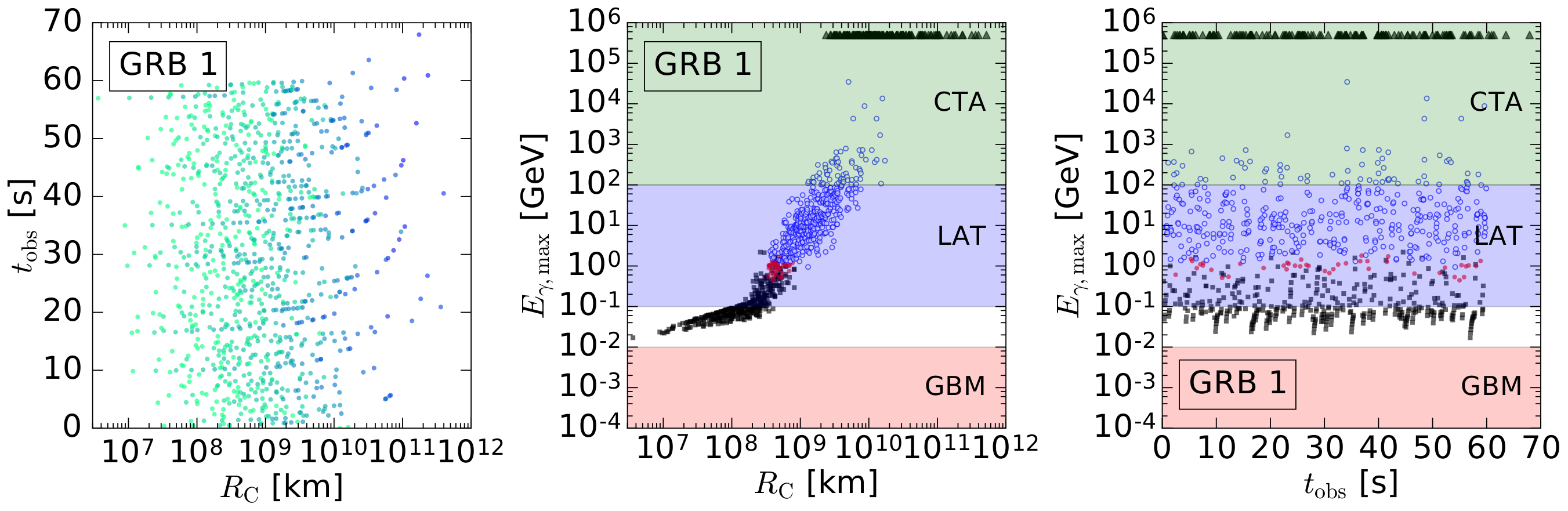} %GRB 1
   \includegraphics[width=\textwidth]{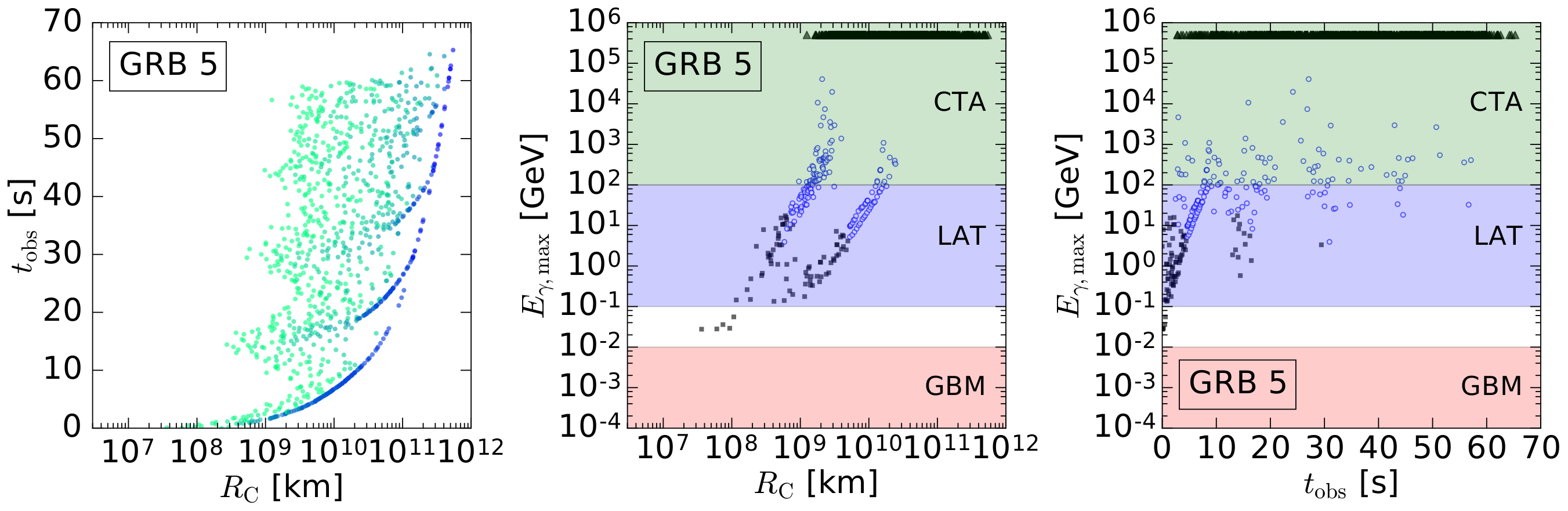} %GRB 5
 \end{center}
 \caption{\label{fig:emaxgamma} Collision time in the observer's frame as a function of collision radius (left column), and maximum gamma-ray energy in the source frame as a function of collision radius (central column) and time (right column), for collisions in GRBs 1 (top row) and 5 (bottom row). In the left column, collisions between older shells --- that have undergone multiple mergers --- are darker, while collisions between younger collisions are lighter; the solid black lines are the average trends. In the central and right columns, collisions are labeled as in \figu{scatter}. Energy ranges accessible by {\it Fermi}-GBM, {\it Fermi}-LAT, and CTA are shaded. Arrows ($\blacktriangle$) represent collisions where the maximum gamma-ray energy is not limited by pair production.}
\end{figure*}

The detection of time delays between gamma-ray signals in different energy bands can provide insight into the dynamics of the GRB central engine and jet.

The maximum energy with which a photon can escape the shell where it is created is limited by the optical depth $\tau_{\gamma\gamma}$ to pair production via $\gamma \gamma \to e^+ e^-$. A photon of a certain energy escapes only if $\tau_{\gamma\gamma} < 1$. Close to the central engine, photon density and, therefore, optical depth, are high. \citet{Bustamante:2014oka} showed that only low-energy gamma rays escape from that region. Higher-energy gamma rays escape at larger radii. In this section, we explore whether the different shell opacities lead to time delays between gamma-ray energy bands.

Figure~\ref{fig:emaxgamma} --- top row, left panel --- shows that, in GRB 1, $t_\text{obs}$ is quite uniformly distributed in $R_\text{C}$, especially between $10^8$ and $10^{10} \, \mathrm{km}$, \ie, above the photosphere. The central panel shows that for many collisions in this range the maximum gamma-ray energy is limited by pair production, while, from $10^9$ km up, an increasing number of collisions is unaffected by it. However, the right panel shows that no separation exists between the arrival times of gamma rays limited and not limited by pair production. In other words, at any time during the burst, low- and high-energy gamma rays arrive indistinctly at Earth from everywhere inside the jet.

For GRB 5 (\figu{emaxgamma}, bottom row), the situation is different. The average $t_\text{obs}$ increases with $R_\text{C}$ between $10^9$ and $10^{11} \, \mathrm{km}$, \ie, some early ($\lesssim 20$ s) gamma-ray detections come from mid-range radii, while all late detections tend come from large radii. As a result, some early gamma rays have lower energies, in the upper {\it Fermi}-LAT and lower CTA bands, limited by pair production. Later detections, coming from larger radii, will have consistently higher energies, not limited by pair production.
There is also an impact on the neutrino light curve: \figu{lcurves} shows that the neutrino flux is much lower for later collisions, which come from larger radii, where neutrino production is inefficient. This behavior is characteristic of bursts with narrow $\Gamma$ distribution, where collisions tend to occur at large radii and late in the jet evolution.

\begin{figure}[tp]
 \begin{center}
  \includegraphics[width = \columnwidth]{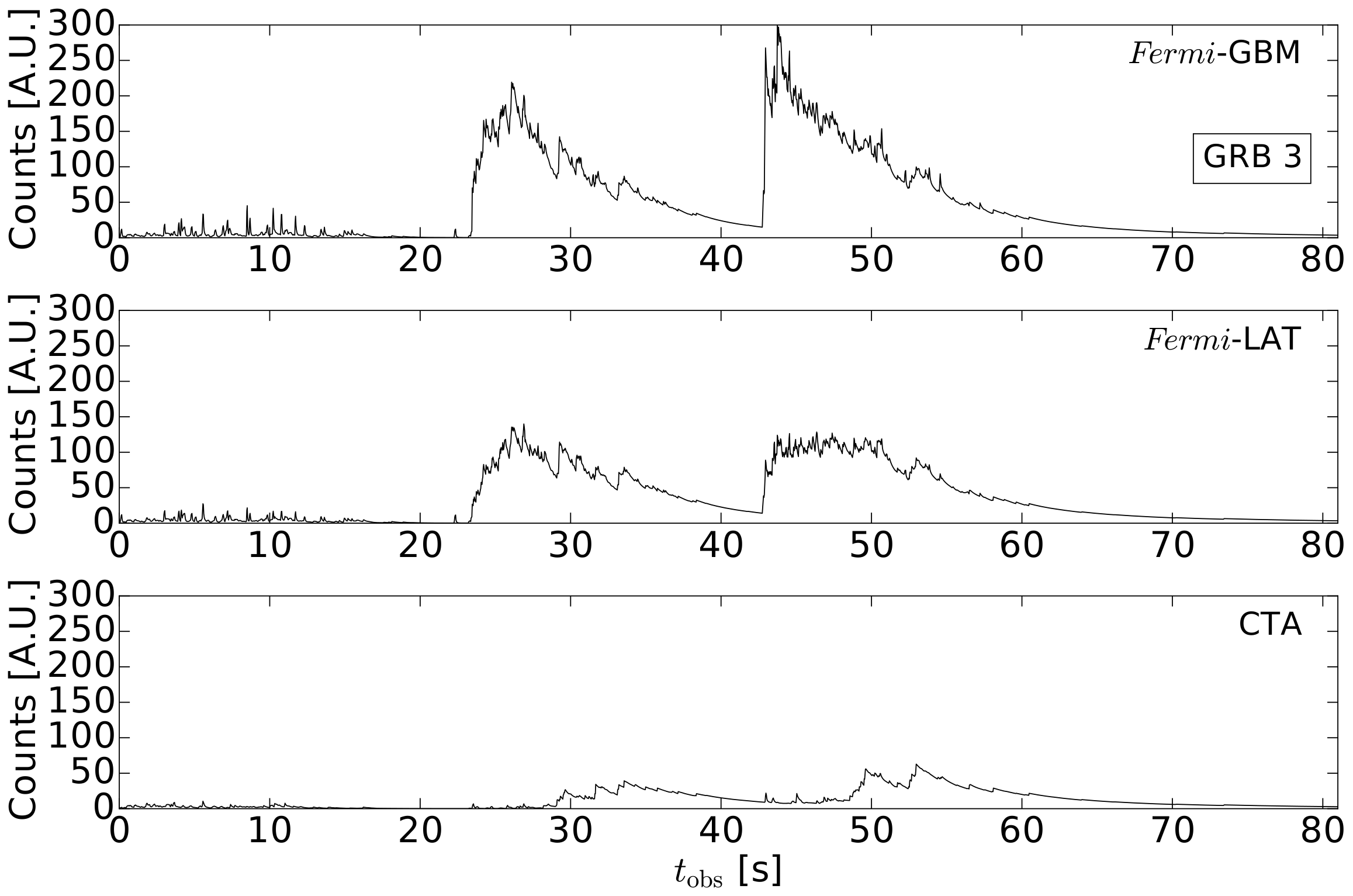}
  \includegraphics[width = \columnwidth]{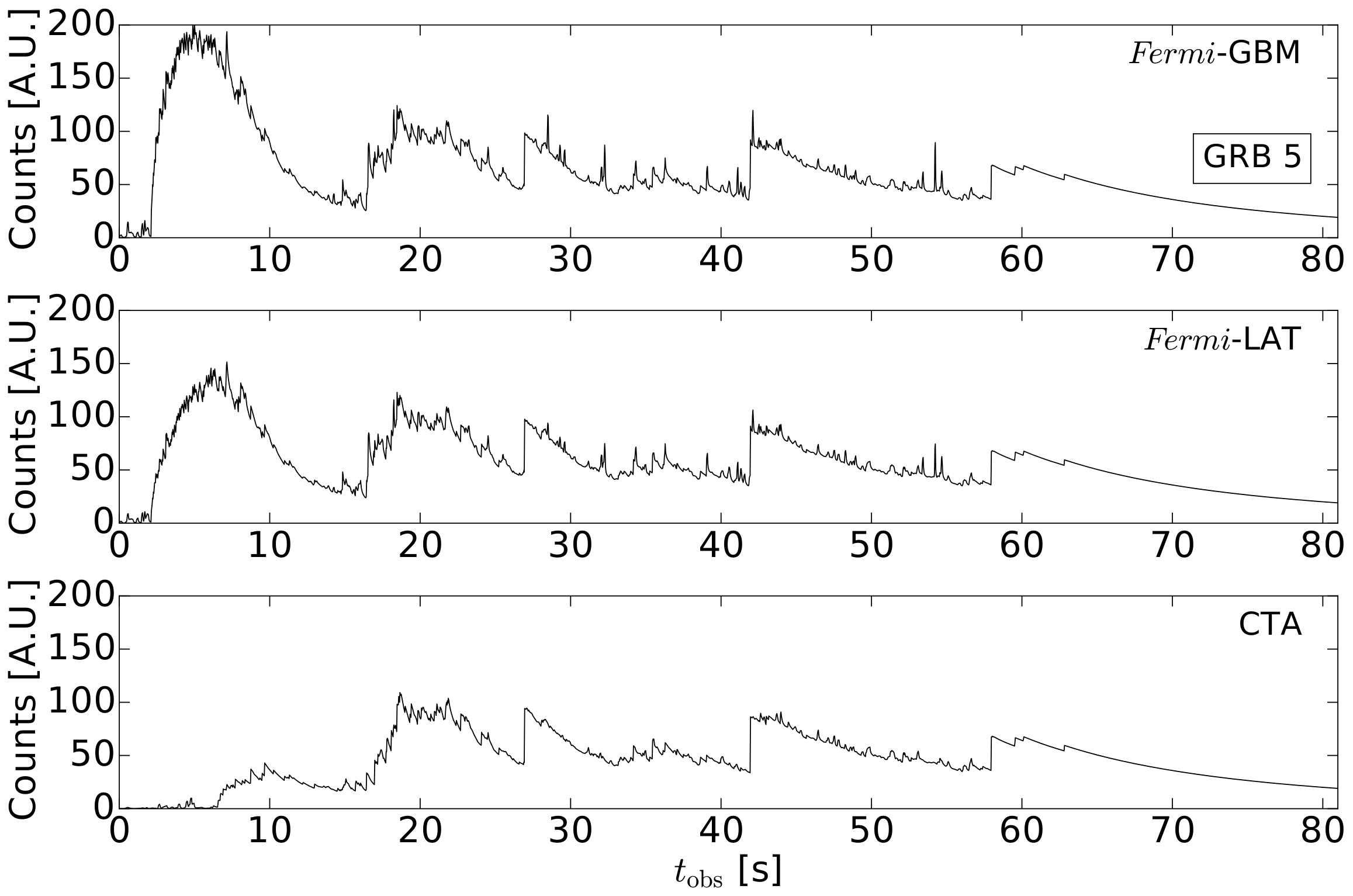}  
 \end{center}
 \caption{\label{fig:bands} Gamma-ray light curves --- using super-photospheric collisions only --- for GRBs 3 and 5, in different energy bands: {\it Fermi}-GBM: $10^{-6}$--$10^{-2}$ GeV; {\it Fermi}-LAT: $10^{-1}$--$10^{2}$ GeV; and CTA: $10^2$--$10^6$ GeV.
 }
\end{figure}

Figure \ref{fig:bands} shows the gamma-ray light curves in different energy bands for GRBs 3 and 5, our two simulations with narrow $\Gamma$ distributions. To produce them, we have assumed that the power-law photon density in the source extends to high enough energies\footnote{This might overestimate the relative height of the light curve in the {\it Fermi}-LAT band.}.  GRB 3 has a delayed start of a few seconds in the LAT band compared to the first detected peak in GBM, and of $\sim 10$ s in the higher-energy CTA band. These delays depend strongly on the energy threshold of the instrument.
In GRB 5, the LAT peak follows the GBM peak after $\sim 2$ s, and the signal in CTA grows to comparable levels $\sim 10$ s later. For this GRB, the suppression affects mainly the first peak in the light curve (the overall normalization of the light curves is arbitrary, but the relative normalization among the different energy bands is fixed). The early suppression of high-energy emission is consistent with observations; see, \eg,~\citet{Castignani:2014gaa}.

Since bursts with time delays between energy bands have narrow $\Gamma$ distributions, they are associated to light curves with broad pulses overlaid with fast variability and possibly weak neutrino emitters (see \Sec~\ref{sec:SyntheticLightCurves}). It is possible, in principle, to use the observation of delays in population studies to find how common these GRBs actually are, which affects the neutrino flux from the full GRB population.

To summarize, our predictions are:
\begin{enumerate}
 \item 
  In GRBs with light curves that have broader pulses overlaid with fast variability, time delays in different wavelength bands are possible.
  \item
  Compared to detection in the GBM energy band ($10^{-6}$--$10^{-2}$ GeV), typical delays are a few seconds long in the LAT band ($10^{-1}$--$10^2$ GeV) and $\sim 10$ s in the CTA band ($10^2$--$10^6$ GeV).
 \item 
  If such delays are observed, the GRB can be a weak neutrino emitter.
 \item
  The fundamental reason for these apparent delays is an early suppression  --- rather than an actual delay --- of high-energy gamma rays coming from smaller $R_\text{C}$, where the radiation densities are higher and the gamma rays cannot escape. This effect is only observable if observation time and $R_\text{C}$ in the collisions are correlated.
\end{enumerate}
An example of a GRB that could match these predictions is GRB 080916C~\citep{Abdo:2009zza}.

These predictions are a qualitative summary based on examples GRB 1--6, which we believe to be representative of a larger set of examples that we have produced. Quantitative results depend on the chosen parameter values, which reflects the observation that no two light curves are identical. \figu{emaxgamma} captures the central feature we observe in all examples that exhibit a 
lag: for these, there is a correlation between $t_\text{obs}$ and $R_\text{C}$ (within the range of $R_\text{C}$ values in which the maximal photon energy depends on $R_\text{C}$).  This 
correlation can be traced back to the engine emitting shells in a relatively narrow range of values of $\Gamma$. The lags in \figu{bands} are clearly visible, although one can see some differences concerning their interpretation. For example, for GRB 3, the first tall peak in the CTA band is clearly delayed by several seconds with respect to the {\it Fermi}-GBM band, whereas, for GRB 5, the precise size of the  lag depends on the instrument threshold of CTA.

%%%%%%%%%%%%%%%%%%%%%%%%%%%%%%%%%%%%%%%%%%%%%%%%%%%%%%%%%%%%%%%%%%%%%%%%%%%%%%%%%%%
%%%%%%%%%%%%%%%%%%%%%%%%%%%%%%%%%%%%%%%%%%%%%%%%%%%%%%%%%%%%%%%%%%%%%%%%%%%%%%%%%%%

\section{Summary and conclusions}\label{sec:Conclusions}

We have simulated the evolution of a baryon-rich GRB jet in the internal shock scenario of the fireball model, by keeping track of all relativistic plasma shells that propagate in it, of the collisions between shells, and of the gamma rays, protons, and neutrinos emitted at the shocks produced during the collisions. Unlike traditional one-zone models that extrapolate the behavior of the whole burst from a single representative collision, our multi-zone simulations consider many such collisions, each happening under different physical conditions.

We have demonstrated that it is possible to generate the various features observed in GRB light curves by varying the behavior of the central emitter, in particular, the initial speeds with which it puts out shells (see \Sec~\ref{sec:SyntheticLightCurves}). The initial speeds determine where collisions between shells will occur during the jet evolution. In this approach, one can relate the properties of the central emitter to observables.

If the initial distribution of shell speeds is ``disciplined'' or narrow --- even if the average speed changes with time --- the associated gamma-ray light curve will have one or more broad pulses overlaid with fast time variability. The associated neutrino flux can be low, because collisions tend to occur at large collision radii. In addition, there is a correlation between observation time and collision radius, which implies that early-time collisions occur at low radii, where radiation densities are high, and high-energy gamma-ray signals are suppressed. As a consequence, we expect delays in the light curves in the {\it Fermi}-LAT energy band with respect to the ones in the {\it Fermi}-GBM band of a few seconds, and in the CTA band with respect to the {\it Fermi}-GBM band of order ten seconds.

If the distribution of speeds is broader, collisions occur over a wide range of collision radii and the light curve is dominated by fast time variability. In this case, neutrino production is always efficient, because, typically, several collisions occur where the radiation densities are high. In this case, we do not expect delays between gamma-ray wavelength bands, because there is not a strong correlation between observation time and collision radius. Inspection of many GRB light curves reveals that most are actually simple, while the ones typically presented in the literature tend to be the more interesting cases with non-trivial features. This means the class of bursts with broad speed distributions may be more representative of the ``typical'' GRB. 

We have also qualitatively confirmed the findings from~\citet{Bustamante:2014oka} for very different parameter sets. Notably, we have shown that, regardless of the initial speed distribution of the shells, different messengers come from different regions of the same GRB: neutrinos predominantly come from regions close to the center, UHECR protons come from intermediate regions, and high-energy gamma rays tend to come from regions further out from the center, where photon densities are low enough that their energies are not limited by pair production on source photons. We have also confirmed the minimal predicted neutrino flux around  $\sim 2 \cdot 10^{-11} \, \mathrm{GeV \, cm^{-2} \, s^{-1} \, sr^{-1}}$ per flavor around 1 PeV,  as long as a significant fraction of the GRBs has broad initial shell speed distributions, which explain observations better; see, \eg, our examples GRB~1,~2,~4 and~6.

Our main results and conclusions are based exclusively on collisions that occur above the photosphere, where photons are not trapped by Thomson scattering. This allows us to assume that the photon spectra in the shells have the same form as the observed spectra at Earth.

Our results  could be exploited in targeted GRB neutrino searches, such as the ones performed by IceCube, to cull a smaller catalog of GRBs that are specially likely to be strong neutrino sources. The non-observation of neutrinos from bursts in such a catalog --- where associated backgrounds are smaller --- could result in stronger upper bounds on prompt high-energy GRB neutrino emission. Our results could be also tested in different gamma-ray wavelength bands: we do not expect significant delays in GRBs with fast time variability without underlying pulse structure.

Due to their high luminosity, short duration, and the high angular precision with which they are detected, GRBs are arguably our best chance at finding a high-energy neutrino counterpart to electromagnetic emission. The study presented here is a step towards a necessary, realistic multi-messenger understanding of GRBs. The observation of these neutrinos would be a smoking-gun signature of high baryonic loading, and thus of the paradigm that GRBs could be the sources of the UHECRs.

%%%%%%%%%%%%%%%%%%%%%%%%%%%%%%%%%%%%%%%%%%%%%%%%%%%%%%%%%%%%%%%%%%%%%%%%%%%%%%%%%%%
%%%%%%%%%%%%%%%%%%%%%%%%%%%%%%%%%%%%%%%%%%%%%%%%%%%%%%%%%%%%%%%%%%%%%%%%%%%%%%%%%%%

\acknowledgments

The authors thank John Beacom, Denise Boncioli, Valerie Connaughton, \v{Z}eljka Bo\v{s}njak, and Eli Waxman for useful discussion and comments on the manuscript. MB acknowledges the hospitality of DESY Zeuthen during the completion of this work. This project has received funding from the European Research Council (ERC) under the European Union's Horizon 2020 research and innovation programme (Grant No.~646623). MB acknowledges support from NSF Grant PHY-1404311.  The work of KM is supported by NSF Grant PHY-1620777.

%%%%%%%%%%%%%%%%%%%%%%%%%%%%%%%%%%%%%%%%%%%%%%%%%%%%%%%%%%%%%%%%%%%%%%%%%%%%%%%%%%%
%%%%%%%%%%%%%%%%%%%%%%%%%%%%%%%%%%%%%%%%%%%%%%%%%%%%%%%%%%%%%%%%%%%%%%%%%%%%%%%%%%%

\newpage
\clearpage

% \bibliographystyle{apj}
% \bibliography{refs.bib}

\onecolumngrid

\newpage
\clearpage
\pagebreak
\vspace*{-0.8cm}

%%%%%%%%%%%%%%%%%%%%%%%%%%%%%%%%%%%%%%%%%%%%%%%%%%%%%%%%%%%%%%%%%%%%%%%%%%%%%%%%%%%
%%%%%%%%%%%%%%%%%%%%%%%%%%%%%%%%%%%%%%%%%%%%%%%%%%%%%%%%%%%%%%%%%%%%%%%%%%%%%%%%%%%

\appendix

\bigskip

\section{Model description}
\label{sec:model}

Here we describe our GRB jet simulation, based on \cite{Kobayashi:1997jk,Daigne:1998xc}.

%%%%%%%%%%%%%%%%%%%%%%%%%%%%%%%%%%%%%%%%%%%%%%%%%%%%%%%%%%%%%%%%%%%%%%%%%%%%%%%%%%%

\subsection{Overview}

The central object in a GRB (\eg, the black hole created by a collapsing massive star) emits collimated, relativistic jets that are rich in baryons, which we assume to be protons. When one of the jets points towards Earth, the gamma-ray emission from it may be detected as a GRB. We simulate the particle emission from this jet.

Since we are located inside the opening angle of the jet, we cannot distinguish between the emission geometry of a collimated jet and that of spherical, isotropic emission. For simplicity, we develop our formalism  under such an isotropically-equivalent scenario. In it, the central engine emits spherical plasma shells that propagate outwards along the jet at relativistic speeds. We simulate their propagation and collisions between them, which produce high-energy gamma rays, protons, and neutrinos. 

The simulation covers only the coasting phase of the GRB, during which shells propagate at constant speed, except when they collide. In the preceding, acceleration phase, they reached their maximum individual speeds, limited by the available kinetic energy and their masses. The coasting phase ends when the shells reach the circumburst medium; there, they decelerate, and might produce an afterglow. The acceleration and deceleration phases are not part of the simulation. 

In our simulation, shells propagate in one dimension. At any time during the simulation, the $k$-th shell is characterized by four basic parameters: $r_k$, the shell radius, as measured from the emitter (\ie, the position of the shell inside the jet); $l_k$, the shell width; $\Gamma_k$, the shell bulk Lorentz factor; and $m_k$, the shell mass.

When two shells collide, they merge into a new shell, with width, speed, and mass calculated from the properties of the shells that collided. The new shell continues propagating in the jet flow and may collide again. Collisions are inelastic. The new shell cools instantly by radiating away its internal energy via particle emission. We compute collisions numerically, following~\citet{Kobayashi:1997jk,Daigne:1998xc}, as detailed below.

Table~\ref{tab:ParameterDescription} describes all the relevant simulation parameters. Unless otherwise noted, quantities therein and in the text are expressed in the source reference frame.

\begin{table*}[t]
 \centering
 \caption{\label{tab:ParameterDescription}Main parameters of the burst simulation}
 \fontsize{7}{7}\selectfont
 \begin{tabular}{clccl}
  \hline
  \hline
  \textbf{Parameter}            & \textbf{Description}                                 & \textbf{Type}           & \textbf{Units}    & \textbf{Notes}                      \\
  \hline
  \multicolumn{5}{c}{\textbf{Burst initialization}} \\
  \hline
  $N_\text{sh}$                 & Initial number of shells                             & Input    & \---   & \\
  $\delta t_{\text{eng}}$       & Uptime of the engine                                 & Input    &  s     &                   \\
  $\Delta t_{\text{eng}}$       & Downtime of the engine                               & Input    &  s     &                   \\
  $r_{N_\text{sh}}$             & Distance from innermost shell to central emitter     & Input    & km     & Default: $10^3$ km \\
  $r_\text{dec}$                & Deceleration radius (circumburst medium starts)      & Input    & km     & Default: $5.5 \cdot 10^{11}$ km \\
  $A_\Gamma$                    & Fluctuation amplitude of $\Gamma_{k,0}$ distribution & Input    & \---   & \\
  $E_{\text{kin},0}^\text{iso}$ & Initial bulk kinetic energy of shells                & Input    & erg    & Common to all initial shells \\
  $z$                           & Redshift of the emitter                              & Input    & \---   & Used to calculate $t_\text{obs}$, \equ{TimeObserverFrame}: $z=2$ by default \\
  $l$                           & Initial shell width                                  & Internal & km     & Common to all initial shells: $l = c \cdot \delta t_{\text{eng}}$ \\
  $d$                           & Initial separation between consecutive shells        & Internal & km     & Common to all initial shell pairs: $d = l$ by default \\
  $r_{k,0}$                     & Initial radius of the $k$-th shell,                  & Internal & km     & $r_{k,0} = r_{N_\text{sh}} + \left( N_\text{sh} - k \right) \left(l+d\right)$ \\
  $\Gamma_{k,0}$                & Initial bulk Lorentz factor of the $k$-th shell      & Internal & \---   & Sampled from the pre-defined $\Gamma$ distribution \\
  $m_{k,0}$                     & Initial mass of the $k$-th shell                     & Internal & GeV    & $m_{k,0} = E_{\text{kin},0}^\text{iso} / \left( \Gamma_{k,0} c^2 \right)$ \\
  \hline
  \multicolumn{5}{c}{\textbf{Burst evolution}} \\
  \hline
  $t$                           & Time in the source frame                             & Internal & s             & \\
  $r_k$                         & Radius of the $k$-th shell                           & Internal & km            & Grows as $r_k = r_{k,0} + c \beta_k t$  \\
  $l_k$                         & Width of the $k$-th shell                            & Internal & km            & Changes only in collisions \\
  $\Gamma_k$                    & Bulk Lorentz factor of the $k$-th shell              & Internal & \---          & Changes only in collisions \\
  $m_k$                         & Mass of the $k$-th shell                             & Internal & GeV           & Changes only in collisions \\
  $\beta_k$                     & Bulk speed of the $k$-th shell                       & Internal & \---          & $\beta_k = \sqrt{1-\Gamma_k^{-2}}$  \\
  $V_{\text{iso},k}$            & Isotropically-equivalent volume of the $k$-th shell  & Internal & km$^3$        & $V_{\text{iso},k} = 4 \pi r_k^2 l_k$  \\
  $E_{\text{kin},k}^\text{iso}$ & Bulk kinetic energy of the $k$-th shell              & Internal & erg           & Changes only in collisions \\
  $\rho_k$                      & Mass density of the $k$-th shell                     & Internal & GeV km$^{-3}$ & $\rho_k = m_k / V_{\text{iso},k}$ \\
  \hline
  \multicolumn{5}{c}{\textbf{Shell collisions}} \\
  \hline
  $m_{\text{f}\left(\text{s}\right)}$         & Mass of the fast (slow) colliding shell       & Internal & GeV & \\
  $\Gamma_{\text{f}\left(\text{s}\right)}$    & Bulk Lorentz factor of the fast (slow) shell  & Internal & \--- & \\
  $\Gamma_{\text{fs}\left(\text{rs}\right)}$  & Lorentz factor of the forward (reverse) shock & Internal & \--- & See \equ{GammaFSRSDef} \\
  $\beta_{\text{fs}\left(\text{rs}\right)}$   & Speed of the forward (reverse) shock          & Internal & \--- & $\beta_{\text{fs}\left(\text{rs}\right)}=\sqrt{1-\Gamma_{\text{fs}\left(\text{rs}\right)}^{-2}}$ \\
  $\beta_\text{m}$                            & Bulk speed of the merged shell                & Internal & \--- & $\beta_\text{m} = \sqrt{1-\Gamma_\text{m}^{-2}}$ \\
  $\rho_m$                                    & Mass density of the merged shell              & Internal & GeV km$^{-3}$ & See \equ{DensityMerged} \\
  $t_\text{coll}$                             & Collision time (source frame)                 & Internal & s    & \\
  $N_\text{coll}$                             & Total number of collisions                    & Output   & \--- & \\
  $t_\text{obs}$                              & Collision time (observer's frame)             & Output   & s    & See \equ{TimeObserverFrame} \\
  $\Gamma_\text{m}$                           & Bulk Lorentz factor of the merged shell       & Output   & \--- & See \equ{GammaMDef} \\
  $E_\text{coll}^\text{iso}$                  & Internal energy liberated in the collision    & Output   & erg  & See \equ{EintMerged} \\
  $l_\text{m}$                                & Width of the merged shell                     & Output   & km   & See \equ{WidthMerged} \\
  $R_C$                                       & Collision radius                              & Output   & km   & \\
  $\delta t_\text{em}$                        & Time for reverse shock to cross fast shell    & Output & s & See \equ{CrossTime} \\
  \hline
  \hline
 \end{tabular}
 \tablecomments{All quantities are expressed in the source (engine) frame, except for $t_\text{obs}$, which is in the observer's frame.}
\end{table*}

%%%%%%%%%%%%%%%%%%%%%%%%%%%%%%%%%%%%%%%%%%%%%%%%%%%%%%%%%%%%%%%%%%%%%%%%%%%%%%%%%%%

\subsection{Burst initialization}
\label{section:FireballInitialisation}

In the simulation, before shell propagation starts, the central engine has already emitted $N_\text{sh}$ shells. Each one is described by the initial tuple $\left(r_{k,0},l_{k,0},\Gamma_{k,0},E_{\text{kin},0}^\text{iso}\right)$. Shells closer to the engine are labeled with higher indices.

The behavior of the engine is described by two timescales\footnote{This is strictly true for the simulated GRBs 1--5. GRB 6 has an overlaid time structure: the engine has an overall active period where it emits $N_\text{up}$ shells, followed by a quiescent period that lasts for $N_\text{down}$ pulses. See \Tab~\ref{tab:input}.}: an ``uptime'', $\delta t_{\text{eng}}$, during which it emits one shell, followed by a ``downtime'', $\Delta t_{\text{eng}}$, during which it is inactive.  The former determines the initial shell width, $l = c \cdot \delta t_{\text{eng}}$, which we assume to be common for all shells, and the latter determines the initial separation between consecutive shells, $d = c \cdot \Delta t_{\text{eng}}$. Thus, each of the initial shells is located at position $r_{k,0} = r_{N_\text{sh}} + \left(N_\text{sh}-k\right)\left(l+d\right)$, where $r_{N_\text{sh}}$ is the distance from the innermost shell to the emitter, which is an input parameter of the simulation. Results do not depend on $r_{N_\text{sh}}$ strongly, unless its value is too large; see \Tab~\ref{tab:ParameterDescription}.

We choose the values of $l$ and $d$ to reproduce the timescale of pulses in observed light curves~\citep{Nakar:2002gd}. If $t_\text{v}$ is the GRB variability timescale, \ie, the characteristic duration of peaks in the light curve, and $t_\text{q}$ is the characteristic quiescent time between consecutive peaks, we expect that, roughly, $l \approx c \cdot t_\text{v}$ and $d \approx c \cdot t_\text{q}$~\citep{Kobayashi:1997jk,Aoi:2009ty}. In the internal shock model, $d$ and $l$ should be comparable. The simulations in~\citet{Kobayashi:1997jk} set $d = l$, while~\citet{Aoi:2009ty} set $d=5 l$. In our simulations, we chose $\delta t_{\text{eng}} = \Delta t_{\text{eng}}$, such that $d = l = c \cdot \delta t_{\text{eng}}$.

The variability timescale of a simulated burst is not an input parameter of the simulation, but a result of it. For our choices of simulation parameter values, we find that the variability timescale, obtained from the post-simulation synthetic light curve (see \Sec~\ref{section:LightCurves}), is close to the input value of $\delta t_{\text{eng}}$.

The initial values of the shell Lorentz factors follow a pre-defined distribution. In the benchmark scenario GRB 1, it is a log-normal distribution; see \Sec~\ref{sec:SyntheticLightCurves}. \Tab~\ref{tab:input} describes the distributions used in GRBs 2--6.

There are two typical schemes to assign initial masses $m_{k,0}$ to the shells: the equal-mass assumption, \ie, $m_{k,0}=m$ for all $k$; and the equal-energy assumption, \ie, $m_{k,0} = E_{\text{kin},0}^\text{iso} / \left( \Gamma_{k,0} c^2 \right)$, with $E_{\text{kin},0}^\text{iso}$ the initial bulk kinetic energy, assumed common to all shells. Our simulation uses the latter, since it appears to match observations more closely~\citep{Nakar:2002gd}.

%%%%%%%%%%%%%%%%%%%%%%%%%%%%%%%%%%%%%%%%%%%%%%%%%%%%%%%%%%%%%%%%%%%%%%%%%%%%%%%%%%%

\subsection{Burst evolution}\label{section:FireballEvolution}

We simulate the coasting phase of the jet, during which shell speeds do not change while they propagate and expand. In our simplified treatment, the shell width and mass also stay constant\footnote{Depending on the internal energy, shell spreading is important especially after collisions. Recent dedicated simulations take into account this effect, but it is neglected in the simplest versions, like the one we have adopted~\citep{Aoi:2009ty}.}. Therefore, the shell volume $V_{\text{iso},k} = 4 \pi r_k^2 l_k$ grows $\propto r_k^2$ and mass density $\rho_k = m_k / V_{\text{iso},k}$ decreases $\propto r_k^{-2}$. Since the shell mass and Lorentz factor are constant during propagation, its bulk kinetic energy $E_{\text{kin},k}^\text{iso} = \Gamma_k m_k c^2$ is constant as well. Speed, width, and mass change only in collisions.

At the start of the simulation, we calculate the collision time for all pairs of neighboring shells, \ie,
\begin{equation}
 \Delta t_{k,k+1} = \frac{d_{k,k+1}}{c(\beta_{k+1}-\beta_k)} ~,
\end{equation}  
where $d_{k,k+1} \equiv r_k - r_{k+1} - l_{k+1}$ is the separation between shells $k$ and $k+1$. The time interval until the next collision occurs is the minimum of these times, \ie,
\begin{equation}
 \Delta t_\text{next} = {\rm min}[\Delta t_{k,k+1}] \;.
\end{equation}
We increase the simulation time to $t \to t + \Delta t_\text{next}$. The collision radius $R_\text{C}$ is set to the radius of the innermost colliding shell. Light emitted from this collision will be detected by a distant observer at time
\begin{equation}\label{equ:TimeObserverFrame}
 t_{\rm obs} = \left(\frac{D\left(z\right)-R_\text{C}}{c} + t\right) \left(1+z\right)  \;,
\end{equation}
with $D\left(z\right)$ the light-travel distance to the emitter with redshift $z$. These equations satisfy well-known relations in the internal shock scenario, $\Delta t_\text{next} \approx 2 \Gamma^2 (d/c)$ and $R_\text{C} \approx 2 \Gamma^2 d$. All shells propagate to their new positions $r_k \to r_k + c \beta_k \Delta t_\text{next}$, the time interval for the next collision is calculated, and the process is repeated. (The term $D\left(z\right)/c$ is just an offset: it will disappear when, in the simulation output, the first emission is set to start at $t_\text{obs} = 0$.)

\begin{figure*}[t!]
 \centering
 \includegraphics[width=0.8\textwidth]{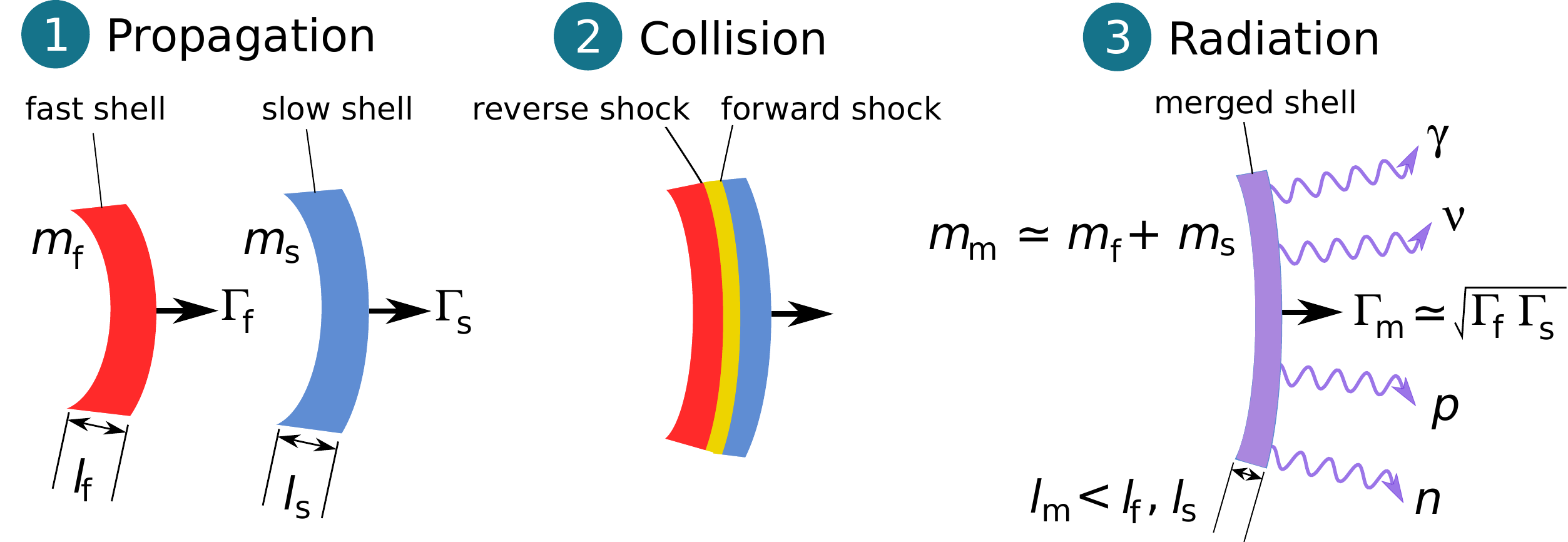}
 \caption{\label{fig:schematic_collision} Schematic illustration of the collision process between two plasma shells, and the ensuing emission of high-energy particles.}
\end{figure*}

We distinguish between the shell bulk kinetic energy, $E_{\text{kin},k}^\text{iso}$, and its internal energy, $E_{\text{int},k}^\text{iso}$. The former is related to the motion of the compact shell, measured in the source rest frame, while the latter is the aggregated kinetic energy of particles moving randomly inside the shell, measured in the shock rest frame.

In a collision, the kinetic energy of the two colliding shells is used partly as bulk kinetic energy for the new shell and partly as its internal energy. For simplicity, we assume that the new shell immediately cools by prompt particle emission; see~\citet{Kobayashi:2001iq} for alternative treatments. While collision details depend on modeling of hydrodynamical properties~\citep{Daigne:2003tp}, here we adopt the simple collision prescription for the relativistic limit introduced in~\citet{Kobayashi:1997jk}, which we outline below.

In the collision of a slow (s) and a fast (f) shell, the internal energy of the merged (m) shell is the difference of kinetic energy before and after the collision, \ie,
\begin{equation}\label{equ:EintMerged}
 E_{\rm coll}^{\rm iso} = (\Gamma_\text{f}  m_\text{f}  + \Gamma_\text{s} m_\text{s}  ) c^2 - \Gamma_\text{m} (m_\text{s} + m_\text{f}) c^2~.
\end{equation}
We assume that this amount of internal energy of the merged shell is radiated away as secondary particles.
From momentum and energy conservation, and assuming $\Gamma_\text{f}, \Gamma_\text{s} \gg 1$, the Lorentz factor of the merged shell is
\begin{equation}\label{equ:GammaMDef}
 \Gamma_\text{m} \simeq \sqrt{\frac{\Gamma_\text{f} m_\text{f} + \Gamma_\text{s} m_\text{s}}{m_\text{f}/\Gamma_\text{f} + m_\text{s}/\Gamma_\text{s}}} \, ,
\end{equation}
which reduces to $\Gamma_\text{m} \simeq \sqrt{\Gamma_\text{f} \Gamma_\text{s}}$ if $m_\text{f} \simeq m_\text{s}$.
Its width is given by~\citep{Kobayashi:1997jk,Aoi:2009ty}
\begin{equation}\label{equ:WidthMerged}
 l_\text{m} \simeq l_\text{s} \frac{\beta_{\rm fs}-\beta_\text{m}}{\beta_{\rm fs}-\beta_\text{s}} + l_\text{f} \frac{\beta_\text{m}-\beta_{\rm rs}}{\beta_\text{f}-\beta_{\rm rs}} \; ,
\end{equation}
where $\beta_\text{fs(rs)} = \sqrt{1-\Gamma_\text{fs(rs)}^{-2}}$ is the speed of the forward (reverse) shock, whose Lorentz factor is
\begin{equation}\label{equ:GammaFSRSDef}
 \Gamma_\text{fs(rs)} = \Gamma_\text{m} \sqrt{\frac{1+2\Gamma_\text{m}/\Gamma_\text{s(f)}}{2+\Gamma_\text{m}/\Gamma_\text{s(f)}}} ~.
\end{equation}
The volume of the new shell is $V_{\text{iso,m}} = 4 \pi R_\text{C}^2 l_\text{m}$.
The density is different between the shocked faster shell and shocked slower shell. 
The new shell has an average density obtained from the assumption of an inelastic collision $m_m \simeq m_\text{f} + m_\text{s}$ already implied in \equ{EintMerged}, \ie,
\begin{equation}\label{equ:DensityMerged}
 \rho_\text{m} \simeq \frac{l_\text{f} \cdot \rho_\text{f}  + l_\text{s} \cdot \rho_\text{s}}{l_\text{m}} \; .
\end{equation}
Therefore, its mass is $m_\text{m} = V_{\text{iso,m}} \rho_\text{m}$, and its kinetic energy is $E_{\text{kin,m}}^\text{iso} = \Gamma_\text{m} m_\text{m} c^2$.
After the collision, the original fast shell is removed from the simulation and the new shell replaces the former slow shell. It is then propagated with the remaining shells in the jet.

If a shell reaches the circumburst medium, where it decelerates, it is removed from the simulation. Following~\citet{Rees:1992ek}, we assume $r_\text{dec} = 5.5 \cdot 10^{11}$ km for the radius at which this happens (see, \eg, Eq.~(15) in~\citet{Meszaros:2006rc}).

The simulation finishes when all shells have reached the circumburst medium, all shells have merged into a single remaining shell, or all remaining shells are ordered outwards with increasing Lorentz factor, so that no more collisions are possible. The output lists $N_\text{coll}$ collisions,
\begin{equation}
 \left( t_{\text{obs},k}, R_{\text{C},k}, l_{\text{m},k}, \Gamma_{\text{m},k}, E_{{\rm coll},k}^{\rm iso} \right) \, , \label{equ:shells}
\end{equation}
where $1 \le k \le N_\text{coll}$. The minimum $t_{{\rm obs},k}$ is taken to be the start of the observation time of the burst and is set to zero. Collisions are arranged so that $t_{\text{obs},1} = 0 \le t_{\text{obs},2} \le \hdots \le t_{\text{obs},N_\text{coll}}$. 

\begin{figure}[t!]
 \centering
 \includegraphics[width=0.45\columnwidth]{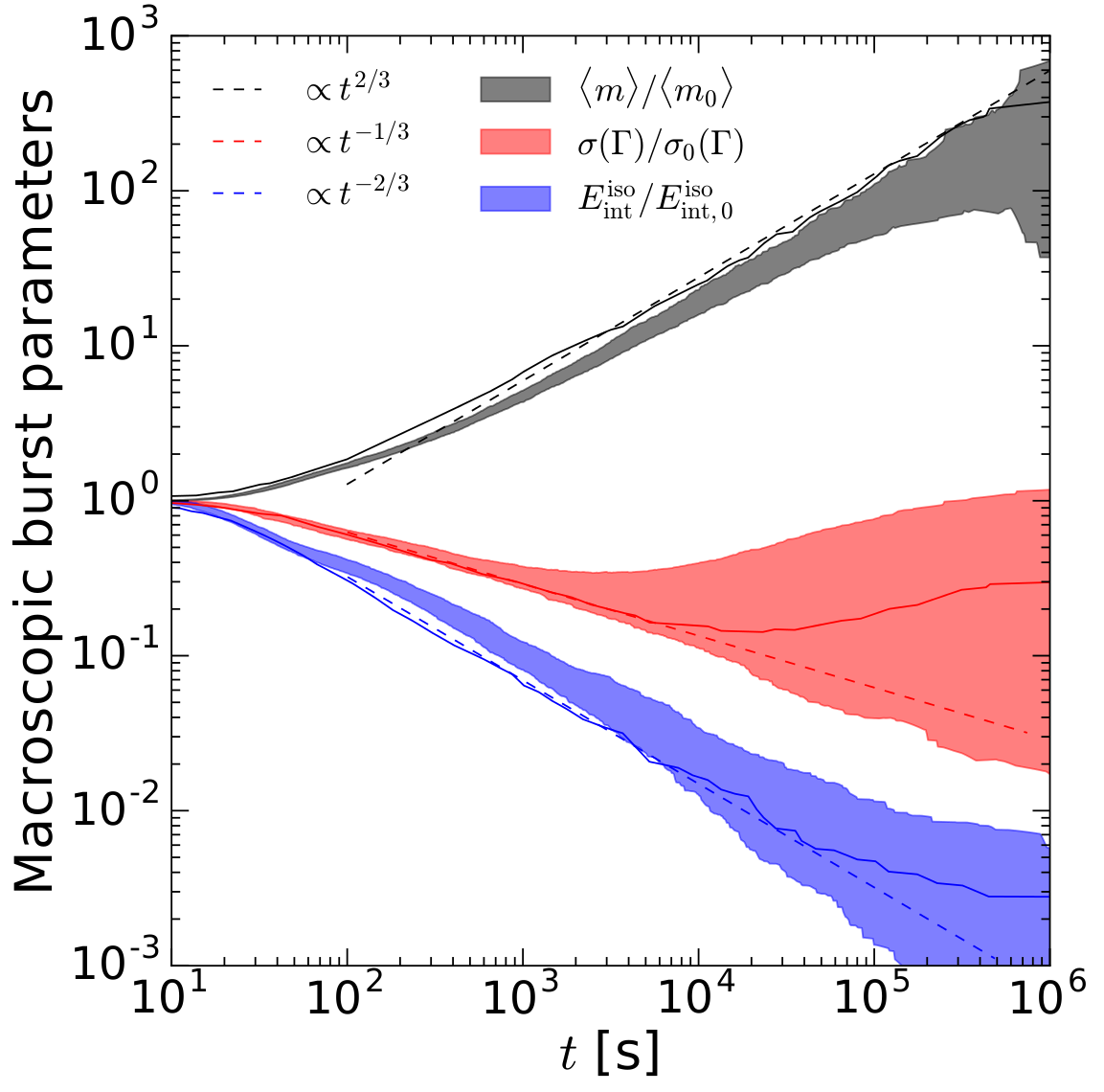}
 \caption{\label{fig:timeevolution}Time evolution, in the source frame, of average shell mass $\langle m \rangle / \langle m_0 \rangle$, standard deviation of the Lorentz factor $\sigma( \Gamma ) / \sigma( \Gamma_0 )$, and total internal energy of a burst, $E_\text{int,tot}^\text{iso}/E_{\text{int,tot},0}^\text{iso}$. The ranges are from our numerical results from a $1000$ simulations run with random setups for the parameter values $N_\text{sh} = 10000$, $\Gamma_0 = 100$, $A_\Gamma = 0.2$, $\delta t_{\text{eng}} = 10^{-3}$ s, $d = l$, $z = 2$, and $E_{\text{kin},0}^\text{iso} = 10^{52}$ erg.
 Solid and dashed lines come from Fig.~1 in~\protect{\citet{Beloborodov:2000nn}} for the same parameter set and refer to numerical calculations and analytical estimates respectively.}
\end{figure}

Figure \ref{fig:timeevolution} shows the time evolution (in the source frame) of macroscopic burst parameters in one of our simulations: average shell mass $\langle m \rangle / \langle m_0 \rangle$ (subscripts of zero indicates values at simulation start), standard deviation of the Lorentz factor\footnote{The reference \citet{Beloborodov:2000nn} refers to this as $\Gamma_\text{rms}$. However we repeated the derivations of their analytical estimates, which are consistent only when interpreting this as the standard deviation.} $\sigma( \Gamma ) / \sigma( \Gamma_0 )$, and total available internal energy of the burst $E_\text{int,tot}^\text{iso}/E_{\text{int,tot},0}^\text{iso}$. The latter is calculated directly as $E_\text{int,tot}^\text{iso} = \Gamma_\text{CM} \sum m_i ( \tilde{\beta}_i^2 / 2 )$, where $\tilde{\beta}$ is the speed in the CM-frame.\footnote{The general relation between the total internal, or free, energy of a gas, $E_\text{int,tot}^\text{iso}$, and its volume is given by $E_\text{int,tot}^\text{iso} \propto V_\text{iso}^{-\left(\gamma-1\right)} \propto r^{-2\left(\gamma-1\right)}$, where $\gamma$ is the adiabatic index. For the relativistic gas in a shell, $\gamma = 4/3$ and $E_\text{int,tot}^\text{iso} \propto r^{-2/3} \propto t^{-2/3}$.} The numerical results of our simulation match the analytical power-law estimates from~\citet{Beloborodov:2000nn}, which assume that fluctuations in the initial Lorentz factors are small, \ie, $A_\Gamma \ll 1$. Deviations occur at late times, when the number of remaining shells is low and the analytical predictions are no longer applicable. This late deviation depends strongly on the random initial setup, so we show ranges obtained after running 1000 different simulations.

Changing the collision dynamics can affect the burst efficiency --- \ie, the fraction of kinetic energy dissipated as secondaries --- and the radii where most of the gamma-ray energy is dissipated. Changes could include re-converting a fraction of collision energy into kinetic energy, partially inelastic collisions, or even fully penetrating shells. We explore simple modifications of our collision model in \App~B.

%%%%%%%%%%%%%%%%%%%%%%%%%%%%%%%%%%%%%%%%%%%%%%%%%%%%%%%%%%%%%%%%%%%%%%%%%%%%%%%%%%%

\subsection{Gamma-ray observables}
\label{section:LightCurves}

The internal energy of a merged shell is split among electrons, magnetic field, and protons. They receive a fraction $\epsilon_e$, $\epsilon_B$, and $\epsilon_p$, respectively. We assume energy equipartition between electrons and photons and fix $\epsilon_e = \epsilon_B = 1/12$ and $\epsilon_p = 5/6$, since this yields the frequently used value of baryonic loading $1/f_e = \epsilon_p/\epsilon_e = 10$~\citep{Waxman:1998yy,Abbasi:2012zw}. Thus, the $k$-th collision dissipates an energy $E_{\gamma,k} = \epsilon_e E_{\text{coll},k}$ as gamma rays, and energy $E_{\gamma,p} = \epsilon_p E_{\text{coll},k}$ as protons, and supports a magnetic energy density of $U_B = \epsilon_B  E_{\text{coll},k} / V_\text{iso}$. The latter translates into a magnetic field intensity, in the shock rest frame, of
\begin{equation}
 B^\prime
 \simeq
 44.7
 \left( \frac{ \Gamma_m } {10^{2.5}} \right)^{-1}
 \left( \frac{ \epsilon_B } { \epsilon_e } \right)^{ \frac{1}{2} }
 \left( \frac{ E_{\gamma,k} } { 10^{50}~\text{erg} } \right)^{ \frac{1}{2} }
 \left( \frac{ R_\text{C} } { 10^9~\text{km} } \right)^{ -1 }
 \left( \frac{ l_\text{m} } { 10^3~\text{km} } \right)^{ -\frac{1}{2} }
 ~\text{kG}~.
\end{equation}

We normalize these individual energies by requiring that the total dissipated energy as gamma rays, an experimentally accessible quantity,
\begin{equation}
 E_{\gamma,{\rm tot}}^{\mathrm{iso}} \equiv \sum_{k=1}^{N_\text{coll}} E_{\gamma,k}^{\rm iso}  \; ,
 \label{equ:fgamma}
\end{equation}
matches a given value $E_{\gamma,\text{norm}}^\text{iso} = 10^{53}$ erg.
This also fixes the energy in protons and magnetic field.

Our simulation does not generate the photon spectrum in the shell. Instead, as in~\citet{Aoi:2009ty,Abbasi:2012zw}, we assume that its shape is that of observed GRB spectra. We parametrize the spectrum as a broken power law, \ie,
\begin{equation}\label{equ:PhotonSpectrum}
 n_\gamma^\prime\left(\varepsilon^\prime\right) \equiv
 \frac{dn_\gamma^\prime}{d\varepsilon^\prime} = C_\gamma^\prime
 \left\{\begin{array}{ll}
  \left(\varepsilon^\prime/\varepsilon_{\gamma,\text{break}}^\prime\right)^{-\alpha_\gamma} & ,\;\; \varepsilon^\prime < \varepsilon_{\gamma,\text{break}}^\prime    \\
  \left(\varepsilon^\prime/\varepsilon_{\gamma,\text{break}}^\prime\right)^{-\beta\gamma}  & ,\;\; \varepsilon^\prime \geq \varepsilon_{\gamma,\text{break}}^\prime
 \end{array}\right. \, .
\end{equation}
Primed quantities are in the shock rest frame. We fix\footnote{It is uncertain how $\varepsilon_{\gamma,\text{break}}^\prime$ changes with $R_\text{C}$, since the scaling expected in the internal shock model has a problem~\citep{Daigne:2003tp}, which is why we prefer to set the photon spectra to observations. However, depending on the origin of the prompt gamma-ray emission --- \eg, synchrotron radiation, inverse Compton scattering --- one can implement model-specific assumptions, as in~\citet{Guetta:2000ye,Guetta:2001cd} (the models therein are already in tension with GRB neutrino searches by IceCube~\citep{Abbasi:2012zw,Aartsen:2014aqy}, though). To correctly calculate the photon break energy for each collision, one needs a time-dependent radiative code, as implemented in~\citet{Bosnjak:2008bd}.  Detailed information on mildly relativistic collisionless physics, such as the injection Lorentz factor (that depends on the number fraction of accelerated particles~\citep{Eichler:2005ug}), is also necessary.  However, such an improvement will not change our conclusions in \Sec~\ref{sec:Conclusions}.} $\alpha_\gamma = 1$, $\beta_\gamma = 2$, and $\varepsilon_{\gamma,\text{break}}^\prime = 1$ keV. 

The photon spectrum in each shell is normalized via
\begin{equation}
 V_{\text{iso},k}^\prime \int_{\varepsilon_{\gamma,\min}^\prime}^{\varepsilon_{\gamma,\max}^\prime} d\varepsilon^\prime \varepsilon^\prime n_\gamma^\prime\left(\varepsilon^\prime\right) = \frac{E_{\gamma,k}^{\text{iso}}}{\Gamma_{\text{m},k}} \, ,
\label{equ:norm}
\end{equation}
where the minimum and maximum energies are $\varepsilon_{\gamma,\min}^\prime = 0.2 \, \mathrm{eV}$ and $\varepsilon_{\gamma,\max}^\prime = 1 \, \mathrm{PeV}$, respectively~\citep{Aoi:2009ty}. Pair production via $\gamma \gamma \to e^+ e^-$ may limit the maximum energy of escaping photons; see \figu{emaxgamma}, right column.

Each collision emits a gamma-ray pulse. The superposition of all pulses, propagated to Earth, is the light curve of the burst; see \Sec~\ref{sec:SyntheticLightCurves}. Following~\citet{Kobayashi:1997jk}, we parametrize the luminosity of the pulse from the $k$-th collision (in the observer's frame) as a peaked profile, with a fast rise and exponential decay (``FRED''), \ie,
\begin{equation}\label{equ:FRED}
 L_{\gamma,k} \left( t_\text{obs} \right) =
 \left\{\begin{array}{ll}
  0                                                                                                & ,\; t_\text{obs} < 0      \\
  h_k \left[ 1 - \frac{ 1 }{ \left( 1 + c t / R_{\text{C},k} \right)^2} \right]                             & ,\; 0 \leq t_\text{obs} < t_{\text{rise},k}           \\
  h_k \left\{ \frac{ 1 }{ \left[ 1 + \left( t - \delta t_{\text{em},k} \right)c/R_{\text{C},k} \right]^2 }  \right. \\
 \left.        \quad     - \frac{ 1 }{ \left( 1 + c t / R_{\text{C},k} \right)^2 }  \right\}              & ,\; t_\text{obs} \geq t_{\text{rise},k}
 \end{array}\right. 
\end{equation} 
where the emission timescale, \ie, the time at which the reverse shock crosses the fast shell, is
\begin{equation}\label{equ:CrossTime}
 \delta t_{\text{em},k} \equiv \frac{l_{\text{m},k}}{c\left(\beta_\text{f}-\beta_\text{rs}\right)} 
\end{equation}
and the ``rise time'',
\begin{equation}
 t_{\text{rise},k} \equiv \frac{\delta t_{\text{em},k}}{2 \Gamma_{\text{m},k}^2} \left(1+z\right) \; ,
\label{equ:trise}
\end{equation}
is the time elapsed since the start of the emission until the peak luminosity is reached. For an illustration, see \Fig~1 in~\citet{Kobayashi:1997jk}. The peak luminosity is 
\begin{equation}
 h_k = \frac{E_{\gamma,k}^\text{iso}}{1+z} \frac{1}{t_{\text{rise},k}} \;.
\end{equation}
The time $t$ in the source frame is related to $t_\text{obs}$ through
\begin{equation}
 t = \frac{2\Gamma_{\text{m},k}^2 t_\text{obs}}{1+z} \;\; .
\end{equation}
Hence, the synthetic light curve $L_\gamma$ is
\begin{equation}\label{equ:LightCurveDef}
 L_{\gamma}\left(t_\text{obs}\right)= \sum_{k=1}^{N_\text{coll}} L_{\gamma,k}\left(t_\text{obs}\right) \;.
\end{equation}
Unless noted otherwise, we only show the light curves for collisions beyond the photosphere (see \Sec~\ref{sec:SyntheticLightCurves}), while we use all collisions for the normalization in \equ{fgamma}. For the examples we show, the fraction of energy dissipated below the photosphere is around $50\%$, so that the total super-photospheric energy output in gamma rays is around half of $E_{\gamma,\text{norm}}^\text{iso}$.

\begin{figure*}[t]
 \centering
 \includegraphics[width=0.7\textwidth]{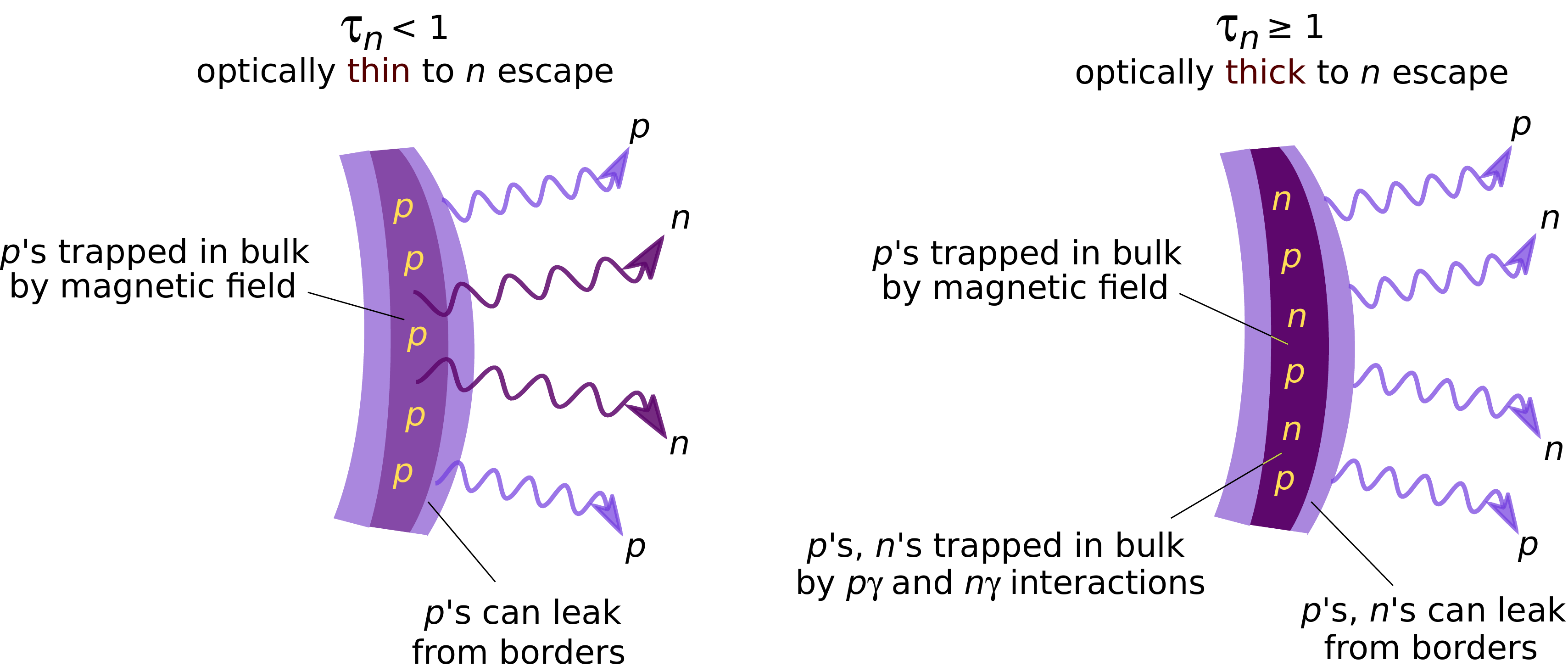}
 \caption{\label{fig:schematic_emission} Schematic illustration of the UHECR emission from a merged shell produced in a collision, when the shell is optically thin (left) and thick (right) to $p\gamma$ and $n\gamma$ interactions.}
\end{figure*}

In \Sec~\ref{sec:delay} we showed light curves in different energy bands. Each band spans the range $\left[ E_{\gamma,\text{low}}^\text{band} , E_{\gamma,\text{high}}^\text{band} \right]$, with $E_{\gamma,\text{low}}^\text{band} = 10^{-6}$, $10^{-1}$, $10^2$ GeV, and $E_{\gamma,\text{high}}^\text{band} = 10^{-2}$, $10^{2}$, $10^6$ GeV, for {\it Fermi}-GBM, {\it Fermi}-LAT, and CTA, respectively. To compute the gamma-ray contribution of the $k$-th collision to each band, we calculate the fraction
\begin{equation}
 f_k^\text{band} =
 \left\{\begin{array}{ll}
  0 & ,\;\; E_{\gamma,\max,k} < E_{\gamma,\text{low}}^\text{band}    \\
  1 & ,\;\; E_{\gamma,\max,k} > E_{\gamma,\text{high}}^\text{band}    \\
  \frac{ \log(E_{\gamma,\max,k}) - \log(E_{\gamma,\text{low}}^\text{band})} {\log(E_{\gamma,\text{high}}^\text{band})-\log(E_{\gamma,\text{low}}^\text{band})} & ,\;\; E_{\gamma,\text{low}}^\text{band} \leq E_{\gamma,\max,k} \leq E_{\gamma,\text{high}}^\text{band}    \\
 \end{array}\right. \, ,
\end{equation}
where $E_{\gamma,\max,k}$ is the maximum gamma-ray energy emitted by the collision. The light curve for each band is then simply calculated using \equ{LightCurveDef}, replacing $L_{\gamma,k} \to f_k^\text{band} L_{\gamma,k}$. Note that this simplified treatment assumes that the product of instrument response times flux is approximately constant within the anticipated energy band --- which is typically a good estimate within the energy bands the instruments are designed for.  A more detailed model for the instrument response or a different shape of the target photon spectrum at TeV energies --- which we assumed to be a power law beyond the break --- will not affect the qualitative shape of the light curves, but may slightly change the relative power in different energy bands or light curve peaks.

The burst duration and variability timescale $t_\text{v}$ are derived from the light curve. For the duration, we use $T_{90}$, the time elapsed between the detection of 5\% and 95\% of the total gamma-ray energy, \ie,
\begin{equation}\label{equ:T90Def}
 T_{90} \equiv t_{95} - t_5 \;\;,
\end{equation}
where
\begin{equation}
 \int_0^{t_f} L_\gamma \left( t_\text{obs} \right) dt_\text{obs} = f \,  \frac{E_{\gamma,\text{tot}}^\text{iso}}{1+z} 
\end{equation}
and $f = 0.05$ or $0.95$. The variability timescale is estimated as
\begin{equation}
 t_\text{v} = T_{90} / N_\text{coll} \, .
\end{equation}
This procedure yields values of $t_\text{v}$ close to $\delta t_{\text{eng}}$.

In some cases (\eg, \figu{neutrinos}), we have compared simulation results to ``standard'' estimators from the one-zone model:
\begin{eqnarray}
 t_\text{v}^\text{std} &=& \frac{d+l}{c} \left(1+z\right)  \; ,\label{equ:TVariabilityStd} \\
 T^\text{std} &=& N_\text{coll} \, t_\text{v}^\text{std} \; ,\label{equ:T90Std} \\
 \Gamma_\text{m}^\text{std} &=& \langle \Gamma_\text{m} \rangle \; ,\\
 R_\text{C}^\text{std} &=& 2 \left(\Gamma_\text{m}^\text{std}\right)^2 \frac{c t_\text{v}^\text{std}}{1+z} \; . \label{equ:RcollStd} 
\end{eqnarray}

%%%%%%%%%%%%%%%%%%%%%%%%%%%%%%%%%%%%%%%%%%%%%%%%%%%%%%%%%%%%%%%%%%%%%%%%%%%%%%%%%%%

\subsection{Neutrinos and cosmic rays}
\label{sec:neutrinos}

In analogy to gamma rays, the proton and neutrino spectra of the complete burst are obtained by summing over the spectra emitted by all the individual collisions. 

We compute secondary particle production using the NeuCosmA software~\citep{Baerwald:2011ee,Hummer:2011ms,Baerwald:2013pu}.
This assumes a proton density $n_p^\prime \propto E^{\prime -2}_p \exp\left( -E_p^\prime / E^\prime_{p,\max} \right)$, with the maximum proton energy $E^\prime_{p,\max}$ obtained from balancing the acceleration rate (with perfect efficiency) with synchrotron, adiabatic, and photohadronic energy loss rates~\citep{Baerwald:2013pu}.
The proton density is normalized like the photon density, \equ{norm}, but replacing $E_{\gamma,k}^\text{iso} \to E_{\gamma,k}^\text{iso} / f_e$. Secondary pion, muon, and kaon spectra, and, consequently, neutrino spectra are computed as in~\citet{Hummer:2011ms}, which includes magnetic field effects on the secondaries, state-of-the-art normalization of the spectra, helicity-dependent muon decays, and flavor mixing. 

Following~\citet{Baerwald:2013pu}, UHECRs escape from each shell as neutrons, produced in photohadronic interactions (``neutron escape''), and as protons that leak out of the shell when their Larmor radius exceeds the shell thickness (``direct escape''). At the highest energies, protons can always leak out --- provided their maximum energy is limited by adiabatic cooling. However, which escape component dominates in each shell depends on the optical depth to photohadronic interactions of the shell in question. \figu{schematic_emission} is a schematic illustration of the components contributing to UHECR emission depending on the optical depth of the merged shell. (If magnetic fields decay fast enough in the bulk of the shell, the direct escape component may be larger. However, this possibility is not contemplated in our framework.)

Our main results, \eg, our light curves and neutrino spectra, are based only on shell collisions that occurred above the photosphere. Below it, Thomson scattering off electrons keeps the photons trapped in the shell. Since our adopted photon spectra are chosen to reproduce observed gamma-ray spectra, we cannot accurately calculate secondary production below the photosphere, where the photon spectra might be different. We  mark ``sub-photospheric'' collisions clearly as such (see, \eg, Figs.~\ref{fig:optdepth} and \ref{fig:scatter}) and exclude them from our flux calculations. Excluding sub-photospheric collisions does not qualitatively change the shape of the light curves.  However, in cases with broader pulse structures, the onset of each pulse is usually dominated by sub-photospheric collisions. Excluding these collisions slightly delays the onset of each peak.

The optical depth to Thomson scattering is calculate from shell properties. Since shells are, on average, electrically neutral, the electron density is equal to the proton density, \ie,
\begin{equation}\label{equ:nbaryonic}
 n'_{e,k} \simeq \frac{m_k}{m_p  \, V'_{\text{iso},k}} \; ,
\end{equation}
where $m_p$ is the proton mass. This assumes that electron-positron pair production does not increase the electron and positron densities significantly. The optical depth to Thomson scattering is then
\begin{equation}\label{equ:lph}
 \tau'_{\text{Th},k} \simeq \frac{1}{n'_{e,k} \, \sigma_\text{Th} \, l'_k} \; ,
\end{equation}
with $\sigma_\text{Th} \approx 66.52$ fm$^2$ the Thomson cross section. A collision is sub-photospheric if $\tau'_{\text{Th}} > 1$. 

For a burst with $E_{\gamma,\text{tot}}^{\text{iso}} \simeq 10^{53} \, \mathrm{erg}$, $\epsilon_e = 1/12$, and a dissipation efficiency of $\varepsilon = 25\%$ (such as GRB 1), the initial kinetic energy per shell is about $10^{51.6} \, \text{erg}$ if 1000 collisions occur. This yields $m_{k,0} = E_{\text{kin},0}^{\text{iso}}/\Gamma_{k,0} c^2 \simeq 10^{49} \, \text{erg}$ for $\Gamma_{k,0} \sim 500$. From \eqs~(\ref{equ:nbaryonic}) and (\ref{equ:lph}), the photospheric radius of the $k$-th shell is
\begin{equation}\label{equ:rph}
 R_{\text{ph},k} \simeq 1.8 \cdot 10^8 \, \mathrm{km} \, \left( \frac{m_k}{10^{49} \, \text{erg}} \right)^{1/2} \; .
\end{equation}

\citet{Bustamante:2014oka} showed that, since the photohadronic optical depth scales similarly to \equ{lph} (replacing $\sigma_\text{Th} \to \sigma_{p\gamma}$), the pion production efficiency at the photosphere is independent of the isotropic volume and only weakly dependent on $\Gamma_k$. The total neutrino flux of the burst is dominated by a few collisions that occur just above the photosphere, where the pion production efficiency is highest.

%%%%%%%%%%%%%%%%%%%%%%%%%%%%%%%%%%%%%%%%%%%%%%%%%%%%%%%%%%%%%%%%%%%%%%%%%%%%%%%%%%%
%%%%%%%%%%%%%%%%%%%%%%%%%%%%%%%%%%%%%%%%%%%%%%%%%%%%%%%%%%%%%%%%%%%%%%%%%%%%%%%%%%%

\section{Additional simulations}
\label{sec:sim_add}

\begin{table*}[t]
 \centering
 \caption{\label{tab:input_add}Description of simulated GRBs 7--10}
 \begin{tabular}{cccccccccp{7.7cm}}
  \hline
  \hline
  Model & $\Gamma_{0,1}$ & $A_{\Gamma,1}$ & $\Gamma_{0,2}$ & $A_{\Gamma,2}$ & $T_p$ &  $N_\text{up}$ & $N_\text{down}$ & $E_{\gamma,\text{norm}}^\text{iso}$ [erg] & Description \\
  \hline
  7  & 5   & \--- & 1500 & \--- & \--- & \--- & \--- & $10^{53}$ & Box-like $\Gamma$ distribution \\
  8  & 50  & 0.1  & 500  & 1.0  & \--- & \--- & \--- & $10^{53}$ & Linear speedup of $\Gamma$ \\
  9  & 500 & 1.0  & \--- & \--- & \--- & 100  & 100  & $10^{52}$ & 100 pulses separated by 100 pulse-times \\
  10 & 500 & 1.0  & \--- & \--- & \--- & 300  & 350  & $10^{52}$ & 300 pulses separated by 350 pulse-times \\
  \hline
  \hline
 \end{tabular}
 \tablecomments{Common values for all models: $N_\text{sh} = 1000$, $\delta t_{\text{eng}} = 10^{-2}$ s, $d = l = c \cdot \delta t_{\text{eng}}$, $r_\text{min} = 10^3$ km, $r_\text{dec} = 5.5 \cdot 10^{11}$ km, $z = 2$,  $\epsilon_e = \epsilon_B = 1/12$, $\epsilon_p = 5/6$, $\eta = 1.0$ (acceleration efficiency~\protect{\citep{Baerwald:2013pu}}). See \Tab~\ref{tab:ParameterDescription} for an explanation of each parameter.}
\end{table*}

\twocolumngrid

\begin{figure*}[t!]
 \begin{center}
  \includegraphics[width=\columnwidth]{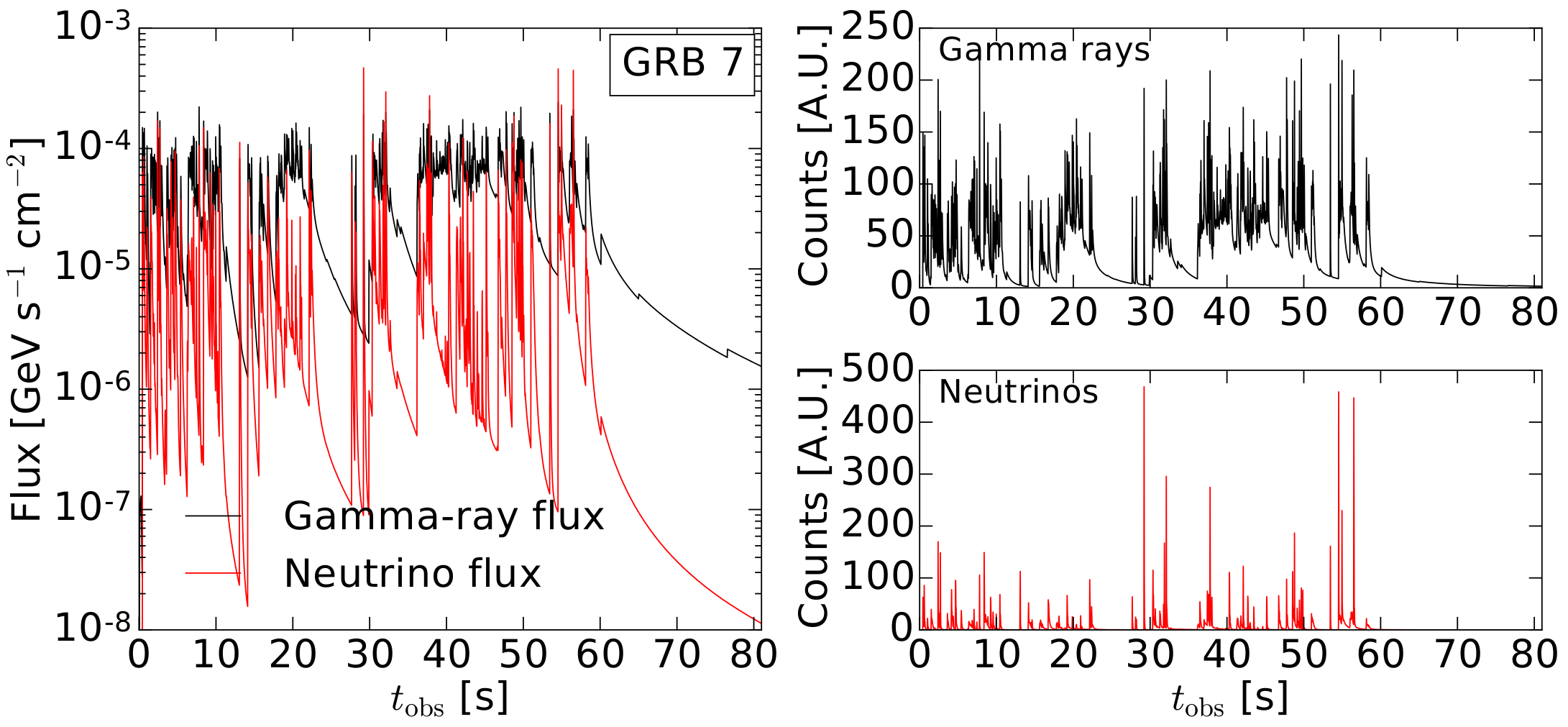}
  \includegraphics[width=\columnwidth]{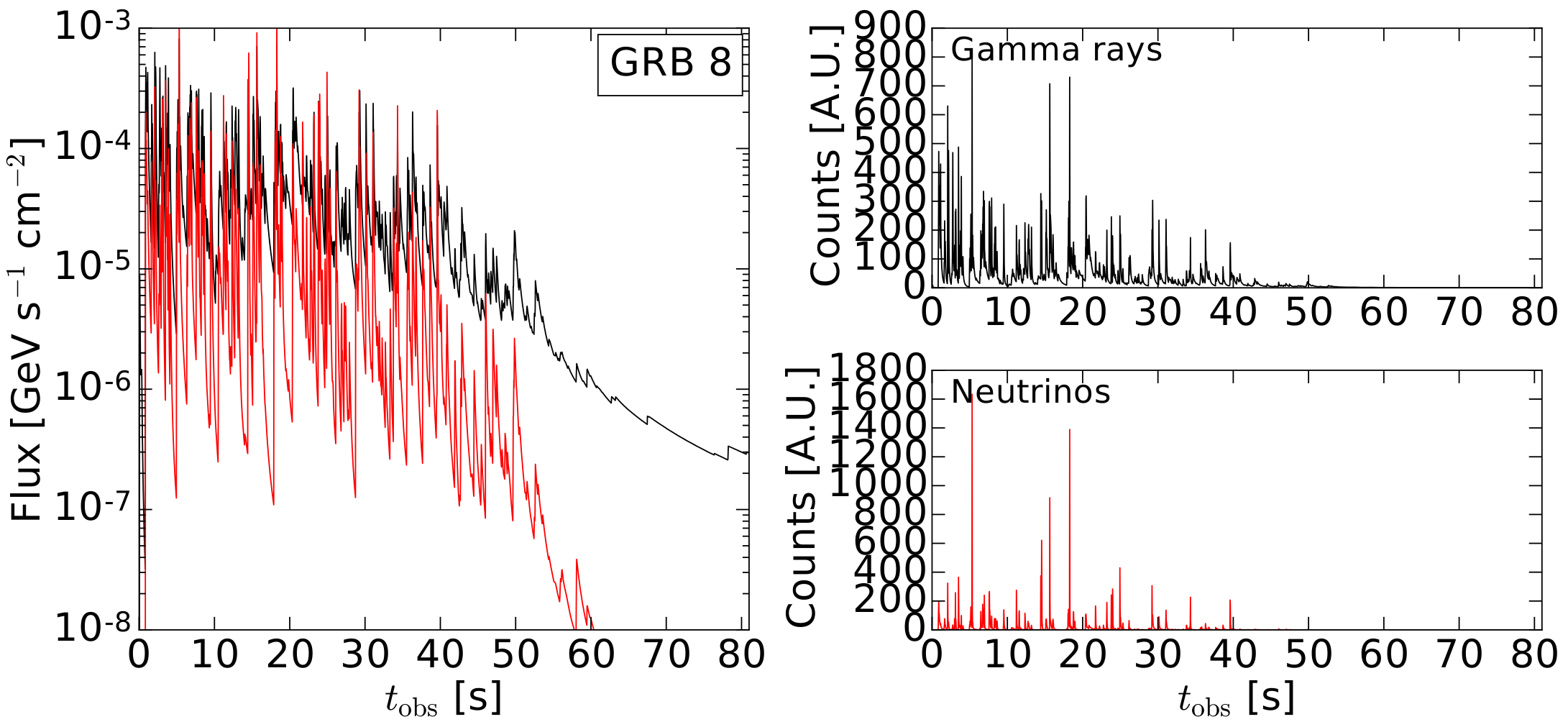}
  \includegraphics[width=\columnwidth]{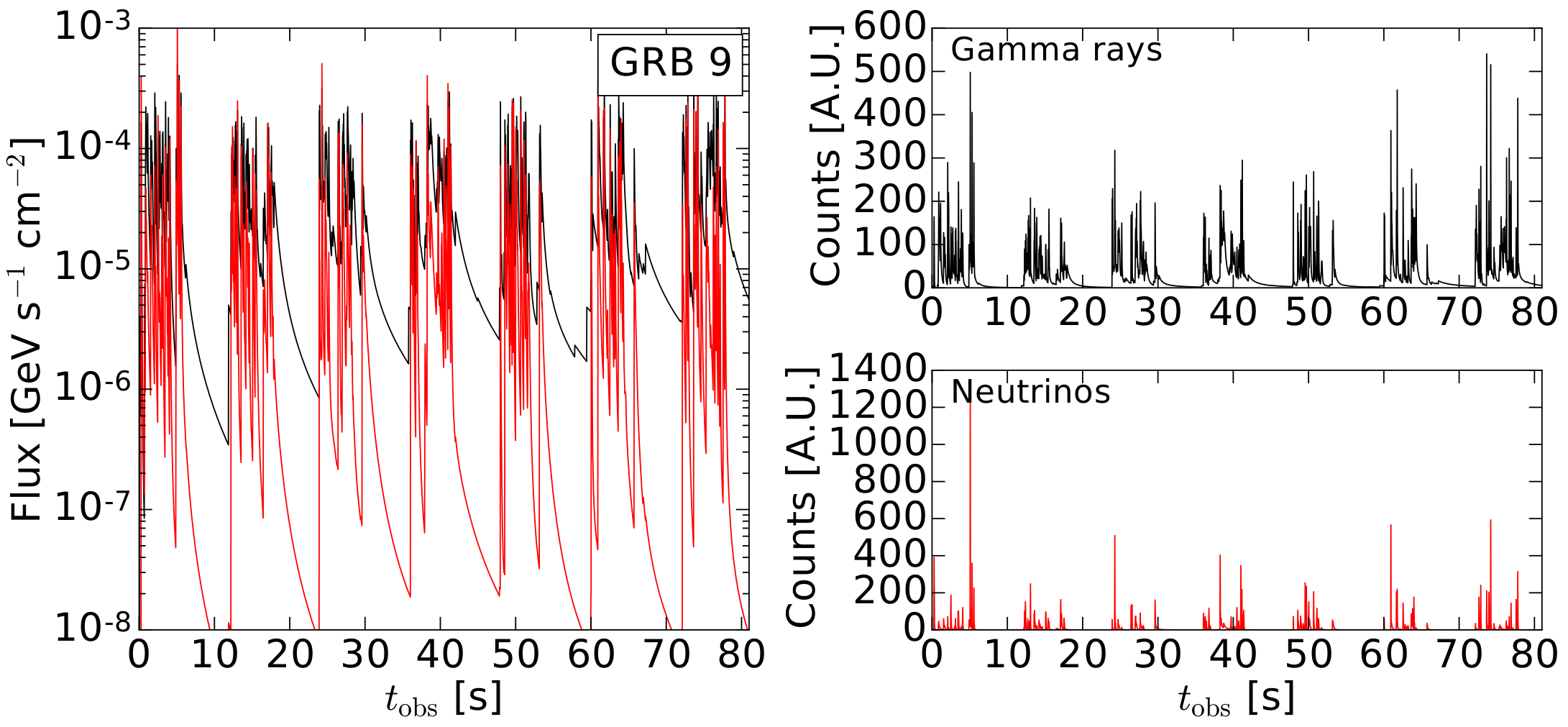}
  \includegraphics[width=\columnwidth]{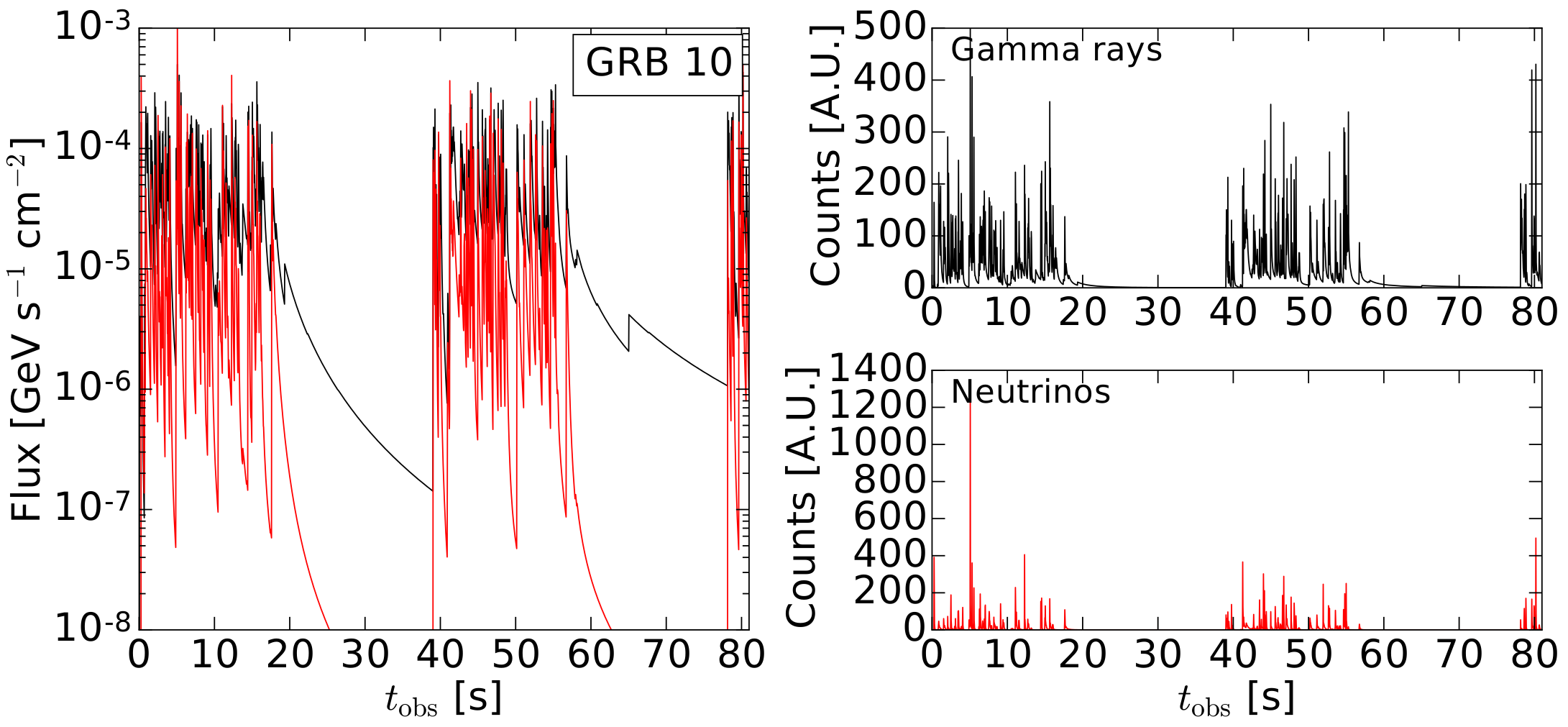}
  \end{center}
 \caption{\label{fig:lcurves_add} Synthetic gamma-ray and neutrino light curves for the simulated GRBs 7--10, from collisions beyond the photosphere. Photon counts are in arbitrary units, obtained by multiplying the flux times a factor of $10^6$ GeV$^{-1}$ cm$^2$ s.}
\end{figure*}

\onecolumngrid

Here we include four additional examples of GRB simulations, GRBs 7--10, that complement the ones showed in the main text.  Table\ \ref{tab:input_add} describes the simulations.  Figure\ \ref{fig:lcurves_add} shows the corresponding light curves.

The light curves of GRB 7 and GRB 8 are similar to that of GRB 1; see \figu{lcurves}.  The light curves of GRB 9 and GRB 10 are similar to that of GRB 4 and GRB 6.

The similarities in the light curves exist in spite of fundamental differences between the behavior of the engine.  This illustrates our point that the qualitative behavior of the examples shown in this work are representative of a larger class of models.

%%%%%%%%%%%%%%%%%%%%%%%%%%%%%%%%%%%%%%%%%%%%%%%%%%%%%%%%%%%%%%%%%%%%%%%%%%%%%%%%%%%
%%%%%%%%%%%%%%%%%%%%%%%%%%%%%%%%%%%%%%%%%%%%%%%%%%%%%%%%%%%%%%%%%%%%%%%%%%%%%%%%%%%

\section{Alternative collision dynamics}
\label{sec:alt}

Here we discuss the impact of modifications to our canonical collision model, which is used in the main text and described in Appendix \ref{sec:model}. We focus on alternative scenarios that can be easily implemented in our framework; that is, we assume that, in each collision, the colliding shells merge and do not consider the case in which they reflect off each other, as in~\citet{Kobayashi:2001iq}. 

One extreme modification is to remove colliding shells from the system after they collide and radiate, which means that multiple collisions are not allowed. This makes simulation results insensitive to details of how shells are treated after colliding. However, removing the shells will modify the whole system, since, in the canonical collision model, collisions among old shells, and between young and old shells, occurred relatively early on.

\begin{figure}[t!]
 \centering
 \includegraphics[width=0.45\columnwidth]{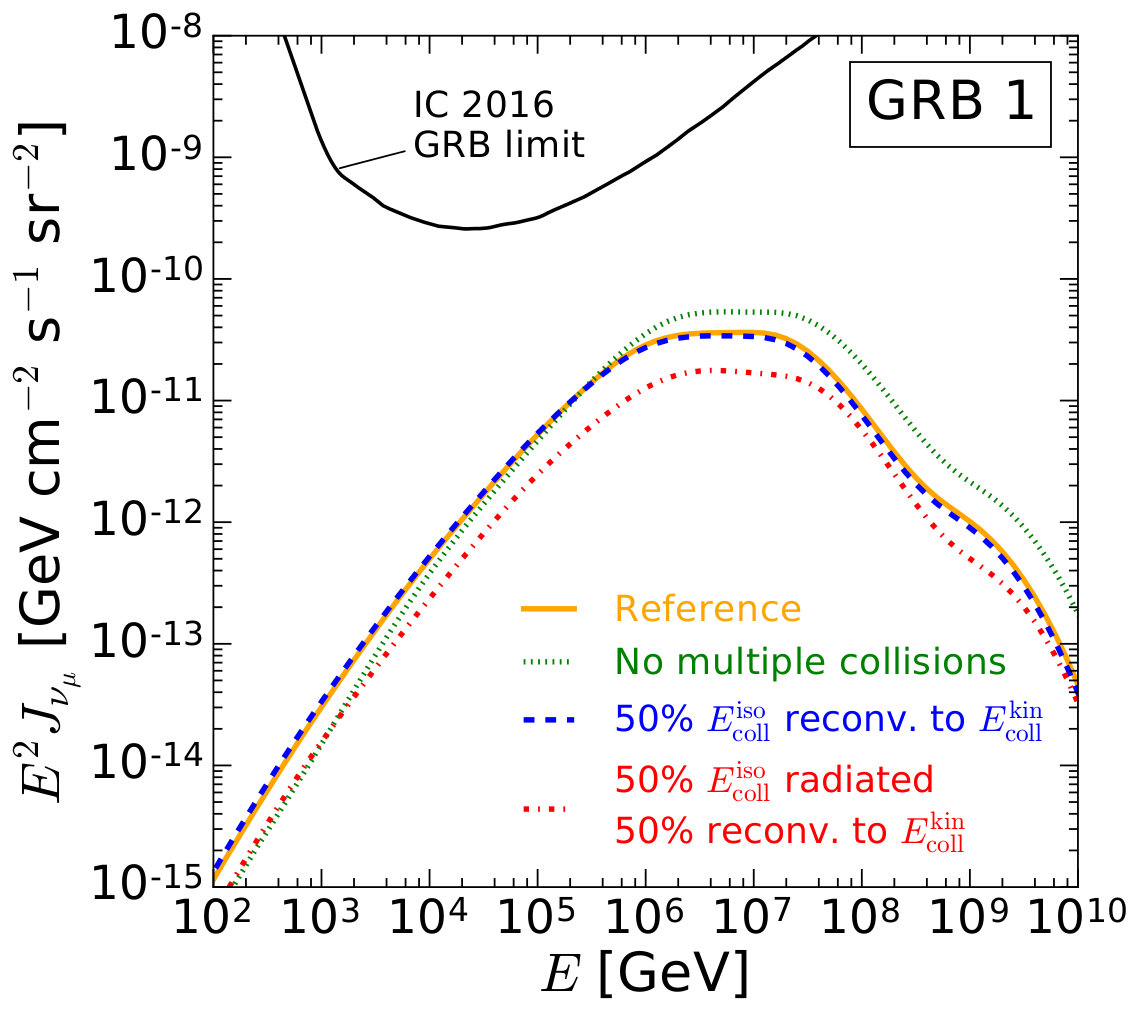}
 \caption{\label{fig:alternative} Three modifications to the canonical collision model described in Appendix \ref{sec:model} and applied to GRB 1. The modified case with no multiplication collisions (green dotted) uses the same per-collision normalization as the reference case (orange, solid). The other two modifications --- 50\% of internal energy reconverted to kinetic energy (blue, dot-dashed) and 50\% radiated with 50\% reconverted (red, dashed) --- are normalized so in each case the burst yields $10^{53}$ erg in gamma rays when adding sub-photospheric and super-photospheric collisions.}
\end{figure}

Figure \ref{fig:alternative} shows the effect on the quasi-diffuse neutrino flux of applying this modification to a simulation that has the same parameters as our reference case, GRB 1. The modified case is labeled ``no multiple collisions''. The collision energies were normalized using $E_{\gamma,\text{norm}}^\text{iso} = 10^{53}$ erg.
The number of collisions is reduced to 499, about half that of the reference GRB 1 model.
The modification results in a higher neutrino flux, because, by forbidding multiple-time collisions, most collisions --- all of them first-time --- occur at low radii, around $10^{8.5}$ km. As we normalize to the same total energy, the per-collision normalization is slightly higher, which further increases the neutrino flux.

Another modification is to assume that only a fraction $\eta$ of the internal energy in each collision, \equ{EintMerged}, is attributed to the non-thermal spectra of the secondaries, while $1-\eta$ is instantly reconverted into kinetic energy of the merged shell. This fraction $1-\eta$ can, for instance, describe a fraction of thermal protons not directly participating in the prompt emission. For simplicity, we assume that the extra kinetic energy translates into an instantaneous increase of the Lorentz factor the merged shell after cooling: $\Gamma_\text{m} = [(1-\eta) E^{\text{iso}}_{\text{coll}} + E_{\text{kin,m}}^{\text{iso}} ] / m_\text{m}$. 

Yet another modification is linked to our assumption that protons can only directly escape the merged shells --- and not leave them by diffusion~\citep{Baerwald:2013pu} or other processes --- which implies that a substantial fraction of the non-thermal baryonic energy will remain trapped by the magnetic fields and eventually reconverted into kinetic energy. The typical fraction $\xi$ of electromagnetic energy and non-thermal baryonic energy which is actually radiated for this escape process is 40--50\%, estimated from energy partition and from the proportion of baryonic energy in the UHECR energy range compared to the full energy range. To account for this, we consider the amount of internal energy used for computing the secondary production is still given by \equ{EintMerged}, but we modify the dynamics so that a fraction of internal energy is reconverted into kinetic energy. This case does not include a fraction of energy going into thermal protons, unlike the previous case.

Figure \ref{fig:alternative} shows the result of both of these modifications to the kinetic and radiated energy, for the case $\eta = \xi =  0.5$. In both cases, the burst was normalized to yield $10^{53}$ erg in gamma rays when adding sub-photospheric and super-photospheric collisions. The number of collisions is similar to that of the reference GRB 1 model. Because the Lorentz factors of the merged shells are higher due to the increased kinetic energy, collisions occur further out in the jet, where particle densities are lower and neutrino production is less efficient. As a result, the neutrino flux associated to these two modifications is slightly lower than the one associated to the reference case, especially if only a fraction $\eta$ of the energy is radiated. Therefore, the minimal super-photospheric flux prediction of $\sim 10^{-11}$ GeV cm$^{-2}$ s$^{-1}$ s$^{-1}$ holds.

The neutrino flux scales with the fraction $\eta$ going into the non-thermal spectra (and magnetic field), which means that, for $\eta = 0.1$, it would be about one order of magnitude lower than our nominal case. On the other hand, the result is rather insensitive to the fraction of reconverted non-thermal energy $1-\xi$. This means that for the combined case --- a fraction $\eta$ into non-thermal spectra and a fraction $1-\xi$ of non-thermal energy reconverted --- we expect that the result is dominated by the effect of $\eta$.

Finally, the three modifications to the collision dynamics that we have explored do not affect our conclusion about the distribution of particle emission with collision radii: neutrinos still come from low radii, UHECR protons come from intermediate radii, and gamma rays come from large radii.

\end{document}